\documentclass[a4paper,UKenglish,cleveref, autoref, thm-restate]{lipics-v2021}

\pdfoutput=1 
\hideLIPIcs  

\usepackage{amsopn}
\usepackage{enumerate}
\usepackage{tikz}
\usetikzlibrary{trees}
\usetikzlibrary{shapes,arrows}
\usetikzlibrary{intersections}
\usepackage[utf8]{inputenc}
\usepackage{amsmath}
\usepackage{amssymb,amsfonts}
\usepackage[T1]{fontenc}
\usepackage{bm}
\usepackage{graphicx}
\usepackage{caption,newfloat}
\usepackage{hyperref}
\usepackage{wrapfig}
\usepackage{blindtext}
\usepackage{array}
\usepackage{makecell}
\usepackage{mathtools}
\usepackage{float}
\usepackage{algpseudocode}
\usepackage[linesnumbered,lined,boxed,commentsnumbered,noend]{algorithm2e}
\SetKwRepeat{Do}{do}{while}%
\usepackage{stmaryrd}
\usepackage{thmtools}
\usepackage{thm-restate}
\usepackage{cleveref}
\usepackage{bussproofs}
\usetikzlibrary{external}
\usepackage{colortbl}
\usepackage{hhline}
\usepackage{multirow}
\usepackage{booktabs}
\newcolumntype{P}[1]{>{\centering\arraybackslash}p{#1}}

\newcommand{\classize}[1]{{\ensuremath{\mathsf{#1}}}\xspace}
\newcommand{\pspace}{\classize{PSPACE}}
\newcommand{\expc}{\classize{EXP}}
\newcommand{\cP}{\classize{P}}
\newcommand{\NP}{\classize{NP}}
\newcommand{\conp}{\classize{coNP}}
\newcommand{\us}{\classize{US}}
\newcommand{\fp}{\classize{FP}}
\newcommand{\dpc}{\classize{DP}}
\newcommand{\fpt}{\classize{FPT}}

\newcommand{\problemize}[1]{{\textsc{#1}}\xspace}

\newcommand{\dnftaut}{\problemize{DNF-Tautology}}

\newcommand{\SAT}{\problemize{Sat}}
\newcommand{\reach}{\problemize{Reachability}}

\newcommand{\logicize}[1]{{\ensuremath{\mathbf{#1}}}\xspace}

\newcommand{\hml}{\logicize{HML}}

\newcommand{\mathL}{\ensuremath{\mathcal{L}}\xspace}
\newcommand{\mathS}{\ensuremath{\mathcal{S}}\xspace}
\newcommand{\mathP}{\ensuremath{\mathcal{P}}\xspace}
\newcommand{\true}{\ensuremath{\mathbf{tt}}\xspace}
\newcommand{\ff}{\ensuremath{\mathbf{ff}}\xspace}

\newcommand{\curle}{\lesssim}
\newcommand{\notcurle}{\not\leq}

\newcommand{\compL}{\ensuremath{\overline{\mathcal{L}}}\xspace}

\newcommand{\eqlen}{\ensuremath{\mathrm{eqlen}}}
\newcommand{\treq}{\ensuremath{\equiv_{trace}}}

\newcommand{\sub}{\ensuremath{\mathrm{Sub}}}

\newcommand{\subcs}{\ensuremath{\mathrm{ConSub_{CS}}}\xspace}
\newcommand{\conscs}{\ensuremath{\mathrm{Cons_{CS}}}\xspace}
\newcommand{\subrs}{\ensuremath{\mathrm{ConSub_{RS}}}\xspace}
\newcommand{\consrs}{\ensuremath{\mathrm{Cons_{RS}}}\xspace}
\newcommand{\algos}{\ensuremath{\mathrm{Prime_S}}\xspace}
\newcommand{\algott}{\ensuremath{\mathrm{Prime^{\diamond}}}\xspace}
\newcommand{\algocs}{\ensuremath{\mathrm{Prime_{CS}}}\xspace}
\newcommand{\algosat}{\ensuremath{\mathrm{Prime^{\mathrm{sat}}}}\xspace}

\newcommand{\algorsu}{\ensuremath{\mathrm{Prime_{RS}^u}}\xspace}
\newcommand{\zero}{\ensuremath{\mathbf{0}}}
\newcommand{\md}{\ensuremath{\mathrm{md}}}
\newcommand{\td}{\ensuremath{\mathrm{td}}}

\newcommand{\depth}{\ensuremath{\mathrm{depth}}}
\newcommand{\sat}{\ensuremath{\textsc{Prop}}}

\newcommand{\enc}{\ensuremath{\mathrm{enc}}}

\newcommand{\simpl}{\ensuremath{\mathrm{simpl}}}

\newcommand{\ruleff}{\ensuremath{\rightarrow_{\mathrm{ff}}}}
\newcommand{\rulezero}{\ensuremath{\rightarrow_{\mathrm{0}}}}
\newcommand{\rulezerosub}{\ensuremath{\rightarrow_{\mathrm{0}}^{sub}}}
\newcommand{\rulezerostar}{\ensuremath{{\rightarrow_{\mathrm{0}}^{sub}}}^*}
\newcommand{\rulett}{\ensuremath{\rightarrow_{\mathrm{tt}}}}
\newcommand{\rulettsub}{\ensuremath{\rightarrow_{\mathrm{tt}}^{sub}}}
\newcommand{\rulettstar}
{\ensuremath{{\rightarrow_{\mathrm{tt}}^{sub}}}^*}
\newcommand{\rulediam}{\ensuremath{\rightarrow_{\mathrm{\diamond}}}}
\newcommand{\rulediamsub}{\ensuremath{\rightarrow_{\mathrm{\diamond}}^{sub}}}
\newcommand{\rulediamstar}
{\ensuremath{{\rightarrow_{\mathrm{\diamond}}^{sub}}}^*}

\newcommand{\reacha}{\ensuremath{\textsc{Reach}_{a}}\xspace}

\newcommand\myarrowa{\mathrel{\stackrel{\makebox[0pt]{\mbox{\normalfont \scriptsize $a$}}}{\longrightarrow}}}
\newcommand\myarrowb{\mathrel{\stackrel{\makebox[0pt]{\mbox{\normalfont \scriptsize $b$}}}{\longrightarrow}}}
\newcommand\myarrowc{\mathrel{\stackrel{\makebox[0pt]{\mbox{\normalfont \scriptsize $c$}}}{\longrightarrow}}}
\newcommand\myarrowd{\mathrel{\stackrel{\makebox[0pt]{\mbox{\normalfont \scriptsize $d$}}}{\longrightarrow}}}
\newcommand\myarrowe{\mathrel{\stackrel{\makebox[0pt]{\mbox{\normalfont \scriptsize $e$}}}{\longrightarrow}}}
\newcommand\myarrowf{\mathrel{\stackrel{\makebox[0pt]{\mbox{\normalfont \scriptsize $f$}}}{\longrightarrow}}}
\newcommand\myarrowai{\mathrel{\stackrel{\makebox[0pt]{\mbox{\normalfont \scriptsize $a^i$}}}{\longrightarrow}}}
\newcommand\myarrowasubi{\mathrel{\stackrel{\makebox[0pt]{\mbox{\normalfont \scriptsize $a_i$}}}{\longrightarrow}}}
\newcommand\myarrowaj{\mathrel{\stackrel{\makebox[0pt]{\mbox{\normalfont \scriptsize $a^j$}}}{\longrightarrow}}}
\newcommand\myarrowan{\mathrel{\stackrel{\makebox[0pt]{\mbox{\normalfont \scriptsize $a_n$}}}{\longrightarrow}}}
\newcommand\notmyarrowan{\mathrel{\stackrel{\makebox[0pt]{\mbox{\normalfont \scriptsize $a_n$}}}{\not\rightarrow}}}
\newcommand\myarrowamax{\mathrel{\stackrel{\makebox[0pt]{\mbox{\normalfont \scriptsize $a^{m}$}}}{\longrightarrow}}}
\newcommand\myarrowaiplusone{\mathrel{\stackrel{\makebox[0pt]{\mbox{\normalfont \scriptsize $a^{i+1}$}}}{\longrightarrow}}}
\newcommand\myarrowaiminusone{\mathrel{\stackrel{\makebox[0pt]{\mbox{\normalfont \scriptsize $a^{i-1}$}}}{\longrightarrow}}}

\newcommand\myarrowtau{\mathrel{\stackrel{\makebox[0pt]{\mbox{\normalfont \scriptsize $t$}}}{\longrightarrow}}}
\newcommand\myarrowepsilon{\mathrel{\stackrel{\makebox[0pt]{\mbox{\normalfont \scriptsize $\varepsilon$}}}{\longrightarrow}}}

\newcommand\notmyarrowa{\mathrel{\stackrel{\makebox[0pt]{\mbox{\normalfont \scriptsize $a$}}}{\not\rightarrow}}}
\newcommand\notmyarrowtau{\mathrel{\stackrel{\makebox[0pt]{\mbox{\normalfont \scriptsize $t$}}}{\not\rightarrow}}}

\newcommand\myarrowz{\mathrel{\stackrel{\makebox[0pt]{\mbox{\normalfont \scriptsize $0$}}}{\longrightarrow}}}
\newcommand\myarrowo{\mathrel{\stackrel{\makebox[0pt]{\mbox{\normalfont \scriptsize $1$}}}{\longrightarrow}}}
\newcommand\vartextvisiblespace[1][.5em]{%
  \makebox[#1]{%
    \kern.07em
    \vrule height.3ex
    \hrulefill
    \vrule height.3ex
    \kern.07em
  }
}

\title{The complexity of deciding characteristic formulae in van Glabbeek's branching-time spectrum} 

\titlerunning{The Complexity of Deciding Characteristic Formulae} 

\author{Luca Aceto}{Department of Computer Science, Reykjavik University, Iceland  \\
Gran Sasso Science Institute, L'Aquila, Italy \and \url{https://en.ru.is/the-university/faculty-and-staff/luca/}}{luca@ru.is}
{https://orcid.org/0000-0001-8554-6907}{}

\author{Antonis Achilleos}{Department of Computer Science, Reykjavik University, Iceland  \and \url{https://sites.google.com/view/antonisachilleos} }{antonios@ru.is}
{https://orcid.org/0000-0002-1314-333X}{}

\author{Aggeliki Chalki}
{Department of Computer Science, Reykjavik University, Iceland \and \url{https://aggelikichal.github.io/}}{angelikic@ru.is}{https://orcid.org/0000-0001-5378-0467}
{}

\author{Anna Ing\'olfsd\'ottir}
{Department of Computer Science, Reykjavik University, Iceland \and \url{https://en.ru.is/the-university/faculty-and-staff/annai}}{annai@ru.is}{https://orcid.org/0000-0001-8362-3075}{}

\authorrunning{Aceto et al.} 

\authorrunning{Anonymous Author(s)}

\Copyright{Anonymous Author(s)}



\category{} 

\relatedversion{} 

\funding{This work has been funded by the projects ``Open Problems in the Equational Logic of Processes (OPEL)'' (grant no.~196050), ``Mode(l)s of Verification and Monitorability'' (MoVeMnt) (grant no.~217987), and  ``Learning and Applying Probabilistic Systems'' (grant  no.~206574-051) of the Icelandic Research Fund.}

\nolinenumbers 

\EventEditors{John Q. Open and Joan R. Access}
\EventNoEds{2}
\EventLongTitle{42nd Conference on Very Important Topics (CVIT 2016)}
\EventShortTitle{CVIT 2016}
\EventAcronym{CVIT}
\EventYear{2016}
\EventDate{December 24--27, 2016}
\EventLocation{Little Whinging, United Kingdom}
\EventLogo{}
\SeriesVolume{42}
\ArticleNo{23}

\begin{document}

\fboxsep=8pt\relax
\fboxrule=2pt\relax

\maketitle

\begin{abstract}
Characteristic formulae give a complete logical description of the behaviour of processes modulo some chosen notion of behavioural semantics. They 
allow one to reduce equivalence or preorder checking to model checking, and are exactly the formulae in the modal logics characterizing classic behavioural equivalences and preorders for which model checking can be reduced to equivalence or preorder checking.

This paper studies the complexity of determining whether a formula is characteristic for some 
process in each of the logics providing modal characterizations of the simulation-based semantics in van Glabbeek's branching-time spectrum. Since characteristic formulae in each of those logics are exactly the satisfiable and prime ones, this article presents complexity results for the satisfiability and primality problems, and investigates the boundary between modal logics for which those problems can be solved in polynomial time and those for which they become computationally hard.

Amongst other contributions, this article also studies the complexity of constructing characteristic formulae in the modal logics characterizing simulation-based semantics, both when such formulae are presented in explicit form and via systems of equations. 
\end{abstract}

\section{Introduction}

Several notions of behavioural relations have been proposed in concurrency theory to describe when one process is a suitable implementation of another. Many such relations have been catalogued by van Glabbeek in his seminal linear-time/branching-time spectrum~\cite{Glabbeek01}, together with a variety of alternative ways of describing them including testing scenarios and axiom systems. To our mind, modal characterizations
of behavioural equivalences and preorders are some of the most classic and pleasing results in concurrency theory---see, for instance,~\cite{HennessyM85} for the seminal Hennessy-Milner theorem and~\cite{BrowneCG88,ERPH13,DeNicolaV95,Glabbeek01} for similar results for other relations in van Glabbeek's
spectrum and other settings. By way of example, in their archetypal modal characterization of bisimilarity, Hennessy and Milner have shown in~\cite{HennessyM85} that, under a mild finiteness condition, two processes are bisimilar if, and only if, they satisfy the same formulae in a multi-modal logic that is now often called Hennessy-Milner logic. Apart from its intrinsic theoretical interest, this seminal logical characterization of bisimilarity means that, when two processes are \emph{not} bisimilar, there is always a formula that distinguishes between them. Such a formula describes a reason why the two processes are not bisimilar, provides useful debugging information and can be algorithmically constructed over finite processes---see, for instance,~\cite{BispingJN22,Cleaveland90} and~\cite{MartensG23}, where Martens and Groote show that, in general, computing minimal distinguishing Hennessy-Milner formulae is \NP-hard. 

On the other hand, the Hennessy-Milner theorem seems to be less useful to show that two processes \emph{are} bisimilar, since that would involve verifying that they satisfy the same formulae, and there are infinitely many of those. However, as shown in works such as~\cite{AcetoMFI19,AcetoILS12,BrowneCG88,GrafS86a,SteffenI94}, the logics that underlie classic modal characterization theorems for equivalences and preorders over processes allow one to express \emph{characteristic formulae}. Intuitively, a characteristic formula $\chi(p)$ for a process $p$ gives a complete logical characterization of the behaviour of $p$ modulo the behavioural semantics of interest $\lesssim$, in the sense that any process is related to $p$ with respect to $\lesssim$ if, and only if, it satisfies $\chi(p)$.\footnote{Formulae akin to characteristic ones first occurred in the study of equivalence of structures using first-order formulae up to some quantifier rank. See, for example, the survey paper~\cite{Thomas93} and the textbook~\cite{EbbinghausFT1994}. The existence of formulae in first-order logic with counting that characterize graphs up to isomorphism has significantly contributed to the study of the complexity of the Graph Isomorphism problem---see, for instance, \cite{CaiFI92,KieferSS15}.} Since the formula $\chi(p)$ can be constructed from $p$, characteristic formulae reduce the problem of checking whether a process $q$ is related to $p$ by $\curle$ to a model checking problem, viz.~whether $q$ satisfies $\chi(p)$. See, for instance, the classic reference~\cite{CleavelandS91} for applications of this approach.

Characteristic formulae, thus, allow one to reduce equivalence and preorder checking to model checking. But what model checking problems can be reduced to equivalence/preorder checking ones? 
To the best of our knowledge, that question was first studied by Boudol and Larsen in~\cite{BoudolL92} in the setting of modal refinement over modal transition systems. 
See~\cite{AcetoMFI19,AFEIP11} for other contributions in that line of research. The aforementioned articles showed that characteristic formulae coincide with those that are \emph{satisfiable} and \emph{prime}. (A formula is prime if whenever it entails a disjunction $\varphi_1 \vee \varphi_2$, then it must entail $\varphi_1$ or $\varphi_2$.) Moreover, characteristic formulae with respect to bisimilarity coincide with the formulae that are satisfiable and \emph{complete}~\cite{Achilleos18}. (A modal formula is complete if, for each formula $\varphi$, it entails either $\varphi$ or its negation.)
The aforementioned results give semantic characterizations of the formulae that are characteristic within the logics that correspond to the behavioural semantics in van Glabbeek's spectrum. Those characterizations tell us for what logical specifications model checking can be reduced to equivalence or preorder checking. However, given a specification expressed as a modal formula, can one decide whether that formula is characteristic and therefore can be model checked using algorithms for behavioural equivalences or preorders? And, if so, what is the complexity of checking whether a formula is characteristic? Perhaps surprisingly, those questions were not addressed in the literature until the recent papers~\cite{AcetoAFI20,Achilleos18}, where it is shown that, in the setting of the modal logics that characterize bisimilarity over natural classes of Kripke structures and labelled transition systems, the problem of checking whether a formula is characteristic for some process modulo bisimilarity is computationally hard and, typically, has the same complexity as validity checking, which is \pspace-complete for Hennessy-Milner logic and \expc-complete for its extension with fixed-point operators~\cite{Holmstrom89,Larsen90} and the $\mu$-calculus~\cite{Kozen83}.

The aforementioned hardness results for the logics characterizing bisimilarity tell us that deciding whether a formula is characteristic in bisimulation semantics is computationally hard. But what about the less expressive logics that characterize the coarser semantics in van Glabbeek's spectrum? And for what logics characterizing relations in the spectrum does computational hardness manifest itself? Finally, what is the complexity of computing a characteristic formula for a process? 

The aim of this paper is to answer the aforementioned questions for some of the simulation-based semantics in the spectrum. 
In particular, we study the complexity of determining whether a formula is characteristic modulo the simulation~\cite{Milner71}, complete simulation and ready simulation preorders~\cite{BloomIM95,LarsenS91}, as well as the trace simulation and the $n$-nested simulation preorders~\cite{GV92}. Since characteristic formulae are exactly the satisfiable and prime ones for each behavioural relation in van Glabbeek's spectrum~\cite{AcetoMFI19}, the above-mentioned tasks naturally break down into studying the complexity of satisfiability and primality checking for formulae in the fragments of Hennessy-Milner logic that characterize those preorders. By using a reduction to the, seemingly unrelated, reachability problem in \emph{alternating graphs}, as defined by Immerman in~\cite[Definition 3.24]{Immerman99}, we discover that both those problems are decidable in polynomial time for the simulation and the complete simulation preorders, as well as for the ready simulation preorder when the set of actions has constant size.  On the other hand,  when the set of actions is unbounded (that is, it is an input of the algorithmic problem at hand), the problems of checking satisfiability and primality for formulae in the logic characterizing the ready simulation preorder are \NP-complete and \conp-complete respectively. We also show that deciding whether a formula is characteristic in that setting is \us-hard~\cite{BlassG82} (that is, it is at least as hard as the problem of deciding whether a given Boolean formula has exactly one satisfying truth assignment) and belongs to \dpc, which is the class of languages that are the intersection of one language in \NP and of one in \conp~\cite{PapadimitriouY84}.\footnote{The class \dpc contains both \NP and {\conp}, and is contained in the class of problems that can be solved in polynomial time with an \NP oracle.} 
These negative results are in stark contrast with the positive results for the simulation and the complete simulation preorder, and indicate that augmenting the logic characterizing the simulation preorder with formulae that state that a process cannot perform a given action suffices to make satisfiability and primality checking computationally hard. In passing, we also prove that, in the presence of at least two actions, (1) for the logics characterizing the trace simulation and 2-nested simulation preorders, satisfiability and primality checking are \NP-complete and \conp-hard respectively, and deciding whether a formula is characteristic is \us-hard, (2) for the logic that characterizes the trace simulation preorder, deciding whether a formula is characteristic is fixed-parameter tractable~\cite{DowneyF95}, with the modal depth of the input formula as the parameter, when the size of the action set is a constant, and (3) deciding whether a formula is characteristic in the modal logic for the 3-nested simulation preorder~\cite{GV92} is \pspace-hard. (The proof of the last result relies on ``simulating'' Ladner’s reduction proving the \pspace-hardness of satisfiability for modal logic~\cite{ladner1977computational} using the limited alternations of modal operators allowed by the logic for the 3-nested simulation preorder.)


We also study the complexity of computing characteristic formulae for finite, loop-free processes modulo the above-mentioned simulation semantics. To do so, we consider two different representations for formulae, namely an explicit form, where formulae are given by strings of symbols generated by their respective grammars, and a declarative form, where formulae are described by systems of equations. We prove that, even for the coarsest semantics we consider, such as the simulation and complete simulation preorders, 
computing the characteristic formula in explicit form for a finite, loop-free process cannot be done in polynomial time, unless $\cP=\NP$. On the other hand, the characteristic formula for a process modulo the preorders we study, apart from the trace simulation preorder, 
can be computed in polynomial time if the output is given in declarative form. Intuitively, this is due to the fact that, unlike the explicit form, systems of equations allow for sharing of subformulae and there are formulae for which this sharing leads to an exponentially more concise representation. Finally, in sharp contrast to that result, we prove that, modulo the trace simulation preorder, even if characteristic formulae are always of polynomial declaration size and polynomial equational length, they cannot be efficiently computed unless $\cP=\NP$. In passing, we remark that all the aforementioned lower and upper bounds hold also for finite processes with loops, provided that, as done in~\cite{AcetoILS12,IngolfsdottirGZ87,SteffenI94}, we add greatest fixed points or systems of equations interpreted as greatest fixed points to the modal logics characterizing the semantics we study in this article.

We summarize our results in Table~\ref{tab:results}. We provide their proofs in the technical appendices.

\section{Preliminaries}

In this paper, we model processes as finite, loop-free \emph{labelled transition systems} (LTS). A finite LTS is a triple $\mathS=(P,A,\longrightarrow)$, where $P$ is a finite set of states (or processes), $A$ is a finite, non-empty set of actions and ${\longrightarrow}\subseteq P\times A\times P$ is a transition relation. As usual, we use $p\myarrowa q$ instead of $(p,a,q)\in {\longrightarrow}$. For each $t\in A^*$, we write $p\myarrowtau q$ to mean that there is a sequence of transitions labelled with $t$ starting from $p$ and ending at $q$. An LTS is \emph{loop-free} iff $p\myarrowtau p$ holds only when $t$ is the empty trace $\varepsilon$. A process $q$ is \emph{reachable} from $p$ if $p\myarrowtau q$, for some $t\in A^*$.
We define the \emph{size}  of an LTS $\mathS=(P,A,\longrightarrow)$, denoted by $|\mathS|$, to be $|P|+|{\longrightarrow}|$. The \emph{size of a process} $p\in P$, denoted by $|p|$, is the cardinality of $\mathrm{reach}(p)=\{q~|~ q \text{ is reachable from } p\}$ plus the cardinality of the set $\longrightarrow$ restricted to $\mathrm{reach}(p)$.
We define the set of \emph{initials} of $p$, denoted $I(p)$, as the set $\{a \in A \mid  p\myarrowa p' \text{ for some } p'\in P\}$. We write $p\myarrowa$ if $a\in I(p)$, $p\notmyarrowa$ if $a\not\in I(p)$, and $p\not\rightarrow$ if $I(p) =\emptyset$. A sequence of actions $t\in A^*$ is a trace of $p$ if there is a $q$ such that $p\myarrowtau q$. We denote 
the set of traces of $p$ by $\mathrm{traces}(p)$. 
The \emph{depth} of a finite, loop-free process $p$, denoted by $\depth(p)$, is the length of a longest trace $t$ of $p$.

In what follows, we shall often describe finite, loop-free processes using the fragment of Milner's CCS~\cite{Milner89} given by the following grammar:
\[
p ::= 0 ~\mid~ a.p  ~\mid~  p+p ,
\]
where $a \in A$. For each action $a$ and terms $p,p'$, we write $p \myarrowa p'$ iff
\begin{enumerate}[(i)]
    \item $p =a.p'$ or
    \item $p =p_1+p_2$, for some $p_1,p_2$, and $p_1\myarrowa p'$ or $p_2\myarrowa p'$ holds.
\end{enumerate}


In this paper, we consider the following relations in van Glabbeek's spectrum: simulation, complete simulation, ready simulation, trace simulation, 2-nested simulation, 3-nested simulation, and bisimilarity. Their definitions are given below.

\begin{definition}[\cite{Milner89,Glabbeek01,AcetoMFI19}]\label{Def:beh-preorders}
We define each of the following preorders as the largest binary relation over $P$ that satisfies the corresponding condition.
\begin{enumerate}[(a)]
    \item \emph{Simulation preorder (S):} $p\curle_{S} q$ $\Leftrightarrow$ for all $p\myarrowa p'$ there exists some $q\myarrowa q'$ such that $p'\curle_S q'$.
    \item \emph{Complete simulation (CS):} $p\curle_{CS} q$ $\Leftrightarrow$
    \begin{enumerate}[(i)]
        \item for all $p\myarrowa p'$ there exists some $q\myarrowa q'$ such that $p'\curle_{CS} q'$, and
        \item $I(p)=\emptyset$ iff $I(q)=\emptyset$.        
    \end{enumerate}
     \item \emph{Ready simulation (RS):} $p\curle_{RS} q$ $\Leftrightarrow$
    \begin{enumerate} [(i)]
        \item for all $p\myarrowa p'$ there exists some $q\myarrowa q'$ such that $p'\curle_{RS} q'$, and
        \item $I(p)=I(q)$.        
    \end{enumerate}
    \item \emph{Trace simulation (TS):} $p\curle_{TS} q$ $\Leftrightarrow$
   \begin{enumerate}[(i)]
    \item for all $p\myarrowa p'$ there exists some  $q\myarrowa q'$ such that $p'\curle_{TS} q'$, and
    \item $\mathrm{traces}(p)=\mathrm{traces}(q)$.
    \end{enumerate}
    \item \emph{$n$-Nested simulation ($n$S)}, where $n\geq 1$, is defined inductively as follows: The $1$-nested simulation preorder $\curle_{1S}$ is $\curle_S$, and the $n$-nested simulation preorder $\curle_{nS}$ for $n > 1$ is the largest relation such that $p\curle_{nS} q$ $\Leftrightarrow$ 
    \begin{enumerate}[(i)]
        \item for all $p\myarrowa p'$ there exists some $q\myarrowa q'$ such that $p'\curle_{nS} q'$, and
        \item $q\curle_{(n-1)S} p$. 
    \end{enumerate}
 \item  \emph{Bisimilarity (BS):} $\curle_{BS}$ is the largest symmetric relation satisfying the condition defining the simulation preorder.
\end{enumerate}
\end{definition}

It is well-known that bisimilarity is an equivalence relation and all the other relations are preorders~\cite{Glabbeek01,Milner89}. We sometimes write $p\sim q$ instead of $p\curle_{BS}$. Moreover, we have that ${\sim}\subsetneq{\curle_{3S}}\subsetneq{\curle_{2S}}\subsetneq {\curle_{TS}}\subsetneq {\curle_{RS}}\subsetneq {\curle_{CS}}\subsetneq {\curle_{S}}$---see~\cite{Glabbeek01}.


\begin{definition}[Kernels of the preorders]
    For each $X\in\{S,CS,RS,TS,2S,3S\}$,  the kernel $\equiv_X$ of $\curle_X$ is the equivalence relation defined thus: for every $p,q\in P$, $p\equiv_X q$ iff $p\curle_X q$ and $q\curle_X p$.
\end{definition}

Each relation $\curle_X$, where $X\in\{S,CS,RS,TS,2S,3S,BS\}$, is characterized by a fragment $\mathL_X$ of Hennessy-Milner logic, \hml, defined as follows~\cite{Glabbeek01,AcetoMFI19}.

\begin{definition}\label{def:mathlx}
For $X\in\{S,CS,RS,TS,2S,3S,BS\}$, $\mathL_X$ is defined by the corresponding grammar given below ($a\in A$):
\begin{enumerate}[(a)] 
\item $\mathL_S$: 
$\varphi_S::= ~ \true ~ \mid ~ \ff ~ \mid ~ \varphi_S\wedge \varphi_S~ \mid ~\varphi_S\vee \varphi_S~ \mid ~ \langle a \rangle\varphi_S.$
\item $\mathL_{CS}$:
$\varphi_{CS}::= ~ \true ~ \mid ~ \ff ~ \mid ~ \varphi_{CS}\wedge \varphi_{CS} ~ \mid ~\varphi_{CS}\vee \varphi_{CS}
~ \mid ~ \langle a \rangle\varphi_{CS} ~ \mid ~ \zero$,
where $\zero=\bigwedge_{a\in A} [a]\ff$.
\item $\mathL_{RS}$:
$\varphi_{RS}::= ~ \true ~ \mid ~ \ff ~ \mid ~ \varphi_{RS}\wedge \varphi_{RS} ~ \mid ~\varphi_{RS}\vee \varphi_{RS}
~ \mid ~ \langle a \rangle\varphi_{RS} ~ \mid ~ [a]\ff.$
\item $\mathL_{TS}$: $\varphi_{TS}::= ~ \true ~ \mid ~ \ff ~ \mid ~ \varphi_{TS}\wedge \varphi_{TS}~ \mid ~\varphi_{TS}\vee \varphi_{TS}~ \mid ~ \langle a \rangle\varphi_{TS} ~ \mid ~ \psi_{TS}$, where $\psi_{TS}::=\ff ~ \mid ~ [a]\psi_{TS}$.
\item $\mathL_{2S}$: 
$\varphi_{2S}::= ~ \true ~ \mid ~ \ff ~ \mid ~ \varphi_{2S}\wedge \varphi_{2S} ~ \mid ~\varphi_{2S}\vee \varphi_{2S}
~ \mid ~ \langle a \rangle\varphi_{2S} ~ \mid ~ \neg\varphi_S.$
\item $\mathL_{3S}$: 
$\varphi_{3S}::= ~ \true ~ \mid ~ \ff ~ \mid ~ \varphi_{3S}\wedge \varphi_{3S} ~ \mid ~\varphi_{3S}\vee \varphi_{3S}
~ \mid ~ \langle a \rangle\varphi_{3S} ~ \mid ~ \neg\varphi_{2S}.$
\item $\hml$ ($\mathL_{BS}$):
    $\varphi_{BS}::= ~ \true ~ \mid ~ \ff ~ \mid ~ \varphi_{BS}\wedge \varphi_{BS} ~ \mid ~\varphi_{BS}\vee \varphi_{BS}~ \mid ~ \langle a \rangle\varphi_{BS} ~ \mid ~ [a]\varphi_{BS} ~ \mid ~ \neg\varphi_{BS}.$
\end{enumerate}
\end{definition}

Note that the explicit use of negation in the grammar for $\mathL_{BS}$ is unnecessary. However, we included the negation operator explicitly so that $\mathL_{BS}$ extends syntactically each of the other modal logics presented in Definition~\ref{def:mathlx}. 

Given a formula $\varphi\in\mathL_{BS}$, the \emph{modal depth} of $\varphi$, denoted by $\md(\varphi)$, is the maximum nesting of modal operators in $\varphi$. (See Definition~\ref{def:modal_depth} in Appendix~\ref{Sect:further-prelims} for the formal definition.)

Truth in an LTS $\mathS=(P,A,\longrightarrow)$ is defined via the satisfaction relation $\models$ as follows:
$$\begin{aligned}
    &p\models\true \text{ and } p\not\models\ff;\\
    &p\models \neg\varphi \text{ iff } p\not\models \varphi;\\
    &p\models \varphi\wedge \psi \text{ iff both } p\models \varphi \text{ and } p\models \psi;\\
    &p\models \varphi\vee \psi \text{ iff } p\models \varphi \text{ or } p\models \psi;\\
    &p\models \langle a\rangle\varphi \text{ iff there is 
 some } p\myarrowa q \text{ such that } q\models\varphi;\\
    &p\models [a]\varphi \text{ iff for all } p\myarrowa q \text{ it holds that } q\models\varphi  .
\end{aligned}$$
If $p\models \varphi$, we say that $\varphi$ is true, or satisfied, in $p$. If $\varphi$ is satisfied in every process in every LTS, we say that $\varphi$ is valid.  
Formula $\varphi_1$ entails $\varphi_2$, denoted by $\varphi_1\models \varphi_2$, if every process
that satisfies  $\varphi_1$ also satisfies   $\varphi_2$. Moreover, $\varphi_1$ and $\varphi_2$ are logically equivalent, denoted by $\varphi_1\equiv\varphi_2$, if $\varphi_1\models \varphi_2$ and $\varphi_2\models \varphi_1$.
A formula $\varphi$ is \emph{satisfiable} if there is a process that satisfies $\varphi$. 
Finally, $\sub(\varphi)$ denotes the set of subformulae of formula $\varphi$.

For $\mathL \subseteq \mathL_{BS}$, we define the dual fragment of $\mathL$ to be $\compL=\{\varphi\mid \neg\varphi\in\mathL\}$, where $\neg\true=\ff$, $\neg\ff=\true$, $\neg (\varphi\wedge\psi)=\neg\varphi\vee\neg\psi$, $\neg (\varphi\vee\psi)=\neg\varphi\wedge\neg\psi$, $\neg [a]\varphi=\langle a\rangle \neg \varphi$, $\neg \langle a\rangle\varphi=[a] \neg \varphi$, and $\neg\neg\varphi = \varphi$. It is not hard to see that  $p\models \neg\varphi$ iff $p\not\models\varphi$, for every process $p$.
Given a process $p$, we define  $\mathL(p)=\{\varphi\in\mathL \mid p\models\varphi\}$. A simplification of the Hennessy-Milner theorem gives a modal characterization of bisimilarity over finite processes. An analogous result is true for every preorder examined in this paper.

\begin{theorem}[Hennessy-Milner theorem~\cite{HennessyM85}]\label{Thm:HMT}
    For all processes $p,q$ in a finite LTS, $p\sim q$ iff $\mathL_{BS}(p)=\mathL_{BS}(q)$.
\end{theorem}

\begin{proposition}[\cite{Glabbeek01,AcetoMFI19}]\label{logical_characterizations}
    Let $X\in\{S,CS,RS,TS,2S,3S\}$. Then  $p\curle_X q$ iff $\mathL_X(p)\subseteq \mathL_X(q)$, for all $p,q\in P$.
\end{proposition}

\begin{remark}
Neither $\ff$ nor disjunction are needed in several of the modal characterizations presented in the above result. The reason for adding those constructs to all the logics is that doing so makes our subsequent results more general and uniform. For example, having $\ff$ and disjunction in all logics allows us to provide algorithms that determine whether a formula in a logic $\mathL$ is prime with respect to a sublogic.
\end{remark}

\begin{definition}[\cite{BoudolL92,AFEIP11}]\label{def:prime-formula}
Let $\mathL\subseteq \mathL_{BS}$. 
A formula $\varphi\in \mathL_{BS}$ is \emph{prime in $\mathL$} if for all $\varphi_1,\varphi_2\in \mathL$, $\varphi\models\varphi_1 \vee \varphi_2$ implies $\varphi\models \varphi_1$ or  $\varphi\models \varphi_2$. 
\end{definition}

When the logic $\mathL$ is clear from the context, we say that $\varphi$ is prime. Note that every unsatisfiable formula is trivially prime in $\mathL$, for every $\mathL$.

\begin{example}\label{ex:prime}
    The formula $\langle a\rangle \true$ is prime in $\mathL_S$. Indeed, let $\varphi_1,\varphi_2\in\mathL_S$ and assume that $\langle a\rangle \true\models\varphi_1\vee\varphi_2$. Since $a.\mathtt{0} \models\langle a\rangle \true$, without loss of generality, we have that $a.\mathtt{0} \models\varphi_1$. We claim that $\langle a\rangle\true\models\varphi_1$. To see this, let $p$ be some process such that $p\models\langle a\rangle\true$---that is, a process such that $p \myarrowa p'$ for some $p'$. It is easy to see that $a.\mathtt{0}\curle_S p$. Since $a.\mathtt{0} \models \varphi_1$, Proposition~\ref{logical_characterizations} yields that $p\models\varphi_1$, proving our claim and the primality of $\langle a\rangle\true$. On the other hand, the formula $\langle a\rangle\true\vee\langle b\rangle\true$ is not prime in $\mathL_S$. Indeed, $\langle a\rangle\true\vee\langle b\rangle\true\models \langle a\rangle\true\vee\langle b\rangle\true$, but neither $\langle a\rangle\true\vee\langle b\rangle\true \models\langle a\rangle\true$  nor $\langle a\rangle\true\vee\langle b\rangle\true \models\langle b\rangle\true$ hold. 
\end{example}

The definition of a characteristic formula within logic \mathL is given next.

\begin{definition}[\cite{AILS07,GrafS86a,SteffenI94}]\label{def:characteristic}
Let $\mathL\subseteq \mathL_{BS}$. 
A formula  $\varphi\in\mathL$ is \emph{characteristic  for $p\in P$ within $\mathL$} iff, for all $q \in P$, it holds that $q \models \varphi\Leftrightarrow\mathL(p) \subseteq \mathL(q)$. We denote by $\chi(p)$ the unique characteristic formula for $p$ with respect to logical equivalence.
\end{definition}

\begin{remark}\label{Remark:charforms}
    Let $X\in\{S,CS,RS,TS,2S,3S,BS\}$. In light of Theorem~\ref{Thm:HMT} and Proposition~\ref{logical_characterizations}, a formula $\varphi\in \mathL_X$  is characteristic for $p$ within $\mathL_X$ iff, for all $q \in P$, it holds that $q \models \varphi\Leftrightarrow p\curle_X q$. This property is often used as an alternative definition of characteristic formula for process $p$ modulo $\curle_X$. In what follows, we shall employ the two definitions interchangeably. 
\end{remark}

In~\cite[Table~1 and Theorem~5]{AcetoMFI19}, Aceto, Della Monica, Fabregas, and Ing\'olfsd\'ottir presented characteristic formulae for each of the semantics we consider in this paper, and showed that characteristic formulae are exactly the satisfiable and prime ones. 

\begin{proposition}[\cite{AcetoMFI19}]\label{prop:charact-via-primality}
For every $X\in\{S,CS,RS,TS,2S\}$, $\varphi\in \mathL_X$ is characteristic for some process within $\mathL_X$ iff $\varphi$ is satisfiable and prime in $\mathL_X$.
\end{proposition}

\begin{remark}
Proposition~\ref{prop:charact-via-primality} is the only result we use from~\cite{AcetoMFI19}  and we employ it as a `black box'. The (non-trivial) methods used in the proof of that result given in that reference do not play any role in our technical developments.

We note, in passing, that the article~\cite{AcetoMFI19} does not deal explicitly with $3S$. However, its results apply to all the $n$-nested simulation preorders.
\end{remark}

We can also consider characteristic formulae modulo equivalence relations as follows.

\begin{definition}\label{def:characteristic-equivalence}
Let $X\in\{S,CS,RS,TS,2S,3S,BS\}$. 
    A formula $\varphi\in \mathL_X$ is characteristic for $p\in P$ modulo $\equiv_X$  iff for all $q\in P$, it holds that $q\models \varphi\Leftrightarrow \mathL_X(p)=\mathL_X(q)$.\footnote{The above definition can also be phrased as follows: A formula $\varphi\in \mathL_X$ is characteristic for $p$ modulo $\equiv_X$ iff, for all $q \in P$, it holds that $q \models \varphi\Leftrightarrow p\equiv_X q$. This version of the definition is used, in the setting of bisimilarity, in references such as~\cite{AcetoAFI20,IngolfsdottirGZ87}.}  
\end{definition} 

When studying the complexity of finding a characteristic formula for some process $p$ with respect to the behavioural relations we have introduced above, we will need some way of measuring the size of the resulting formula as a function of $|p|$. A formula in $\mathL_X$, where $X\in\{S,CS,RS,TS,2S,3S,BS\}$, can be given in \emph{explicit form} as in Definition~\ref{def:mathlx} or
by means of a system of 
equations. 
In the latter case, we say that the formula is given in \emph{declarative form}. For example, formula $\phi=\langle a \rangle(\langle a \rangle \true \wedge \langle b \rangle \true)\wedge\langle b \rangle(\langle a \rangle \true \wedge \langle b \rangle \true)$ can be represented by the equations $\phi= \langle a \rangle \phi_1 \wedge\langle b \rangle \phi_1$ and $\phi_1=\langle a \rangle \true \wedge \langle b \rangle \true$. We define:
\begin{itemize}
    \item the \emph{size} of formula $\varphi$, denoted by $|\varphi|$, to be the number of symbols that appear in the explicit form of $\varphi$,
    \item the \emph{declaration size} of formula $\varphi$, denoted by $\mathrm{decl}(\varphi)$, to be the number of equations that are used in the declarative form of $\varphi$, and
    \item the \emph{equational length} of formula $\varphi$, denoted by $\eqlen(\varphi)$, to be the maximum number of symbols that appear in an equation in the declarative form of $\varphi$.
\end{itemize}
For example, for the aforementioned formula $\phi$, we have that $|\phi|=13$, $\mathrm{decl}(\phi)=2$, and $\eqlen(\phi)=5$. Note that $\mathrm{decl}(\varphi) \leq |\sub(\varphi)| \leq |\varphi|$, for each $\varphi$.

\section{The complexity of deciding characteristic formulae modulo preorders}\label{section:decide-characteristic}

In this section, we address the complexity of deciding whether  formulae in $\mathL_S$, $\mathL_{CS}$, $\mathL_{RS}$, $\mathL_{TS}$, $\mathL_{2S}$, and $\mathL_{3S}$ are characteristic.
Since characteristic formulae in those logics are exactly the satisfiable and prime ones~\cite[Theorem~5]{AcetoMFI19}, we 
study the complexity of checking satisfiability and primality separately in Subsections~\ref{section:deciding-satisfiability} and~\ref{section:deciding-primality}.

\subsection{The complexity of satisfiability}\label{section:deciding-satisfiability}

To address the complexity of the satisfiability problem  in $\mathL_S$, $\mathL_{CS}$, or $\mathL_{RS}$, we associate a set $I(\varphi)\subseteq 2^A$ to every formula $\varphi\in\mathL_{RS}$. Intuitively, $I(\varphi)$ describes all possible sets of initial actions that a process $p$ can have, when $p\models \varphi$.

\begin{definition}\label{def:cs-rs-I(phi)}
    Let $\varphi\in\mathL_{RS}$. We define $I(\varphi)$ inductively as follows:
    \begin{enumerate}[(a)]
        \item $I(\true)=2^A$,
        \item $I(\ff)=\emptyset$,
        \item $I([a]\ff)= \{X ~\mid ~ X\subseteq A \text{ and } a\not\in X\}$,
        \item $I(\langle a\rangle \varphi)=\begin{cases}
           \emptyset, &\text{if } I(\varphi)=\emptyset,\\
           \{X ~\mid ~ X\subseteq A \text{ and } a\in X\}, &\text{otherwise}
        \end{cases}$
        \item $I(\varphi_1\vee\varphi_2)=I(\varphi_1)\cup I(\varphi_2)$,
        \item $I(\varphi_1\wedge\varphi_2)=I(\varphi_1)\cap I(\varphi_2)$.
    \end{enumerate}
    Note that $I(\zero)=\{\emptyset\}$.
\end{definition}

\begin{restatable}{lemma}
{Iphiproperty}\label{lem:I(phi)-property}
For every $\varphi\in\mathL_{RS}$, the following statements hold:
\begin{enumerate}[(a)]
    \item  for every $S\subseteq A$, $S\in I(\varphi)$ iff there is a process $p$ such that $I(p)=S$ and $p\models\varphi$.
    \item  $\varphi$ is unsatisfiable iff $I(\varphi)=\emptyset$.
\end{enumerate}
\end{restatable}

When the number of actions is constant, $I(\varphi)$ can be computed in linear time for every $\varphi\in\mathL_{RS}$. For $\mathL_{CS}$, we need even less information; indeed, it is sufficient to define $I(\varphi)$  so that it encodes whether $\varphi$ is unsatisfiable, or is satisfied only in deadlocked states (that is, states with an empty set of initial actions), or is satisfied only in processes that are not deadlocked, or is satisfied both in some deadlocked and non-deadlocked states. This information can be computed in linear time for every $\varphi\in\mathL_{CS}$, regardless of the size of the action set.

\begin{restatable}{corollary}{polysat}\label{cor:sat-s-cs-rs-poly}\quad
    \begin{enumerate}[(a)]
     \item Satisfiability of formulae in $\mathL_{CS}$ and $\mathL_S$ is  decidable in linear time.
     \item Let $|A|=k$, where $k\geq 1$ is a constant. Satisfiability of formulae in $\mathL_{RS}$ is  decidable in linear time.
    \end{enumerate}
\end{restatable}

On the other hand, if we can use an unbounded number of actions, the duality of $\langle a\rangle$ and $[a]$ can be employed to define a polynomial-time reduction from  {\SAT}, the satisfiability problem for propositional logic, to satisfiability in $\mathL_{RS}$. Moreover, if we are allowed to nest $[a]$ modalities ($a\in A$) and have at least two actions, we can encode $n$ propositional literals using formulae of $\log n$ size and reduce \SAT to satisfiability in  $\mathL_{TS}$ in polynomial time. Finally, satisfiability in $\mathL_{2S}$ is in \NP, which can be shown by an appropriate tableau construction.

\begin{restatable}{proposition}{twostssat}\label{prop:sat-rs-ts-2s-np-complete}
Let either $X=RS$ and $|A|$ be unbounded or $X\in\{TS,2S\}$ and $|A|>1$. Satisfiability of formulae in $\mathL_X$ is \NP-complete.
\end{restatable}

Deciding satisfiability of formulae in  $\compL_{2S}$ when $|A|>1$, turns out to be \pspace-complete. (A proof is provided in Appendix~\ref{subsection:3s-satisfiability}.) This means that satisfiability of $\mathL_{3S}$ is also \pspace-complete, since $\compL_{2S}\subseteq\mathL_{3S}$. 

\begin{restatable}{proposition}
{threessat}\label{prop:3-s-sat}
    Let $|A|>1$. Satisfiability of formulae in $\mathL_{3S}$ is \pspace-complete.
\end{restatable}

\subsection{The complexity of primality}\label{section:deciding-primality}

We now study the complexity of checking whether a formula is prime in the logics that characterize some of the relations in Definition~\ref{Def:beh-preorders}. 

\medskip 
\noindent\textbf{Primality in $\mathL_S$.} Unsatisfiable formulae are trivially prime. Note also that in the case that $|A|=1$, all satisfiable formulae in $\mathL_S$ are prime. To address the problem for any action set, for every satisfiable formula $\varphi\in\mathL_S$ we can efficiently compute a logically equivalent formula $\varphi'$  given by the grammar $\varphi::= \true ~|~ \langle a\rangle \varphi ~|~ \varphi\wedge\varphi  ~|~ \varphi\vee\varphi$. We examine the complexity of deciding primality of such formulae.

\begin{table}
\fbox{\begin{minipage}{0.45 \textwidth}
\begin{prooftree}
    \AxiomC{$\varphi_1\vee\varphi_2,\varphi\Rightarrow \psi$ }
    \RightLabel{\scriptsize(L$\vee_1$)}
    \UnaryInfC{$\varphi_1,\varphi\Rightarrow \psi~|_{\forall}~ \varphi_2,\varphi\Rightarrow \psi$}
\end{prooftree}
\vspace{2mm}
\begin{prooftree}
    \AxiomC{$\varphi_1\wedge\varphi_2,\varphi\Rightarrow \langle a\rangle\psi$}
    \RightLabel{\scriptsize(L$\wedge_1$)}
    \UnaryInfC{$\varphi_1,\varphi\Rightarrow \langle a\rangle \psi~|_{\exists}~ \varphi_2,\varphi\Rightarrow \langle a\rangle \psi$}
\end{prooftree}
\vspace{2mm}
\begin{prooftree}
    \AxiomC{$\varphi_1,\varphi_2\Rightarrow\psi_1\wedge \psi_2$ }
    \RightLabel{\scriptsize(R$\wedge$)}
    \UnaryInfC{$\varphi_1,\varphi_2\Rightarrow \psi_1~|_{\forall}~\varphi_1,\varphi_2\Rightarrow \psi_2$}
\end{prooftree}
\vspace{2mm}
\begin{prooftree}
    \AxiomC{$\langle a\rangle \varphi_1, \langle a\rangle \varphi_2 \Rightarrow \langle a\rangle \psi$ }
    \RightLabel{\scriptsize($\diamond$)}
    \UnaryInfC{$\varphi_1,\varphi_2\Rightarrow \psi$}
\end{prooftree}
\end{minipage}
\hfill
\begin{minipage}{0.45\textwidth}
\begin{prooftree}
    \AxiomC{$\varphi,\varphi_1\vee\varphi_2\Rightarrow \psi$ }
    \RightLabel{\scriptsize(L$\vee_2$)}
    \UnaryInfC{$\varphi_1,\varphi\Rightarrow \psi~|_{\forall}~ \varphi_2,\varphi\Rightarrow \psi$}
\end{prooftree}
\vspace{2mm}
\begin{prooftree}
    \AxiomC{$\varphi,\varphi_1\wedge\varphi_2\Rightarrow \langle a\rangle\psi$}
    \RightLabel{\scriptsize(L$\wedge_2$)}
    \UnaryInfC{$\varphi_1,\varphi\Rightarrow \langle a\rangle \psi~|_{\exists}~ \varphi_2,\varphi\Rightarrow \langle a\rangle \psi$}
\end{prooftree}
\vspace{2mm}
\begin{prooftree}
    \AxiomC{$\varphi_1,\varphi_2\Rightarrow\psi_1\vee \psi_2$ }
    \RightLabel{\scriptsize(R$\vee$)}
    \UnaryInfC{$\varphi_1,\varphi_2\Rightarrow \psi_1~|_{\exists}~\varphi_1,\varphi_2\Rightarrow \psi_2$}
\end{prooftree}
\vspace{2mm}
\begin{prooftree}
    \AxiomC{$\varphi_1,\varphi_2\Rightarrow\true$ }
    \RightLabel{\scriptsize(tt)}
    \UnaryInfC{\textsc{True}}
\end{prooftree}
\end{minipage}}   
\vspace{2mm}
    \caption{Rules for the simulation preorder. If $|_{\forall}$ is displayed  in the conclusion of a rule, then the rule is called universal. Otherwise, it is called existential.}
    \label{tab:S-rules}
\end{table}

\begin{restatable}{proposition}{simprime}\label{prop:sim-primality}
    Let $\varphi\in\mathL_S$ such that $\ff\not\in\sub(\varphi)$. Deciding whether $\varphi$ is prime is in \cP.
\end{restatable}
\begin{proof}
We describe algorithm $\algos$ that, on input  $\varphi$, decides primality of $\varphi$. 
$\algos$ constructs a rooted directed acyclic graph, denoted by $G_\varphi$, from the formula $\varphi$ as follows. Every vertex of the graph is either of the form $\varphi_1,\varphi_2\Rightarrow\psi$---where $\varphi_1$, $\varphi_2$ and $\psi$ are sub-formulae of $\varphi$---, or \textsc{True}. The algorithm starts from vertex $x=(\varphi,\varphi\Rightarrow\varphi)$ and applies some rule in Table~\ref{tab:S-rules} to $x$ in top-down fashion to generate one or two new vertices that are given at the bottom of the rule. These vertices are the children of $x$ and the vertex $x$ is labelled with either $\exists$ or $\forall$, depending on which one is displayed at the bottom of the applied rule. If $x$ has only one child, \algos labels it with $\exists$. The algorithm recursively continues this procedure on the children of $x$. If no rule can be applied on a vertex, then this vertex has no outgoing edges. For the sake of clarity and consistency, we assume that right rules, i.e.\ (R$\vee$) and (R$\wedge$),  are applied before the left ones, i.e.\ (L$\vee_i$) and (L$\wedge_i$), $i=1,2$, by the algorithm.
The graph generated in this way 
is an \emph{alternating graph}, as defined by Immerman in~\cite[Definition 3.24]{Immerman99} (see Definition~\ref{def:alternating-graph}). In $G_\varphi$,  the source vertex $s$ is  $\varphi,\varphi\Rightarrow\varphi$, and the target vertex $t$ is \textsc{True}. Algorithm \algos solves the problem {\reacha} on input $G_\varphi$, where \reacha is \reach on alternating graphs and is defined in~\cite[pp. 53--54]{Immerman99}. It accepts $\varphi$ iff \reacha accepts $G_\varphi$. Intuitively, the source vertex $(\varphi,\varphi\Rightarrow \varphi)$ can reach the target vertex \textsc{True} in the alternating graph $G_\varphi$ exactly when for each pair of disjuncts  $\psi_1$ and $\psi_2$ in the disjunctive normal form of  $\varphi$ there is a disjunct  $\psi_3$ in the disjunctive normal form of  $\varphi$ that is entailed by both $\psi_1$ and $\psi_2$. It turns out that this is a necessary and sufficient condition for the primality of $\varphi$.
For example, consider the formula $\langle a\rangle \true\vee\langle b\rangle\true$. There is no disjunct of $\langle a\rangle \true\vee\langle b\rangle\true$ that is entailed by both  $\langle a\rangle \true$ and $\langle b\rangle\true$. This is because that formula is not prime,  as we observed in Example~\ref{ex:prime}. On the other hand, the formula $\langle a\rangle\true\vee\langle a\rangle\langle b\rangle\true$ is prime since each of its disjuncts entails $\langle a\rangle\true$.  The full technical details are included in Appendix~\ref{subsection:sim-primality-appendix}. 
Note that graph $G_\varphi$ is of polynomial size and there is a linear-time algorithm solving {\reacha}~\cite{Immerman99}.
\end{proof}

\noindent\textbf{Primality in $\mathL_{CS}$.} Note that, in the case of $\mathL_{CS}$, the rules in Table~\ref{tab:S-rules} do not work any more because, unlike $\mathL_{S}$, the logic $\mathL_{CS}$ can express some `negative information' about the behaviour of processes. For example, let $A=\{a\}$ and $\varphi=\langle a\rangle \true$. Then, \algos accepts $\varphi$, even though  $\varphi$ is not prime in $\mathL_{CS}$. Indeed, $\varphi\models\langle a\rangle\langle a\rangle \true\vee\langle a\rangle\zero$, but $\varphi\not\models\langle a\rangle\langle a\rangle \true$ and $\varphi\not\models\langle a\rangle\zero$. However, we can overcome this problem as described in the proof sketch of Proposition~\ref{prop:cs-primality} below.

\begin{restatable}{proposition}{csprime}\label{prop:cs-primality}
    Let $\varphi\in\mathL_{CS}$ be a formula such that every $\psi\in\sub(\varphi)$ is satisfiable. Deciding whether $\varphi$ is prime is in \cP.
\end{restatable}
\begin{proof}
   Consider the algorithm that first computes the formula $\varphi^\diamond$ by applying rule $\langle a\rangle\true\rulediam\true$, and rules $\true\vee\psi\rulett\true$ and $\true\wedge\psi\rulett\psi$ modulo commutativity on $\varphi$. It holds that $\varphi$ is prime iff $\varphi^\diamond$ is prime and $\varphi^\diamond\models\varphi$. Next, the algorithm decides primality of $\varphi^\diamond$ by solving reachability on a graph constructed as in the case of simulation using the rules in Table~\ref{tab:S-rules}, where rule (tt) is replaced by rule (0), whose premise is $\zero,\zero\Rightarrow\zero$ and whose  conclusion is \textsc{True}. To verify $\varphi^\diamond\models\varphi$, the algorithm computes a process $p$ for which $\varphi^\diamond$ is characteristic within $\mathL_{CS}$ and checks whether $p\models\varphi$.  In fact, the algorithm has also a preprocessing phase during which it applies a set of rules on $\varphi$ and obtains an equivalent formula with several desirable properties. See Appendix~\ref{subsection:cs-primality-appendix} for full details, where Remark~\ref{rem:type-ordering-rules} comments on the type and ordering of the rules applied.
\end{proof}


\noindent{\textbf{Primality in $\mathL_{RS}$.}} The presence of formulae of the form $[a]\ff$ in $\mathL_{RS}$ means that a prime formula $\varphi\in\mathL_{RS}$ has at least to describe which actions are necessary or forbidden for any process that satisfies $\varphi$. For example, let $A=\{a,b\}$. Then, $\langle a\rangle \zero$ is not prime, since $\langle a\rangle \zero\models (\langle a\rangle \zero\wedge [b]\ff)\vee (\langle a\rangle \zero\wedge \langle b\rangle\true)$, and $\langle a\rangle \zero$ entails neither $\langle a\rangle \zero\wedge [b]\ff$ nor $\langle a\rangle \zero\wedge \langle b\rangle\true$. Intuitively, we call a formula $\varphi$ \emph{saturated} if $\varphi$ describes exactly which actions label the outgoing edges of any process $p$ such that $p\models\varphi$. Formally, $\varphi$ is saturated iff $I(\varphi)$ is a singleton.

If the action set is bounded by a constant, given $\varphi$, we can efficiently construct a formula $\varphi^s$ such that (1) $\varphi^s$ is saturated and for every $\langle a\rangle\varphi'\in\sub(\varphi^s)$, $\varphi'$ is saturated, (2) $\varphi$ is prime iff $\varphi^s$ is prime and $\varphi^s\models\varphi$, and (3) primality of $\varphi^s$ can be efficiently reduced to $\reacha(G_{\varphi^s})$.

\begin{restatable}{proposition}{rsprime}\label{prop:rs-primality}
    Let $|A|=k$, where $k\geq 1$ is a constant, and $\varphi\in\mathL_{RS}$ be such that if $\psi\in\sub(\varphi)$ is unsatisfiable, then $\psi=\ff$ and $\psi$ occurs in the scope of some $[a]$. Deciding whether $\varphi$ is prime is in \cP.
\end{restatable}

As the following result indicates, primality checking for formulae in $\mathL_{RS}$ becomes computationally hard when $|A|$ is not a constant.

\begin{restatable}{proposition}
{rsprimehard}\label{prop:decide-prime-rs-infinite-actions-hard}
Let $|A|$ be unbounded.  Deciding primality of  formulae in $\mathL_{RS}$ is \conp-complete.
\end{restatable}
\begin{proof}
We
give a polynomial-time reduction from \SAT to deciding whether a formula in $\mathL_{RS}$ is not prime. Let $\varphi$ be a propositional formula over $x_0,\dots,x_{n-1}$. We set $\varphi'=(\varphi\wedge \neg x_n)\vee (x_n\wedge \bigwedge_{i=1}^{n-1} \neg x_i)$
and $\varphi''$ to be $\varphi'$ where $x_i$ is substituted with $\langle a_i \rangle \zero$ and $\neg x_i$ with $[a_i] \ff$,
where $A=\{a_0,\dots, a_n\}$. As $\varphi''$ is satisfied in $a_n.\mathtt{0}$, it is a satisfiable formula, and so $\varphi''$ is prime in $\mathL_{RS}$ iff $\varphi''$ is characteristic within $\mathL_{RS}$.
    We show that $\varphi$ is satisfiable iff $\varphi''$ is not characteristic  within $\mathL_{RS}$. Let $\varphi$ be satisfiable and let $s$ denote a satisfying assignment of $\varphi$. Consider $p_1,p_2\in P$ such that: 
    \begin{itemize}
        \item $p_1\myarrowasubi \mathtt{0}$ iff $s(x_i)=\text{true}$, for $0\leq i\leq n-1$, and $p_1\notmyarrowan$, and
        \item $p_2\myarrowan \mathtt{0}$ and $p_2\notmyarrowa$ for every $a\in A\setminus\{a_n\}$.
    \end{itemize} 
    It holds that $p_i\models \varphi''$, $i=1,2$, 
    $p_1\notcurle_{RS} p_2$, and $p_2\notcurle_{RS} p_1$. Suppose that there is a process $q$, 
    such that $\varphi''$ is characteristic for $q$ within $\mathL_{RS}$. If $q\myarrowan$, then $q\notcurle_{RS} p_1$. On the other hand, if $q\notmyarrowan$, then $q\notcurle_{RS} p_2$. So, both cases lead to a contradiction, which means that $\varphi''$ is not characteristic within $\mathL_{RS}$. For the converse implication, assume that $\varphi$ is unsatisfiable. This implies that there is no process satisfying the first disjunct of $\varphi''$. Thus, $\varphi''$ is characteristic for $p_2$, described above, within $\mathL_{RS}$.
    
    Proving the matching upper bound is non-trivial. There is a \conp algorithm that uses properties of prime formulae and rules of Table~\ref{tab:S-rules}, carefully adjusted to the case of ready simulation. We describe the algorithm and prove its correctness in Appendix~\ref{subsection:rs-primality-unbounded-appendix}.
\end{proof}

\noindent\textbf{Primality in $\mathL_{TS}$.} If we have more than one action, a propositional literal can be encoded by using the restricted nesting of modal operators that is allowed by the grammar for $\mathL_{TS}$. This observation is the crux of the proof of the following result. 

\begin{restatable}{proposition}
{tsprimehard}\label{prop:decide-prime-ts-infinite-actions-hard}
 Let $|A|>1$. Deciding primality of formulae in $\mathL_{TS}$ is \conp-hard.
\end{restatable}
\begin{proof}
    Let $A=\{0,1\}$. The proof follows the steps of the proof of Proposition~\ref{prop:decide-prime-rs-infinite-actions-hard}. The initial and basic idea is that given an instance $\varphi$ of \SAT over $x_1,\dots,x_n$, every $x_i$ is substituted with $[\overline{{b_i}_1}]\ff\wedge \langle {b_i}_1\rangle ([\overline{{b_i}_2}]\ff\wedge \langle {b_i}_2\rangle(\dots ([\overline{{b_i}_k}]\ff\wedge \langle {b_i}_k\rangle\zero)\dots))$ and $\neg x_i$ with $[{b_i}_1] [{b_i}_2]\dots [{b_i}_k]\ff$, where ${b_i}_1\dots{b_i}_k$ is the binary representation of $i$ and $\overline{b}=0$, if $b=1$, and  $\overline{b}=1$, if $b=0$. For more technical details, see Appendix~\ref{subsection:ts-primality-unbounded}.
\end{proof}

In contrast to the case for $\mathL_{RS}$, bounding the size of the action set is not sufficient for deciding primality of formulae in $\mathL_{TS}$ in polynomial time. However, we show that both satisfiability and primality  become efficiently solvable if we bound both $|A|$ and the modal depth of the input formula.

 \begin{restatable}{proposition}
 {tsboundeddepth}\label{prop:ts-bounded-depth}
  Let $|A|= k$ and $\varphi\in\mathL_{TS}$ with $\md(\varphi)=d$, where $k,d\geq 1$ are constants. Then, there is an algorithm that decides whether $\varphi$ is satisfiable and prime in linear  time.
\end{restatable}
\begin{proof}
    It is necessary and sufficient to check that there is a process $p$ with $\depth(p)\leq d$ such that (1) $p\models\varphi$ and (2) for every $q$ with $\depth(q)\leq d+1$, if $q\models\varphi$ then $p\curle_{TS} q$. Since $k$ and $d$ are considered to be constants, there is an algorithm that does so and requires linear time in $|\varphi|$. In particular, the algorithm runs in $\mathcal{O}(2^{2k^{d+1}}\cdot k^{d+1}\cdot |\varphi|)$.
\end{proof}

To classify the problem of deciding whether formulae in $\mathL_{TS}$ are characteristic when $|A|$ is bounded, let us briefly introduce fixed-parameter tractable problems---see, for instance,~\cite{DowneyF13,FlumG06} for textbook accounts of this topic. Let $L\subseteq \Sigma^*\times\Sigma^*$ be a parameterized problem. We denote by $L_y$ the associated fixed-parameter problem $L_y=\{x ~|~ (x,y)\in L\}$, where $y$ is the parameter. Then, $L\in\fpt$ (or $L$ is fixed-parameter tractable) if there are a constant $\alpha$ and an algorithm to determine if $(x,y)$ is in $L$ in time $f(|y|)\cdot |x|^\alpha$, where $f$ is a computable function~\cite{DowneyF95}.  

\begin{corollary}\label{cor:ts-char-fpt}
 Let $|A|= k$, where $k\geq 1$ is a constant. The problems of deciding whether formulae in $\mathL_{TS}$ are satisfiable, prime, and characteristic are in \fpt, with the modal depth of the input formula as the parameter.
\end{corollary}

We note that the \conp-hardness argument from Proposition~\ref{prop:decide-prime-ts-infinite-actions-hard} 
applies also to logics that include  $\mathL_{TS}$. 
Since $\mathL_{TS}\subseteq\mathL_{2S}$, the \conp-hardness of deciding primality of formulae in $\mathL_{TS}$ with $|A|>1$ implies the same lower bound for deciding primality of formulae in $\mathL_{2S}$ when $|A|>1$.  Next, we show that in $\mathL_{3S}$ with $|A|>1$ the problem becomes \pspace-hard.

\medskip

\noindent\textbf{Primality in $\mathL_{3S}$.} 
Let $|A|>1$. \pspace-hardness of $\mathL_{3S}$-satisfiability implies \pspace-hardness of $\compL_{3S}$-validity. 
Along the lines of the proof of \cite[Theorem 26]{AcetoAFI20}, we prove the following result.

\begin{restatable}{proposition}
{decidethreesprime}\label{prop:decide-prime-3s}
  Let $|A|>1$. Deciding prime formulae within $\mathL_{3S}$ is \pspace-hard.
\end{restatable}

\begin{remark}
    Note that primality within $\mathL_{BS}$ coincides with primality modulo $\sim$. In~\cite{AcetoAFI20}, primality modulo  $\sim$ is called completeness and it is shown to be decidable in \pspace. However, the algorithm used in~\cite{AcetoAFI20} does not immediately imply that primality within $\mathL_{3S}$ is in \pspace.
\end{remark}

Interestingly, \pspace-hardness of $\mathL_{2S}$-validity implies the following theorem.

\begin{theorem}\label{thm:2scompl-prime}
    Let $X \in \{CS, RS, TS, 2S,3S\}$ and $|A|>1$. The problem of deciding whether a formula in $\compL_{2S}$
    is prime in $\mathL_X$ is \pspace-hard.
\end{theorem}
\begin{proof}
    We reduce $\mathL_{2S}$-validity to this problem.
    Let $\varphi \in \mathL_{2S}$.
    The reduction will return a formula $\varphi'$, such that $\varphi$ is $\mathL_{2S}$-valid if and only if $\varphi'$ is prime in $\mathL_X$.
    If $\mathtt{0} \not\models \varphi$, then let $\varphi' = \true$; in this case, $\varphi$ is not valid and $\true$ is not prime in $\mathL_{X}$.
    Otherwise, let $\varphi' = \zero \vee \neg \varphi$.
    If $\varphi$ is valid, then $\varphi' \equiv \zero$ and therefore $\varphi'$ is prime in $\mathL_X$.
    On the other hand, if $\varphi$ is not valid, then there is some process $p \models \neg \varphi$. 
    From $\mathtt{0} \models \varphi$, it holds that $p\myarrowa$.
    Then, $\varphi'\models\zero\vee\bigvee_{a\in A} \langle a\rangle\true$, but $\varphi'\not\models\zero$ and $\varphi'\not\models\bigvee_{a\in A} \langle a\rangle\true$. Since $\zero\vee\bigvee_{a\in A} \langle a\rangle\true\in\mathL_{CS}$, $\varphi'$ is not prime in $\mathL_{X}$, where $X\in\{CS,RS,TS,2S,3S\}$.
\end{proof}

Theorem~\ref{thm:2scompl-prime} shows that when deciding primality in $\mathL_X$, if we allow the input to be in a logic $\mathL$ that is more expressive than $\mathL_X$, the computational complexity of the problem can increase. It is then reasonable to constrain the input of the problem to be in $\mathL_X$ in order to obtain tractable problems as in the case of $\mathL_S$ and $\mathL_{CS}$.

 Before we give our main result summarizing the complexity of deciding characteristic formulae, we introduce two classes that play an important role in pinpointing the complexity of deciding characteristic formulae within $\mathL_{RS}$, $\mathL_{TS}$, and $\mathL_{2S}$. The first class is $\dpc=\{L_1\cap L_2 \mid L_1\in\NP \text{ and } L_2\in \conp\}$~\cite{PapadimitriouY84} and the second one is \us \cite{BlassG82}, which is defined thus: A language $L\in\us$ iff there is a non-deterministic Turing machine $T$ such that, for every instance $x$ of $L$, $x\in L$ iff $T$ has exactly one accepting path. The problem $\textsc{UniqueSat}$, viz.~the problem of deciding whether a given Boolean formula has exactly one satisfying truth assignment, is \us-complete and $\us\subseteq\dpc$~\cite{BlassG82}.


\begin{restatable}{theorem}{maintheorem}\label{Thm:main-US}\quad
    \begin{enumerate}[(a)]
        \item Deciding characteristic formulae within $\mathL_S$, $\mathL_{CS}$, or $\mathL_{RS}$ with a bounded action set is in \cP.
        \item Deciding characteristic formulae within $\mathL_{RS}$ with an unbounded action set is \us-hard and belongs to \dpc.
        \item Deciding characteristic formulae within $\mathL_{TS}$ or $\mathL_{2S}$ is \us-hard.
        \item Deciding characteristic formulae within $\mathL_{3S}$ is \pspace-hard.
    \end{enumerate}
\end{restatable}


\begin{table}[t]
\begin{center}
\begin{tabular}{ | P{1.7cm} | P{0.7cm} | P{0.7cm} | P{0.7cm} | P{1cm} | P{1cm} | P{0.9cm} | P{1.3cm} | P{1.3cm} | } 
\hline
 & $\mathL_S$ & $\mathL_{CS}$ & $\mathL_{RS}^{=k}$ &  $\mathL_{RS}^{>k}$ & $\mathL_{TS}^{>1}$ & $\mathL_{2S}^{>1}$ & $\mathL_{3S}^{>1}$ & $\mathL_{BS}$\\  
 \hline
Satisfiability & \cP & \cP & \cP &  \NP-comp. & \NP-comp. & \NP-comp.  & \pspace-comp.  & \cellcolor[gray]{0.9}\pspace-comp.\\  
 \hline
Primality & \cP & \cP & \cP &  \conp-comp. & \conp-hard & \conp-hard & \pspace-hard & \cellcolor[gray]{0.9}\pspace-comp.\\  
 \hline
$\text{Finding}_{decl}$ & \fp & \fp & \fp &  \fp & \NP-hard & \fp & \fp & \fp  \\  
 \hline
$\text{Finding}_{expl}$ & \multicolumn{8}{c}{\NP-hard} \vline \\  
 \hline
\end{tabular}
\end{center}
\caption{The complexity of deciding satisfiability and primality, and of finding characteristic formulae for different logics. $\text{Finding}_{decl}$ (resp.\ $\text{Finding}_{expl}$) denotes the problem of finding the characteristic formula for a given finite loop-free process, when the output is given in declarative (resp.\ explicit) form. Superscripts $=k$, $>k$, and $>1$ mean that the action set is bounded by a constant, unbounded, and has more than one action, respectively. \fp is the class of functions computable in polynomial time. All the results shown in white cells have been proven in this paper, whereas results in light gray are from~\cite{AcetoAFI20}. 
}
\label{tab:results}
\end{table}

\section{Finding characteristic formulae: The gap between trace simulation and the other preorders}\label{section:find}


Let $X\in\{S,CS\}$ or $X=RS$ and $|A|$ is bounded by a constant. The complexity of finding characteristic formulae within $\mathL_X$ depends on the representation of the output. If the characteristic formula has to be given in explicit form, then the following result holds.

\begin{restatable}{proposition}
{findingcharhard}\label{prop:trace-findchar-reduction}
Let $X\in\{S,CS\}$ or $X=RS$ and $|A|$ is bounded by a constant. If finding the characteristic formula within $\mathL_X$  for a given finite loop-free process can be done in polynomial time when the output is given in explicit form, then $\cP=\NP$.
\end{restatable}
\begin{proof}
    If the assumption of the proposition is true, the results of this paper allow us to decide trace equivalence of two finite loop-free processes in polynomial time. (For details, the reader can see Appendix~\ref{subsection:find-easy-appendix}.) Since trace equivalence for such processes is \conp-complete~\cite[Theorem 2.7(1)]{HRS76}, this implies that $\cP=\NP$.
\end{proof}

However, if output formulae are given in declarative form, then finding characteristic formulae within $\mathL_X$, where $X\in\{nS,CS,RS,BS\}$, $n\geq 1$, can be done in polynomial time. 

\begin{restatable}{proposition}
{findingchareasy}\label{prop:find-char-complexity-easy}
    For every $X\in\{nS,CS,RS,BS\}$, where $n\geq 1$, there is a polynomial-time algorithm that, given a finite loop-free process $p$, outputs a formula in declarative form that is characteristic  for $p$ within $\mathL_X$. 
\end{restatable}
\begin{proof}
    The proof relies on inductive definitions of characteristic formulae within $\mathL_X$, where $X\in\{S,CS,RS,2S,BS\}$, given in~\cite{IngolfsdottirGZ87,AcetoILS12}, and within $\mathL_{nS}$, $n\geq 3$, given in Appendix~\ref{subsection:find-easy-appendix}. These definitions guarantee that there are polynomial-time recursive procedures which construct characteristic formulae within $\mathL_X$. We prove the proposition for $X=2S$ below.

    Given a finite loop-free process $p$, the characteristic formula for $p$ within $\mathL_{2S}$ is defined
    as follows: 
    $\displaystyle\chi_{2S}(p)=\bar{\chi}_S(p)\wedge\bigwedge_{a\in A}~\bigwedge_{p\myarrowa p'} \langle a \rangle\chi_{2S}(p'), \text{ where } \displaystyle\bar{\chi}_{S}(p)=\bigwedge_{a\in A}[a]~\bigvee_{p\myarrowa p'} \bar{\chi}_S(p').$
    Consider the algorithm that recursively constructs $\chi_{2S}(p)$.
    The algorithm has to construct $\chi_{2S}(p')$ and $\bar{\chi}_S(p')$ for every $p'\in\mathrm{reach}(p)$, yielding a linear number of equations. Moreover, for every $p'\in\mathrm{reach}(p)$, $\bar{\chi}_S(p')$ is of linear size in $|p'|$. If $p'=\mathtt{0}$, then $\bar{\chi}_S(p')=\bigwedge_{a\in A} [a]\ff$. Otherwise, $|\bar{\chi}_{S}(p')|=\mathcal{O}(|\{p''\mid p'\myarrowa p''\}|+|A|)$, where $|A|$ is added because for every $a\in A$ such that $p'\notmyarrowa$, $[a]\ff$ is a conjunct of $\bar{\chi}_S(p')$. Note that for every $p''$, if $\bar{\chi}_S(p'')$ occurs in $\bar{\chi}_{S}(p')$, it is considered to  add $1$ to the size of $\bar{\chi}_{S}(p')$. Therefore, $|\bar{\chi}_{S}(p')|$ is of  linear size in $|p'|$. Using a similar argument, we can show that $\chi_{2S}(p')$ is of linear size. Thus, the algorithm constructs a linear number of equations, each of which is of linear size in $|p|$. The proofs for $X\in\{nS,CS,RS,BS\}$, $n\neq 2$, are analogous.
\end{proof}

\begin{remark}
    Note that the recursive procedures given in~\cite{IngolfsdottirGZ87,AcetoILS12} and Appendix~\ref{subsection:find-easy-appendix} provide characteristic formulae for finite processes with loops provided that we enrich the syntax of our logics by adding greatest fixed points. See, for example, \cite{AcetoILS12}. Consequently, constructing characteristic formulae for finite processes within $\mathL_X$, $X\in\{nS,CS,RS,BS\}$, $n\geq 1$, can be done in polynomial time.
\end{remark}


We now present the complexity gap between finding characteristic formulae for preorders $CS, RS, BS,$ and $nS$, $n\geq 1$, and the same search problem for preorder $TS$. In the former case, there are characteristic formulae with both declaration size and equational length that are polynomial in the size of the processes they characterize, and they can be efficiently computed. On the contrary, for $TS$, even if characteristic formulae are always of polynomial declaration size and polynomial equational length, they \emph{cannot} be efficiently computed unless $\cP=\NP$. 

\begin{restatable}{proposition}
{tsfindcharhard}\label{prop:find-char-ts-hard}
    Assume that for every finite loop-free process $p$, there is a characteristic formula  within $\mathL_{TS}$ for $p$, denoted by $\chi(p)$, such that both $\mathrm{decl}(\chi(p))$ and $\eqlen(\chi(p))$ are in $\mathcal{O}(|p|^k)$ for some $k\in\mathbb{N}$. Given a finite loop-free process $p$, if $\chi(p)$ can be computed in polynomial time, then $\cP=\NP$. 
\end{restatable}

Next, we prove that we do not expect that a finite loop-free process $p$ has always a short characteristic formula within $\mathL_{TS}$ when this is combined with a second condition. To show that statement, we need the following lemma.

\begin{lemma}\label{lem:trace-equiv-ts-preorder}
  For every finite $p$ and $q$,  $\mathrm{traces}(p)=\mathrm{traces}(q)$ iff $p \curle_{TS} p+q$ and  $q \curle_{TS} p+q$.
\end{lemma}
\begin{proof}
    If $\mathrm{traces}(p)=\mathrm{traces}(q)$, then $p \curle_{TS} p+q$. Indeed, for every $p\myarrowa p'$, it holds that $p+q\myarrowa p'$ and, trivially, $p'\curle_{TS} p'$. Moreover, $\mathrm{traces}(p+q)=\mathrm{traces}(p)\cup \mathrm{traces}(q)=\mathrm{traces}(p)$. Symmetrically, $q \curle_{TS} p+q$. Conversely, if $p \curle_{TS} p+q$ and  $q \curle_{TS} p+q$, then $\mathrm{traces}(p+q)=\mathrm{traces}(p)=\mathrm{traces}(q)$, and we are done. 
\end{proof}

\begin{proposition}\label{prop:ts-char-short-char}
    Assume that the following two conditions hold:
    \begin{enumerate}
        \item  For every finite loop-free process $p$, there is a characteristic formula  within $\mathL_{TS}$ for $p$, denoted by $\chi(p)$, such that both $\mathrm{decl}(\chi(p))$ and $\eqlen(\chi(p))$ are in $\mathcal{O}(|p|^k)$ for some $k\in\mathbb{N}$.
        \item Given a finite loop-free process $p$ and a formula $\varphi$ in declarative form, deciding whether $\varphi$ is characteristic for $p$  within $\mathL_{TS}$ is in \NP.
    \end{enumerate}
 Then $\NP=\conp$.
\end{proposition}
\begin{proof}
    We describe an \NP algorithm $\mathcal{A}$ that decides non-membership in \textsc{Sat} and makes use of conditions 1 and 2 of the proposition. Let $\phi$ be an input CNF formula to \textsc{Sat}. Algorithm $\mathcal{A}$ computes the DNF formula $\neg \phi$ for which it needs to decide \textsc{DNF-Tautology}. Then, $\mathcal{A}$ reduces \textsc{DNF-Tautology} to deciding trace equivalence of processes $p_0$ and $q$ constructed as described in the proof of~\cite[Theorem 2.7(1)]{HRS76} (or Proposition~\ref{prop:trace-complexity-b} in Appendix~\ref{section:decide-preorders-appendix}).  $\mathcal{A}$ can decide if $\mathrm{traces}(p_0)=\mathrm{traces}(q)$ by checking $p_0\curle_{TS} p_0+q$ and $q\curle_{TS} p_0+q$ because of Lemma~\ref{lem:trace-equiv-ts-preorder}. Finally, $\mathcal{A}$ reduces $p_0\curle_{TS} p_0+q$ (resp.\ $q\curle_{TS} p_0+q$) to model checking: it needs to check whether $p_0+q\models\chi(p_0)$ (resp.\ $p_0+q\models\chi(q)$) (by Proposition~\ref{prop:preorder-char-reduction}(a)). To this end, $\mathcal{A}$ guesses two formulae $\varphi_{p_0}$ and $\varphi_{q}$ in declarative form of polynomial declaration size and equational length, and two witnesses that verify that $\varphi_{p_0}$ and $\varphi_{q}$ are characteristic within $\mathL_{TS}$ for $p_0$ and $q$, respectively. This can be done due to conditions 1 and 2. $\mathcal{A}$ rejects the input iff both $p_0+q\models\chi(p_0)$ and $p_0+q\models\chi(q)$ are true.
\end{proof}

\section{A note on deciding characteristic formulae modulo equivalence relations}\label{section:decide-char-mod-equiv-relations}

So far, we have studied the complexity of algorithmic problems related to characteristic formulae in the modal logics that characterize the simulation-based preorders in van Glabbeek's spectrum. As shown in~\cite{AcetoMFI19}, those logics are powerful
enough to describe characteristic formulae for each finite, loop-free process up to the preorder they characterize. It is therefore natural to wonder whether they can also express characteristic formulae modulo the kernels of those preorders. The following result indicates that the logics $\mathL_X$, where $X\in\{S,CS,RS\}$, have very weak expressive power when it comes to defining characteristic formulae modulo $\equiv_X$. 

\begin{proposition}\label{prop:S-CS-RS-equiv-char}
    No formula in $\mathL_S$ is characteristic for some process $p$ with respect to $\equiv_S$. For $X\in\{CS,RS\}$, a formula $\varphi$ is characteristic for some process $p$ with respect to $\equiv_X$ iff it is logically equivalent to $\bigwedge_{a\in A} [a] \ff$. 
\end{proposition}
\begin{proof}
    Assume, towards contradiction, that there is a formula $\varphi_c^S$ in $\mathL_S$ that is characteristic for some process $p$ with respect to $\equiv_S$. Let $\ell$ be the depth of $p$ and $a\in A$. Define process $q = p + a^{\ell + 1}\mathtt{0}$---that is, $q$ is a copy of $p$ with an additional path that has exactly $\ell+1$ $a$-transitions. It is easy to see that $p \curle_S q$, but $q \not\curle_S p$.  Since $p\models\varphi_c^S$, it holds that $q\models \varphi_c^S$. However, $q\not\equiv_S p$, which contradicts our assumption that $\varphi_c^S$ is characteristic for $p$ with respect to $\equiv_S$. 
    For $X\in\{CS, RS\}$, note that a formula $\varphi$  is logically equivalent to $\bigwedge_{a\in A} [a] \ff$ iff it is satisfied only by processes without outgoing transitions, and so it is characteristic for 
    any such process modulo $\equiv_X$. To prove that no formula is  characteristic for some process $p$ with positive depth  modulo  $\equiv_{CS}$ or $\equiv_{RS}$, a similar argument to the one for $\equiv_S$ can be used. For $\equiv_{RS}$, the action $a$ should be chosen such that $p\myarrowa p'$ for some $p'$.
\end{proof}

For $TS$ and $2S$, there  are non-trivial characteristic formulae modulo $\equiv_{TS}$ and $\equiv_{2S}$, respectively. For example, if $A=\{a,b\}$, the formula $\varphi_a = \langle a\rangle([a]\ff \wedge [b]\ff)\wedge [b]\ff\wedge [a][a]\ff\wedge [a][b]\ff$ is satisfied only by processes that are equivalent, modulo those equivalences, to process $p_a = a.\mathtt{0}$ that has a single transition labelled with $a$. Thus, $\varphi_a$ is characteristic for $p_a$ modulo both $\equiv_{TS}$ and $\equiv_{2S}$. 
We can use the following theorem as a tool to prove hardness of deciding characteristic formulae modulo some equivalence relation.
Theorem~\ref{thm:reduction-validity-char} below is an extension of~\cite[Theorem 26]{AcetoAFI20}, so that it holds for every $X$ such that a characteristic formula modulo $\equiv_X$  exists, namely $X\in\{CS,RS,TS,2S,3S, BS\}$.
 

\begin{restatable}{theorem}
{reduction}\label{thm:reduction-validity-char}
Let $X\in\{CS,RS,TS,2S,3S, BS\}$. Validity in $\compL_X$ reduces in polynomial time to deciding characteristic formulae with respect to $\equiv_X$.
\end{restatable}

Note that, from the results of Subsection~\ref{section:deciding-satisfiability}, validity in $\compL_{RS}$ with an unbounded action set, $\compL_{TS}$  with $|A|>1$, and $\compL_{2S}$ with $|A|>1$ is \conp-complete, whereas 
validity in $\compL_{3S}$ with $|A|>1$ is \pspace-complete. Consequently, from Theorem~\ref{thm:reduction-validity-char}, deciding whether a formula is characteristic modulo $\equiv_{RS}$ with an unbounded action set, $\equiv_{TS}$  with $|A|>1$, and $\equiv_{2S}$  with $|A|>1$ is \conp-hard. That problem is \pspace-hard modulo $\equiv_{3S}$ with $|A|>1$. 


\section{Conclusions}\label{Sect:conclusions}

In this paper, we studied the complexity of determining whether a formula is characteristic for some finite, loop-free process in each of the logics providing modal characterizations of the simulation-based semantics in van Glabbeek's branching-time spectrum~\cite{Glabbeek01}. Since, as shown in~\cite{AcetoMFI19}, characteristic formulae in each of those logics are exactly the satisfiable and prime ones, we gave complexity results for the satisfiability and primality problems, and investigated the boundary between logics for which those problems can be solved in polynomial time and those for which they become computationally hard. Our results show that computational hardness already manifests itself in ready simulation semantics~\cite{BloomIM95,LarsenS91} when the size of the action set is not a constant. Indeed, in that setting, the mere addition of formulae of the form $[a]\ff$ to the logic that characterizes the simulation preorder yields a logic whose satisfiability and primality problems are \NP-hard and \conp-hard respectively. Moreover, we show that deciding primality in the logic characterizing 3-nested simulation is \pspace-hard in the presence of at least two actions. 

Amongst others, we also studied the complexity of constructing characteristic formulae in each of the logics we consider, both when such formulae are presented in explicit form and in declarative form. In particular, one of our results identifies a sharp difference between trace simulation and the other semantics when it comes to constructing characteristic formulae. For all the semantics apart from trace simulation, there are characteristic formulae that have  declaration size and equational length that are polynomial in the size of the processes they characterize and they can be efficiently computed. On the contrary, for trace simulation, even if characteristic formulae are always of polynomial declaration size and polynomial equational length, they \emph{cannot} be efficiently computed, unless $\cP=\NP$. 

Our results are summarized in Table~\ref{tab:results} and open several avenues for future research that we are currently pursuing. First of all, 
the precise complexity of primality checking is still open for the logics characterizing the $n$-nested simulation semantics. We conjecture that checking primality in $\mathL_{2S}$  is \conp-complete and that \pspace-completeness holds for $n$-nested simulation when $n\geq 3$. Next, we plan to study the complexity of deciding whether formulae are characteristic in the extensions of the modal logics we have considered in this article with greatest fixed points. Indeed, in those extended languages, one can define characteristic formulae for finite processes. It is known that deciding whether a formula is characteristic is \pspace-complete for \hml, but becomes \expc-complete for its extension with fixed-point operators---see reference~\cite{AcetoAFI20}. It would be interesting to see whether similar results hold for the other logics. Finally, building on the work presented in~\cite{AcetoMFI19}, we plan to study the complexity of the algorithmic questions considered in this article for (some of) the linear-time semantics in van Glabbeek's spectrum.

\bibliography{lipics-v2021-sample-article}

\begin{thebibliography}{10}

\bibitem{AcetoAFI20}
Luca Aceto, Antonis Achilleos, Adrian Francalanza, and Anna Ing{\'{o}}lfsd{\'{o}}ttir.
\newblock The complexity of identifying characteristic formulae.
\newblock {\em J. Log. Algebraic Methods Program.}, 112:100529, 2020.
\newblock \href {https://doi.org/10.1016/j.jlamp.2020.100529} {\path{doi:10.1016/j.jlamp.2020.100529}}.

\bibitem{AcetoMFI19}
Luca Aceto, Dario Della~Monica, Ignacio F{\'{a}}bregas, and Anna Ing{\'{o}}lfsd{\'{o}}ttir.
\newblock When are prime formulae characteristic?
\newblock {\em Theor. Comput. Sci.}, 777:3--31, 2019.
\newblock URL: \url{https://doi.org/10.1016/j.tcs.2018.12.004}.

\bibitem{AFEIP11}
Luca Aceto, Ignacio F{\'{a}}bregas, David de~Frutos{-}Escrig, Anna Ing{\'{o}}lfsd{\'{o}}ttir, and Miguel Palomino.
\newblock Graphical representation of covariant-contravariant modal formulae.
\newblock In Bas Luttik and Frank Valencia, editors, {\em Proceedings 18th International Workshop on Expressiveness in Concurrency, {EXPRESS} 2011, Aachen, Germany, 5th September 2011}, volume~64 of {\em {EPTCS}}, pages 1--15, 2011.
\newblock \href {https://doi.org/10.4204/EPTCS.64.1} {\path{doi:10.4204/EPTCS.64.1}}.

\bibitem{AILS07}
Luca Aceto, Anna Ing\'{o}lfsd\'{o}ttir, Kim~Guldstrand Larsen, and Jiri Srba.
\newblock {\em Reactive Systems: Modelling, Specification and Verification}.
\newblock Cambridge University Press, USA, 2007.

\bibitem{AcetoILS12}
Luca Aceto, Anna Ing{\'{o}}lfsd{\'{o}}ttir, Paul~Blain Levy, and Joshua Sack.
\newblock Characteristic formulae for fixed-point semantics: a general framework.
\newblock {\em Math. Struct. Comput. Sci.}, 22(2):125--173, 2012.
\newblock URL: \url{https://doi.org/10.1017/S0960129511000375}.

\bibitem{Achilleos18}
Antonis Achilleos.
\newblock The completeness problem for modal logic.
\newblock In {\em Proc. of Logical Foundations of Computer Science - International Symposium, {LFCS} 2018}, volume 10703 of {\em Lecture Notes in Computer Science}, pages 1--21. Springer, 2018.
\newblock URL: \url{https://doi.org/10.1007/978-3-319-72056-2\_1}.

\bibitem{BispingJN22}
Benjamin Bisping, David~N. Jansen, and Uwe Nestmann.
\newblock Deciding all behavioral equivalences at once: {A} game for linear-time-branching-time spectroscopy.
\newblock {\em Log. Methods Comput. Sci.}, 18(3), 2022.
\newblock URL: \url{https://doi.org/10.46298/lmcs-18(3:19)2022}.

\bibitem{BlassG82}
Andreas Blass and Yuri Gurevich.
\newblock On the unique satisfiability problem.
\newblock {\em Inf. Control.}, 55(1-3):80--88, 1982.
\newblock \href {https://doi.org/10.1016/S0019-9958(82)90439-9} {\path{doi:10.1016/S0019-9958(82)90439-9}}.

\bibitem{BloomIM95}
Bard Bloom, Sorin Istrail, and Albert~R. Meyer.
\newblock Bisimulation can't be traced.
\newblock {\em J. {ACM}}, 42(1):232--268, 1995.
\newblock URL: \url{https://doi.org/10.1145/200836.200876}.

\bibitem{BoudolL92}
G{\'{e}}rard Boudol and Kim~Guldstrand Larsen.
\newblock Graphical versus logical specifications.
\newblock {\em Theor. Comput. Sci.}, 106(1):3--20, 1992.
\newblock \href {https://doi.org/10.1016/0304-3975(92)90276-L} {\path{doi:10.1016/0304-3975(92)90276-L}}.

\bibitem{BrowneCG88}
Michael~C. Browne, Edmund~M. Clarke, and Orna Grumberg.
\newblock Characterizing finite {Kripke} structures in propositional temporal logic.
\newblock {\em Theor. Comput. Sci.}, 59:115--131, 1988.
\newblock URL: \url{https://doi.org/10.1016/0304-3975(88)90098-9}.

\bibitem{CaiFI92}
Jin{-}yi Cai, Martin F{\"{u}}rer, and Neil Immerman.
\newblock An optimal lower bound on the number of variables for graph identification.
\newblock {\em Comb.}, 12(4):389--410, 1992.
\newblock \href {https://doi.org/10.1007/BF01305232} {\path{doi:10.1007/BF01305232}}.

\bibitem{Cleaveland90}
Rance Cleaveland.
\newblock On automatically explaining bisimulation inequivalence.
\newblock In Edmund~M. Clarke and Robert~P. Kurshan, editors, {\em Computer Aided Verification, 2nd International Workshop, {CAV} '90}, volume 531 of {\em Lecture Notes in Computer Science}, pages 364--372. Springer, 1990.
\newblock URL: \url{https://doi.org/10.1007/BFb0023750}.

\bibitem{CleavelandS91}
Rance Cleaveland and Bernhard Steffen.
\newblock Computing behavioural relations, logically.
\newblock In Javier~Leach Albert, Burkhard Monien, and Mario Rodr{\'{\i}}guez{-}Artalejo, editors, {\em Automata, Languages and Programming, 18th International Colloquium, ICALP91, Madrid, Spain, July 8-12, 1991, Proceedings}, volume 510 of {\em Lecture Notes in Computer Science}, pages 127--138. Springer, 1991.
\newblock \href {https://doi.org/10.1007/3-540-54233-7\_129} {\path{doi:10.1007/3-540-54233-7\_129}}.

\bibitem{ERPH13}
David de~Frutos{-}Escrig, Carlos Gregorio{-}Rodr{\'{\i}}guez, Miguel Palomino, and David Romero{-}Hern{\'{a}}ndez.
\newblock Unifying the linear time-branching time spectrum of process semantics.
\newblock {\em Log. Methods Comput. Sci.}, 9(2), 2013.
\newblock \href {https://doi.org/10.2168/LMCS-9(2:11)2013} {\path{doi:10.2168/LMCS-9(2:11)2013}}.

\bibitem{DeNicolaV95}
Rocco De~Nicola and Frits~W. Vaandrager.
\newblock Three logics for branching bisimulation.
\newblock {\em J. {ACM}}, 42(2):458--487, 1995.
\newblock URL: \url{https://doi.org/10.1145/201019.201032}.

\bibitem{DowneyF95}
Rodney~G. Downey and Michael~R. Fellows.
\newblock Fixed-parameter tractability and completeness {I}: Basic results.
\newblock {\em {SIAM} J. Comput.}, 24(4):873--921, 1995.
\newblock \href {https://doi.org/10.1137/S0097539792228228} {\path{doi:10.1137/S0097539792228228}}.

\bibitem{DowneyF13}
Rodney~G. Downey and Michael~R. Fellows.
\newblock {\em Fundamentals of Parameterized Complexity}.
\newblock Texts in Computer Science. Springer, 2013.
\newblock URL: \url{https://doi.org/10.1007/978-1-4471-5559-1}.

\bibitem{EbbinghausFT1994}
Heinz{-}Dieter Ebbinghaus, J{\"{o}}rg Flum, and Wolfgang Thomas.
\newblock {\em Mathematical logic {(2.} ed.)}.
\newblock Undergraduate texts in mathematics. Springer, 1994.

\bibitem{FlumG06}
J{\"{o}}rg Flum and Martin Grohe.
\newblock {\em Parameterized Complexity Theory}.
\newblock Texts in Theoretical Computer Science. An {EATCS} Series. Springer, 2006.
\newblock URL: \url{https://doi.org/10.1007/3-540-29953-X}.

\bibitem{Glabbeek01}
Rob J.~{van} Glabbeek.
\newblock The linear time - branching time spectrum {I}.
\newblock In Jan~A. Bergstra, Alban Ponse, and Scott~A. Smolka, editors, {\em Handbook of Process Algebra}, pages 3--99. North-Holland / Elsevier, 2001.
\newblock \href {https://doi.org/10.1016/b978-044482830-9/50019-9} {\path{doi:10.1016/b978-044482830-9/50019-9}}.

\bibitem{GrafS86a}
Susanne Graf and Joseph Sifakis.
\newblock A modal characterization of observational congruence on finite terms of {CCS}.
\newblock {\em Inf. Control.}, 68(1-3):125--145, 1986.
\newblock URL: \url{https://doi.org/10.1016/S0019-9958(86)80031-6}.

\bibitem{GV92}
Jan~Friso Groote and Frits~W. Vaandrager.
\newblock Structured operational semantics and bisimulation as a congruence.
\newblock {\em Inf. Comput.}, 100(2):202--260, 1992.
\newblock URL: \url{https://doi.org/10.1016/0890-5401(92)90013-6}.

\bibitem{HS96}
Huttel H. and Shukla S.
\newblock On the complexity of deciding behavioural equivalences and preorders.
\newblock Technical report, State University of New York at Albany, USA, 1996.

\bibitem{HalpernM92}
Joseph~Y. Halpern and Yoram Moses.
\newblock A guide to completeness and complexity for modal logics of knowledge and belief.
\newblock {\em Artif. Intell.}, 54(2):319--379, 1992.
\newblock \href {https://doi.org/10.1016/0004-3702(92)90049-4} {\path{doi:10.1016/0004-3702(92)90049-4}}.

\bibitem{HennessyM85}
Matthew Hennessy and Robin Milner.
\newblock Algebraic laws for nondeterminism and concurrency.
\newblock {\em J. {ACM}}, 32(1):137--161, 1985.
\newblock URL: \url{https://doi.org/10.1145/2455.2460}.

\bibitem{Holmstrom89}
S{\"{o}}ren Holmstr{\"{o}}m.
\newblock A refinement calculus for specifications in {Hennessy-Milner} logic with recursion.
\newblock {\em Formal Aspects Comput.}, 1(3):242--272, 1989.
\newblock URL: \url{https://doi.org/10.1007/BF01887208}.

\bibitem{HT94}
Dung~T. Huynh and Lu~Tian.
\newblock On deciding some equivalences for concurrent processes.
\newblock {\em {RAIRO} Theor. Informatics Appl.}, 28(1):51--71, 1994.
\newblock \href {https://doi.org/10.1051/ita/1994280100511} {\path{doi:10.1051/ita/1994280100511}}.

\bibitem{HRS76}
Harry B.~Hunt III, Daniel~J. Rosenkrantz, and Thomas~G. Szymanski.
\newblock On the equivalence, containment, and covering problems for the regular and context-free languages.
\newblock {\em J. Comput. Syst. Sci.}, 12(2):222--268, 1976.
\newblock \href {https://doi.org/10.1016/S0022-0000(76)80038-4} {\path{doi:10.1016/S0022-0000(76)80038-4}}.

\bibitem{Immerman99}
Neil Immerman.
\newblock {\em Descriptive Complexity}.
\newblock Springer, 1999.
\newblock \href {https://doi.org/10.1007/978-1-4612-0539-5} {\path{doi:10.1007/978-1-4612-0539-5}}.

\bibitem{IngolfsdottirGZ87}
Anna Ingolfsdottir, Jens~Christian Godskesen, and Michael Zeeberg.
\newblock Fra {Hennessy-Milner} logik til {CCS}-processer.
\newblock Master's thesis, Aalborg University, 1987.
\newblock In Danish.

\bibitem{KS90}
Paris~C. Kanellakis and Scott~A. Smolka.
\newblock {CCS} expressions, finite state processes, and three problems of equivalence.
\newblock {\em Inf. Comput.}, 86(1):43--68, 1990.
\newblock \href {https://doi.org/10.1016/0890-5401(90)90025-D} {\path{doi:10.1016/0890-5401(90)90025-D}}.

\bibitem{KieferSS15}
Sandra Kiefer, Pascal Schweitzer, and Erkal Selman.
\newblock Graphs identified by logics with counting.
\newblock In Giuseppe~F. Italiano, Giovanni Pighizzini, and Donald Sannella, editors, {\em Mathematical Foundations of Computer Science 2015 - 40th International Symposium, {MFCS} 2015, Milan, Italy, August 24-28, 2015, Proceedings, Part {I}}, volume 9234 of {\em Lecture Notes in Computer Science}, pages 319--330. Springer, 2015.
\newblock \href {https://doi.org/10.1007/978-3-662-48057-1\_25} {\path{doi:10.1007/978-3-662-48057-1\_25}}.

\bibitem{Kozen83}
Dexter Kozen.
\newblock Results on the propositional $\mu$-calculus.
\newblock {\em Theor. Comput. Sci.}, 27:333--354, 1983.
\newblock URL: \url{https://doi.org/10.1016/0304-3975(82)90125-6}.

\bibitem{ladner1977computational}
Richard~E. Ladner.
\newblock The computational complexity of provability in systems of modal propositional logic.
\newblock {\em {SIAM} J. Comput.}, 6(3):467--480, 1977.
\newblock URL: \url{https://doi.org/10.1137/0206033}.

\bibitem{Larsen90}
Kim~Guldstrand Larsen.
\newblock Proof systems for satisfiability in {Hennessy-Milner} logic with recursion.
\newblock {\em Theor. Comput. Sci.}, 72(2{\&}3):265--288, 1990.
\newblock URL: \url{https://doi.org/10.1016/0304-3975(90)90038-J}.

\bibitem{LarsenS91}
Kim~Guldstrand Larsen and Arne Skou.
\newblock Bisimulation through probabilistic testing.
\newblock {\em Inf. Comput.}, 94(1):1--28, 1991.
\newblock URL: \url{https://doi.org/10.1016/0890-5401(91)90030-6}.

\bibitem{MartensG23}
Jan Martens and Jan~Friso Groote.
\newblock Computing minimal distinguishing {Hennessy-Milner} formulas is {NP}-hard, but variants are tractable.
\newblock In Guillermo~A. P{\'{e}}rez and Jean{-}Fran{\c{c}}ois Raskin, editors, {\em 34th International Conference on Concurrency Theory, {CONCUR} 2023}, volume 279 of {\em LIPIcs}, pages 32:1--32:17. Schloss Dagstuhl - Leibniz-Zentrum f{\"{u}}r Informatik, 2023.
\newblock URL: \url{https://doi.org/10.4230/LIPIcs.CONCUR.2023.32}.

\bibitem{Milner71}
Robin Milner.
\newblock An algebraic definition of simulation between programs.
\newblock In D.~C. Cooper, editor, {\em Proceedings of the 2nd International Joint Conference on Artificial Intelligence, IJCAI 1971}, pages 481--489. William Kaufmann, 1971.
\newblock URL: \url{http://ijcai.org/Proceedings/71/Papers/044.pdf}.

\bibitem{Milner89}
Robin Milner.
\newblock {\em Communication and Concurrency}.
\newblock Prentice Hall, 1989.

\bibitem{PT87}
Robert Paige and Robert~Endre Tarjan.
\newblock Three partition refinement algorithms.
\newblock {\em {SIAM} J. Comput.}, 16(6):973--989, 1987.
\newblock \href {https://doi.org/10.1137/0216062} {\path{doi:10.1137/0216062}}.

\bibitem{Pap94}
Christos~H. Papadimitriou.
\newblock {\em Computational Complexity}.
\newblock Addison-Wesley, 1994.
\newblock URL: \url{https://books.google.is/books?id=JogZAQAAIAAJ}.

\bibitem{PapadimitriouY84}
Christos~H. Papadimitriou and Mihalis Yannakakis.
\newblock The complexity of facets (and some facets of complexity).
\newblock {\em J. Comput. Syst. Sci.}, 28(2):244--259, 1984.
\newblock \href {https://doi.org/10.1016/0022-0000(84)90068-0} {\path{doi:10.1016/0022-0000(84)90068-0}}.

\bibitem{ShuklaRHS96}
Sandeep~K. Shukla, Daniel~J. Rosenkrantz, Harry B.~Hunt III, and Richard~Edwin Stearns.
\newblock The polynomial time decidability of simulation relations for finite processes: {A} {HORNSAT} based approach.
\newblock In Ding{-}Zhu Du, Jun Gu, and Panos~M. Pardalos, editors, {\em Satisfiability Problem: Theory and Applications, Proceedings of a {DIMACS} Workshop, Piscataway, New Jersey, USA, March 11-13, 1996}, volume~35 of {\em {DIMACS} Series in Discrete Mathematics and Theoretical Computer Science}, pages 603--641. {DIMACS/AMS}, 1996.
\newblock \href {https://doi.org/10.1090/dimacs/035/17} {\path{doi:10.1090/dimacs/035/17}}.

\bibitem{SteffenI94}
Bernhard Steffen and Anna Ing{\'{o}}lfsd{\'{o}}ttir.
\newblock Characteristic formulae for processes with divergence.
\newblock {\em Inf. Comput.}, 110(1):149--163, 1994.
\newblock URL: \url{https://doi.org/10.1006/inco.1994.1028}.

\bibitem{Thomas93}
Wolfgang Thomas.
\newblock On the {Ehrenfeucht}-{Fra{\"{\i}}ss{\'{e}}} game in theoretical computer science.
\newblock In Marie{-}Claude Gaudel and Jean{-}Pierre Jouannaud, editors, {\em TAPSOFT'93: Theory and Practice of Software Development, International Joint Conference CAAP/FASE}, volume 668 of {\em Lecture Notes in Computer Science}, pages 559--568. Springer, 1993.
\newblock URL: \url{https://doi.org/10.1007/3-540-56610-4\_89}.

\end{thebibliography}

\appendix

\section{Further preliminaries}\label{Sect:further-prelims}

We include here the material we need to use when proving in detail the results of this paper.

\begin{remark}\label{rem:remark-on-cs-rs}
    We use $\zero$ and $\bigwedge_{a\in A}[a]\ff$ interchangeably, especially when it comes to $\mathL_{CS}$ or $\mathL_{RS}$ formulae. Note that in the case of $|A|=1$, ready simulation coincides with complete simulation and $\mathL_{CS}=\mathL_{RS}$.
\end{remark}

\begin{remark}\label{rem:preorders-preserved-under-plus}
 As is well-known~\cite{Glabbeek01}, the  preorders defined in Definition~\ref{Def:beh-preorders} are preserved by action prefixing, i.e.\ if $p\curle_X q$, then $a.p\curle_X a.q$, and by the operator $+$, i.e.\ if $p_1\curle_X q_1$ and $p_2\curle_X q_2$, then $p_1+p_2\curle_X q_1+q_2$.
\end{remark}



Formally, the  modal depth of a formula is defined as follows.

\begin{definition}\label{def:modal_depth}
Given a formula $\varphi\in\mathL_{BS}$, the modal depth of $\varphi$, denoted by $\md(\varphi)$, is defined inductively as follows:
\begin{itemize}
    \item $\md(\true)=\md(\ff)=0$,
    \item $\md(\langle a\rangle\varphi)=\md([a]\varphi)=\md(\varphi)+1$,
    \item $\md(\varphi_1\wedge\varphi_2)=\md(\varphi_1\vee\varphi_2)=\max\{\md(\varphi_1),\md(\varphi_2)\}$,
    \item $\md(\neg\varphi)=\md(\varphi)$.
\end{itemize}
\end{definition}





The following is a corollary of Proposition~\ref{logical_characterizations} and Definition~\ref{def:characteristic}.

\begin{corollary}\label{cor:characteristic}
For every $X\in\{S,CS,RS,TS,2S,3S,BS\}$, a formula  $\varphi$ is characteristic within $\mathL_X$ for $p\in P$ if, for all $q \in P$, it holds that $q \models \varphi$ iff $\mathL_X(p) \subseteq \mathL_X(q)$ iff $p\curle_X q$. 
\end{corollary}

\begin{remark}
    We sometimes say that $\varphi$ is characteristic with respect to $\curle_X$ for $p$ when $\varphi$ is characteristic within $\mathL_X$ for $p$.
    Similarly to Corollary~\ref{cor:characteristic}, $\varphi$ is characteristic modulo $\equiv_X$ for $p$ if, for all $q \in P$, it holds that $q \models \varphi\Leftrightarrow\mathL_X(p) = \mathL_X(q)\Leftrightarrow p\equiv_X q$. 
\end{remark}

We derive the following corollary from Corollary~\ref{cor:characteristic}.

\begin{corollary}\label{cor:unique-process}
Let $\varphi\in\mathL_X$, $X\in\{S,CS,RS,TS,2S,3S,BS\}$. If $\varphi$ is characteristic for both $p$ and $q$, then $p\equiv_X q$.
\end{corollary}

\begin{corollary}\label{cor:logical-char-implication}
Let $\varphi,\psi\in\mathL_X$ be characteristic within $\mathL_X$ for $p,q$, respectively, where $X\in\{S,CS,RS,TS,2S,3S,BS\}$. Then, $\varphi\models\psi$ iff $q\curle_X p$.
\end{corollary}

The following proposition states that for characteristic formulae, preorder-checking and model-checking are interreducible.

\begin{proposition}\label{prop:preorder-char-reduction}
\begin{enumerate}[(a)]
    \item Given processes $p$ and $q$, deciding $p\curle_X q$, where $X\in\{S,CS,RS,TS,$\\$2S,3S,BS\}$, reduces to finding $\chi(p)$ within $\mathL_X$ and checking whether $q\models \chi(p)$.
    \item Given a formula $\varphi\in\mathL_X$, where $X\in\{S,CS,RS,TS,2S,3S,BS\}$, and a process $q$, if $\varphi$ is characteristic within $\mathL_X$, deciding $q\models\varphi$ reduces to finding a process $p$ for which $\varphi$ is characteristic within  $\mathL_X$ and checking whether $p\curle_X q$.
\end{enumerate}
\end{proposition}
\begin{proof} 
\begin{enumerate}[(a)]
    \item If $q\models \chi(p)$, then $\mathL_X(p)\subseteq \mathL_X(q)$ by Definition~\ref{def:characteristic}, which implies that $p\curle_X q$ by Proposition~\ref{logical_characterizations}.
    \item If $p\curle_X q$, then $\mathL_X(p)\subseteq \mathL_X(q)$ by Proposition~\ref{logical_characterizations}. Since $\varphi$ is characteristic for $p$, by Definition~\ref{def:characteristic}, $\varphi\in\mathL_S(p)$, and so $\varphi\in\mathL_S(q)$.\qedhere
\end{enumerate}
\end{proof}

We extensively use the disjunctive normal form (DNF) of formulae. A formula in $\mathL_{BS}$ is in DNF if it is in the form $\bigvee_{i=1}^k\varphi_i$, $k\geq 1$, where every $\varphi_i$, $1\leq i\leq k$, contains no disjunctions. To transform a formula $\varphi\in\mathL_{BS}$ into DNF, we can call Algorithm~\ref{algo:dnf}. The next lemma is true for formulae in $\mathL_X$, where $X\neq \{2S,3S,BS\}$.

\begin{algorithm}
\caption{Computing the DNF of a formula in $\mathL_{BS}$}
\begin{algorithmic}[1]
\DontPrintSemicolon
\Procedure{dnf}{$\varphi$}
    \State \lIf{$\varphi=\varphi_1\vee\varphi_2$}{$\varphi\gets\Call{dnf}{\varphi_1}\vee\Call{dnf}{\varphi_2}$}
    \State \lIf{$\varphi=\varphi_1\wedge(\varphi_2\vee\varphi_3)$}{$\varphi\gets(\Call{dnf}{\varphi_1\wedge\varphi_2})\vee(\Call{dnf}{\varphi_1\wedge\varphi_3})$} 
    \State \If{$\varphi=\langle a\rangle\varphi'$ and $\varphi'\neq\varphi_1\vee\varphi_2$}
    {$\varphi'\gets\Call{dnf}{\varphi'}$\;
    $\varphi\gets\Call{dnf}{\langle a\rangle \varphi'}$\;}
    \State \lIf{$\varphi=\langle a\rangle(\varphi_1\vee\varphi_2)$}
    {$\varphi\gets\Call{dnf}{\langle a\rangle\varphi_1}\vee\Call{dnf}{\langle a\rangle\varphi_2}\;$}
    \State \If{$\varphi=[a]\varphi'$ and $\varphi'\neq\varphi_1\vee\varphi_2$}
    {$\varphi'\gets\Call{dnf}{\varphi'}$\;
    $\varphi\gets\Call{dnf}{[a] \varphi'}$\;}
    \State \lIf{$\varphi=[a](\varphi_1\vee\varphi_2)$}
    {$\varphi\gets\Call{dnf}{[a]\varphi_1}\vee\Call{dnf}{[a]\varphi_2}\;$}
    \State {return $\varphi$\;}
\EndProcedure
\end{algorithmic}
\label{algo:dnf}
\end{algorithm}

\begin{lemma}\label{lem:DNF-equiv}
    Let $\varphi\in\mathL_X$, where $X\in\{S,CS,RS,TS\}$. Then, the DNF of $\varphi$ is logically equivalent to $\varphi$.
\end{lemma}
\begin{proof}
  To transform a formula in $\mathL_X$, where $X\in\{S,CS,RS,TS\}$, into its DNF, we do not need to distribute $[a]$ over disjunctions. The proof of the lemma is immediate from the following facts: 
   $\langle a\rangle (\varphi_1\vee\varphi_2)\equiv\langle a\rangle\varphi_1\vee \langle a\rangle\varphi_2$,
  $\varphi_1\wedge(\varphi_2\vee\varphi_3)\equiv(\varphi_1\wedge \varphi_2) \vee (\varphi_1\wedge \varphi_3)$, and 
for every $\varphi,\psi,\chi\in\mathL_X$, $X\in\{S,CS,RS,TS\}$, if $\psi\equiv \chi$, then $\varphi[\psi/\chi]\equiv\varphi$.
\end{proof}

\begin{remark}
    Note that Lemma~\ref{lem:DNF-equiv} does not hold when $X\in\{2S,3S,BS\}$, since, for example, $[a](\langle a\rangle \true\vee\langle b\rangle \true)\not\models[a]\langle a\rangle \true\vee[a]\langle b\rangle \true$.
\end{remark}



The following two basic lemmas that are true for formulae in $\mathL_{BS}$.

\begin{lemma}\label{lem:disjunction_lemma}
For every $\varphi_1,\varphi_2,\psi\in \mathL_{BS}$, $\varphi_1\vee \varphi_2\models \psi$ iff $\varphi_1 \models\psi$ and $\varphi_2 \models\psi$.
\end{lemma}
\begin{proof}
($\Rightarrow$) Let $p_1\models\varphi_1$. Then, $p_1\models\varphi_1\vee \varphi_2$, and so $p_1\models\psi$. The same argument is true for any $p_2$ that satisfies $\varphi_2$. Thus, $\varphi_1 \models\psi$ and $\varphi_2 \models\psi$.\\
($\Leftarrow$) Let $p\models\varphi_1\vee \varphi_2$. Then, $p\models\varphi_1$ or $p\models\varphi_2$. In either case, $p\models\psi$. So, $\varphi_1\vee \varphi_2\models\psi$.
\end{proof}

\begin{lemma}\label{lem:diamond_lemma}
For every $\varphi,\psi\in \mathL_{BS}$, $\langle a\rangle\varphi\models \langle a\rangle \psi$ iff $\varphi \models\psi$.
\end{lemma}
\begin{proof}
($\Rightarrow$) Suppose that $\varphi\not\models\psi$ and let $p$ be a process such that $p\models\varphi$ and $p\not\models\psi$. Consider process $q$ such that $q\myarrowa p$ and there is no other $p'$ such that $q\myarrowb p'$, $b\in A$. Then $q\models\langle a\rangle\varphi$ and $q\not\models\langle a\rangle \psi$, contradiction.\\
($\Leftarrow$) Let $p\models\langle a\rangle\varphi$. Then, there is $p\myarrowa p'$ such that $p'\models \varphi$ and so $p'\models\psi$. Hence, $p\models\langle a\rangle \psi$.
\end{proof}

Finally, $\sub(\varphi)$ denotes the set of subformulae of formula $\varphi$ and  $\varphi[\psi/\psi']$ is formula $\varphi$ where every occurrence of $\psi$ is replaced by $\psi'$. Then, the following lemma is true. 

\begin{lemma}\label{lem:substitution_bs}
    Let $\varphi,\psi,\chi\in \mathL_{BS}$.  If $\varphi\models\psi$, then $\chi\models\chi[\varphi/\psi]$.
\end{lemma}
\begin{proof}
  We prove the lemma by induction on the structure of $\chi$.
  \begin{itemize}
      \item If $\varphi\not\in\sub(\chi)$ or $\chi=\varphi$, then the lemma is trivially true. 
       \item If $\chi=\phi\vee\phi'$, by inductive hypothesis, $\phi\models\phi[\varphi/\psi]$ and $\phi'\models\phi'[\varphi/\psi]$. So, $\phi\models\phi[\varphi/\psi]\vee\phi'[\varphi/\psi]$ and $\phi'\models\phi[\varphi/\psi]\vee\phi'[\varphi/\psi]$. From Lemma~\ref{lem:disjunction_lemma}, $\phi\vee\phi'\models\phi[\varphi/\psi]\vee\phi'[\varphi/\psi]$.
      \item If $\chi=\phi\wedge\phi'$, by inductive hypothesis $\phi\models\phi[\varphi/\psi]$ and $\phi'\models\phi'[\varphi/\psi]$. Then, $\phi\wedge\phi'\models\phi[\varphi/\psi]$ and $\phi\wedge\phi'\models\phi'[\varphi/\psi]$. It holds that for every $p$ such that $p\models\phi\wedge\phi'$, $p\models\phi[\varphi/\psi]$ and $p\models\phi'[\varphi/\psi]$. So, $p\models\phi[\varphi/\psi]\wedge \phi'[\varphi/\psi]$. As a result, $\phi\wedge\phi'\models\phi[\varphi/\psi]\wedge \phi'[\varphi/\psi]$. 
      \item If $\chi=\langle a \rangle\phi$, by inductive hypothesis, $\phi\models\phi[\varphi/\psi]$. Then, $\langle a \rangle\phi\models\langle a \rangle\phi[\varphi/\psi]$ from Lemma~\ref{lem:disjunction_lemma}.
      \item If $\chi=[a]\phi$, by inductive hypothesis, $\phi\models\phi[\varphi/\psi]$. It suffices to show that for every $\varphi,\psi\in\mathL_{BS}$, $\varphi\models\psi$ implies $[a]\varphi\models [a]\psi$. Assume $\varphi\models\psi$ is true and let $p\models[a]\varphi$. This means that for every $p\myarrowa p'$, $p'\models\varphi$. By hypothesis, $p'\models\psi$, and so $p\models [a]\psi$.\qedhere
  \end{itemize}
\end{proof}

We use two kinds of reductions: Karp reductions and polytime Turing reductions. We provide their definitions below, which can be found, for instance, in~\cite{Pap94}.

\begin{definition}
We say that a problem $\Pi_1$ reduces to $\Pi_2$ (under Karp reductions), denoted by $\Pi_1\leq_m^p \Pi_2$, if there is an
algorithm $A$ that has the following properties:
\begin{itemize}
    \item given an instance $x$ of $\Pi_1$, $A$ produces an instance $y$ of $\Pi_2$,
    \item $A$ runs in time polynomial in $|x|$, which implies that $|y|$ is polynomial in $|x|$,
    \item $x$ is a $yes$ instance of $\Pi_1$ iff $y$ is a yes instance of $\Pi_2$.
\end{itemize}
\end{definition}

\begin{definition}
We say that problem $\Pi_1$ reduces to $\Pi_2$ under polytime Turing reductions, denoted by $\Pi_1\leq_T^p \Pi_2$, if there is an algorithm $A$ for $\Pi_1$ that has the following properties:
\begin{itemize}
    \item on any given instance $x$ of $\Pi_1$, $A$ uses polynomial steps in $|x|$,
    \item a step is either a standard computation step or an oracle call to  $\Pi_2$.
\end{itemize}
\end{definition}

A polytime Turing reduction from $\Pi_1$ to $\Pi_2$ means that $\Pi_1$ can be efficiently solved, assuming that $\Pi_2$ is easy to solve. Hence, the following proposition is true.

\begin{proposition}\label{prop:conp-complete-under-turing}
    If a  problem that is \conp-hard under polytime Turing reductions can be solved in polynomial time, then $\cP=\NP$.
\end{proposition}
\begin{proof}
    If a \conp-hard problem can be decided in \cP, then $\cP=\conp$, which in turn implies that $\conp=\NP=\cP$.
\end{proof}

For formulae in $\mathL_{BS}$, model checking is tractable.

\begin{proposition}[\cite{HalpernM92}]\label{model-checking-complexity}
    Given a formula $\varphi\in\mathL_{BS}$ and a process $p$, there is an algorithm for checking if $p$ satisfies $\varphi$ that runs in time $\mathcal{O}(|p|\cdot |\varphi|)$.
\end{proposition}

\begin{proposition}[\cite{CleavelandS91}]\label{model-checking-complexity-decl}
    Given  a process $p$ and a formula $\varphi\in\mathL_{BS}$ in declarative form, there is an algorithm for checking if $p$ satisfies $\varphi$ that runs in time $\mathcal{O}(|p|\cdot \mathrm{decl}(\varphi))$.
\end{proposition}

Let $\varphi\in\mathL_{BS}$. The DNF $\bigvee_{i=1}^k \varphi_i$ of $\varphi$ can have exponential size with respect to $|\varphi|$. However, we can non-determinically choose $\varphi_i$ for some $1\leq i\leq k$ in polynomial time.

\begin{lemma}\label{lem:DNF-poly-guess}
    Let $\varphi\in\mathL_{BS}$ and $\bigvee_{i=1}^k \varphi_i$ be the  DNF of $\varphi$. We can compute $\varphi_i$ for some $1\leq i\leq k$ in non-deterministic polynomial time.
\end{lemma}
\begin{proof}
    We prove the lemma by induction on the structure of $\varphi$.
    \begin{itemize}
        \item If $\varphi$ does not contain disjunctions, then it is in DNF and $k=1$. So, we return $\varphi$.
        \item Assume $\varphi=\varphi_1\vee\varphi_2$ and let $\bigvee_{i=1}^{k_1}{\varphi_1}_i$,  $\bigvee_{i=1}^{k_2}{\varphi_2}_i$ be the DNFs of $\varphi_1$ and $\varphi_2$, respectively. Then, we non-deterministically choose between $j=1$ and $j=2$. By inductive hypothesis, we can compute ${\varphi_j}_i$ for some $1\leq i\leq k_j$, and return this formula.
        \item Assume $\varphi=\varphi_1\wedge\varphi_2$ and let $\bigvee_{i=1}^{k_1}{\varphi_1}_i$,  $\bigvee_{i=1}^{k_2}{\varphi_2}_i$ be the DNFs of $\varphi_1$ and $\varphi_2$, respectively. By inductive hypothesis, we can compute ${\varphi_1}_{i_1}$, ${\varphi_2}_{i_2}$  for some $1\leq i_1\leq k_1$, $1\leq i_2\leq k_2$. We do this computation and we return ${\varphi_1}_{i_1}\wedge{\varphi_1}_{i_1}$.
        \item Assume $\varphi=\langle a\rangle \varphi'$ and let $\bigvee_{i=1}^{k}\varphi'_i$ be the DNF of $\varphi'$. By inductive hypothesis, we can compute $\varphi'_{i}$, for some $1\leq i\leq k$. We return $\langle a\rangle \varphi'_{i}$.
        \item The case of $\varphi=[a] \varphi'$ is similar to the previous one. \qedhere
    \end{itemize}
\end{proof}

The definitions of an alternating graph and the problem of reachability on alternating graphs are provided below.

\begin{definition}[\cite{Immerman99}]\label{def:alternating-graph}
An \emph{alternating graph} $G=(V,E,A,s,t)$ is a directed graph whose vertices are either existential or  universal, i.e.\ they are labelled with $\exists$ or $\forall$, respectively. $V$ and $E$ are the sets of vertices and edges, respectively, $A$ is the set of universal vertices, and $s$ and $t$ are two vertices that are called source and target, respectively.  
\end{definition}

\begin{definition}[\cite{Immerman99}]\label{def:alternating_reachability}
Let $G=(V,E,A,s,t)$ be an alternating graph. Let $P^G$ be the smallest relation on $V\times V$ that satisfies the following clauses:
\begin{enumerate}
    \item $P^G(x,x)$.
    \item If $x$ is existential and for some $(x,z)\in E$ it holds that $P^G(z,y)$, then $P^G(x,y)$.
    \item If $x$ is universal, there is at least one edge leaving $x$, and $P^G(z,y)$ holds for all edges $(x,z)$, then $P^G(x,y)$.
\end{enumerate}
If $P^G(x,y)$, we say that there is an alternating path from $x$ to $y$.

We define 
$\reacha=\{G ~ | ~ G \text{ is an alternating graph and }P^G(s,t)\}.
$
\end{definition}

\section{The complexity of satisfiability}\label{section:deciding-satisfiability-appendix}

Proofs of Subsection~\ref{section:deciding-satisfiability} are given in detail in  this part
of the appendix.

\Iphiproperty*
\begin{proof}
 (a)   We prove the lemma by induction on the structure of $\varphi$. 
    \begin{itemize}
        \item Assume $\varphi=\ff$. The lemma trivially holds.
        \item Assume $\varphi=\true$. For every set $S\subseteq A$, there is a process $p$ with $I(p)=S$ and $p\models\true$ and $S\in I(\true)$.  
        \item Assume $\varphi=[a]\ff$. For every $S\subseteq A$, $S\in I(\varphi)$ iff $a\not\in S$ iff there is a process $p$ such $I(p)=S$ and $p\models [a]\ff$.
        \item Assume $\varphi=\langle a\rangle\varphi'$. Let $S\subseteq A$. If $S\in I(\varphi)$, then from Definition~\ref{def:cs-rs-I(phi)}(d), $a\in I(\varphi)$ and $I(\varphi')\neq\emptyset$. So, there is some $T\subseteq A$ and $T\in I(\varphi')$ and from inductive hypothesis there is a process $q$ with $I(q)=T$ such that $q\models\varphi'$. Consider the process $p$ such that $p\myarrowa q$, and $p\myarrowb\mathtt{0}$ for every $b\in S$, $b\neq a$. Then, $I(p)=S$ and $p\models \langle a \rangle \varphi'$. Conversely, if there is a process $p$ such that $p\models \langle a \rangle \varphi'$ and $I(p)=S$, then $p\myarrowa p'$ for some $p'$ such that $p'\models\varphi'$. From inductive hypothesis, $I(p')\in I(\varphi')$, so $I(\varphi')\neq \emptyset$, and from Definition~\ref{def:cs-rs-I(phi)}(d), $I(\varphi)=\{X~\mid~ X\subseteq A \text{ and } a\in X\}$. Consequently, $I(p)\in I(\varphi)$, since $a\in I(p)$. 
        \item Assume $\varphi=\varphi_1\vee\varphi_2$. $S\in I(\varphi)=I(\varphi_1)\cup I(\varphi_2)$ iff $S\in I(\varphi_1)$ or  $S\in I(\varphi_2)$ iff there is $p$ with $I(p)=S$ such that $p\models\varphi_1$ or there is $p$ with $I(p)=S$ such that $p\models\varphi_2$ iff there is $p$ with $I(p)=S$ such that $p\models\varphi_1\vee\varphi_2$, where to prove the equivalences we used the inductive hypothesis, Definition~\ref{def:cs-rs-I(phi)}(e), and the definition of truth in a process.
        \item Assume $\varphi=\varphi_1\wedge\varphi_2$. $S\in I(\varphi)=I(\varphi_1)\cap I(\varphi_2)$ iff $S\in I(\varphi_1)$ and  $S\in I(\varphi_2)$ iff there is $p_1$ with $I(p_1)=S$ such that $p\models\varphi_1$ and there is $p_2$ with $I(p_2)=S$ such that $p_2\models\varphi_2$. We prove that the last statement is equivalent to the existence of some $q$ for which $I(q)=S$ and $q\models\varphi$. Let the last statetment in the above equivalences be true and  consider the process $p_1+p_2$. It holds $I(p_1)=I(p_2)=I(p_1+p_2)=S$, so $p_1\curle_{RS} p_1+p_2$ and $p_2\curle_{RS} p_1+p_2$. From Proposition~\ref{logical_characterizations}, $p_1+p_2\models\varphi_1\wedge\varphi_2$. Conversely, if there is $q$ with $q=S$ such that $q\models\varphi_1\wedge\varphi_2$, the last statement above follows from the fact that $q\models \varphi_1$ and $q\models\varphi_2$.
    \end{itemize}
    (b) This is an immediate corollary of (a).
\end{proof}

\subsection{Satisfiability in $\mathL_{RS}$}\label{subsection:rs-satisfiability}

In the case of $\mathL_{RS}$, we distinguish between the case of $|A|=k$, where $k\geq 1$ is a constant, and the case of $A$ being unbounded. In the former case, the satisfiability problem is linear-time solvable, whereas in the latter case, the number of actions appearing in $\varphi$ can be linear in $|\varphi|$, and satisfiability becomes \NP-complete.

\subsubsection{The case of $|A|=k$, $k\geq 1$}\label{subsubsection:rs-satisfiability-bounded}

\begin{proposition}\label{prop:rs-satisfiability}
    Let $|A|=k$, where $k\geq 1$ is a constant. Satisfiability of formulae in $\mathL_{RS}$ is linear-time decidable.
\end{proposition}
\begin{proof}
    Let $\varphi\in\mathL_{RS}$. Consider algorithm \consrs that recursively computes $I(\varphi)$. Note that since $k$ is a constant, for every $\psi\in\mathL_{RS}$, $I(\varphi)$ is of constant size. We can prove by induction on the structure of $\varphi$, that \consrs requires linear time in $|\varphi|$. \consrs accepts $\varphi$ iff $I(\varphi)\neq\emptyset$. The correctness of the algorithm is immediate from Lemma~\ref{lem:I(phi)-property}.
\end{proof}

\begin{proposition}\label{prop:rs-satisfiability-satisfiable}
Let $|A|=k$, where $k\geq 1$ is a constant. There is a polynomial-time algorithm that on input a satisfiable $\varphi\in\mathL_{RS}$, it returns $\varphi'$ such that (a) $\varphi\equiv\varphi'$, and (b) if  $\psi\in\sub(\varphi')$ is unsatisfiable, then $\psi=\ff$ and occurs in $\varphi'$ in the scope of some $[a]$.
\end{proposition}
\begin{proof}
    Let $\varphi\in\mathL_{RS}$ be a satisfiable formula. Consider algorithm \subrs that computes $I(\psi)$ for every $\psi\in\sub(\varphi)$ and stores  $I(\psi)$ in memory. For every $\psi\in\sub(\varphi)$ such that $I(\psi)=\emptyset$, \subrs substitutes $\psi$ with $\ff$ in $\varphi$. We denote by $\varphi^{\mathrm{ff}}$ the obtained formula. Then, \subrs repeatedly applies the rules $\langle a\rangle\ff\ruleff \ff$, $\ff\vee\psi\ruleff\psi$, $\psi\vee\ff\ruleff\psi$, $\ff\wedge\psi\ruleff \ff$, and $\psi\wedge\ff\ruleff \ff$ on $\varphi^{\mathrm{ff}}$ until no rule can be applied, and returns the resulting formula, which we denote by $\varphi'$. Since every substitution has replaced a formula $\psi$ with some $\psi'\equiv\psi$, from Lemma~\ref{lem:substitution_bs}, $\varphi\equiv\varphi'$. From the type of the substitutions made on $\varphi$, every unsatisfiable formula has been substituted with $\ff$, and all occurrences of $\ff$ have been eliminated except for the ones that are in the scope of some $[a]$. Moreover, it is not hard to see that the algorithm requires polynomial time.
\end{proof}

\subsubsection{The case of $|A|$ being unbounded}\label{subsubsection:rs-satisfiability-unbounded}

Let $\varphi\in\mathL_{RS}$. Assume now that the number of actions appearing in $\varphi$ can be linear in $|\varphi|$. Then, satisfiability 
in $\mathL_{RS}$ becomes \NP-hard.

\begin{proposition}\label{prop:sat-validity-RS-modal-depth-one}
 Let $|A|$ be unbounded and $\varphi\in\mathL_{RS}$ with $\md(\varphi)=1$. Deciding whether $\varphi$ is satisfiable is
 $\NP$-complete. 
\end{proposition}
\begin{proof} Let $\varphi\in\mathL_{RS}$ with $\md(\varphi)=1$ and let  $|\varphi|=n$.  
  If $\varphi$ is satisfiable, there is a process $p$ of depth at most $1$ that satisfies $\varphi$. Let $A_\varphi$ denote the set of different actions that appear in $\varphi$. It holds that $|A_\varphi|=\mathcal{O}(n)$, and there are $\mathcal{O}(2^n)$ different processes of depth $1$ over $A_\varphi$. So, the problem can be solved in non-deterministic polynomial time by guessing a process of depth $1$ and checking whether it satisfies $\varphi$. To prove that the problem is $\NP$-hard, consider a propositional formula $\psi$ in CNF with variables $x_1,\dots, x_n$. Construct formula $\psi'$ by substituting every literal $x_i$ with $\langle a_i\rangle\true$ and every literal $\neg x_i$ with $[a_i]\ff$. Then, it is not hard to see that $\psi$ is satisfiable iff there is a process that satisfies $\psi'$. 
\end{proof}

We show that satisfiability of formulae in the more general case of $\mathL_{RS}$  still belongs to \NP, and so it is $\NP$-complete. We first prove the following lemma.

\begin{lemma}\label{lem:RS-poly-sat-no-disj}
 Let $|A|$ be unbounded and $\varphi$ be a formula in $\mathL_{RS}$ that does not contain disjunctions. Then, the satisfiability of $\varphi$ can be decided in polynomial time.
\end{lemma}
\begin{proof}
Consider the algorithm that first repeatedly applies the rules $\langle a\rangle\ff\ruleff \ff$, $\ff\wedge\psi\ruleff \ff$, and $\psi\wedge\ff\ruleff \ff$ on $\varphi$ until no rule can be applied. If the resulting formula $\varphi'$ is $\ff$, it rejects. Otherwise, it checks that for every $\varphi_1\wedge\varphi_2\in \sub(\varphi')$, it is not the case that $\varphi_1=\langle a\rangle \varphi_1'$ and $\varphi_2=[a]\ff$, for some $a\in A$. This is a necessary and sufficient condition for acceptance.
\end{proof}

\begin{corollary}\label{cor:RS-sat-NP}
Let $|A|$ be unbounded. Satisfiability of formulae in $\mathL_{RS}$  is $\NP$-complete.
\end{corollary}
\begin{proof}
  Let $\varphi\in\mathL_{RS}$ and $\bigvee_{i=1}^k\varphi_i$ be the DNF of $\varphi$.   Consider the algorithm that decides satisfiability of  $\varphi$ by guessing $\varphi_i$ for some $1\leq i\leq k$, checking whether $\varphi_i$ is satisfiable, and accepting iff $\varphi_i$ is satisfiable. The correctness of the algorithm relies on the fact  that $\varphi_1\vee\varphi_2$ is satisfiable iff at least one of $\varphi_1, \varphi_2$ is satisfiable, and Lemma~\ref{lem:DNF-equiv}. The polynomial-time complexity of the algorithm is because of Lemmas~\ref{lem:DNF-poly-guess} and~\ref{lem:RS-poly-sat-no-disj}. Moreover, \NP-hardness of the problem is a corollary of Proposition~\ref{prop:sat-validity-RS-modal-depth-one}.
\end{proof}

\subsection{Satisfiability in $\mathL_{CS}$}\label{subsection:cs-satisfiability}

To decide satisfiability in the case of complete simulation, we show that  we need much less information than $I(\varphi)$ gives for a formula $\varphi\in\mathL_{CS}$. Alternatively, we associate $\varphi$ to a set $J(\varphi)$, which is one of $\emptyset, \{\emptyset\}, \{\alpha\},\{\emptyset,\alpha\}$. Note that the main difference here is that we let $\alpha$ symbolize every possible set of actions.

\begin{definition}\label{def:cs-J(phi)}
Let $\varphi\in\mathL_{CS}$. We define $J(\varphi)$ inductively as follows:
    \begin{enumerate}[(a)]
        \item $J(\true)=\{\emptyset,\alpha\}$,
        \item $J(\ff)=\emptyset$,
        \item $J(\zero)= \{\emptyset\}$,
        \item $J(\langle a\rangle \varphi)=\begin{cases}
           \emptyset, &\text{if } J(\varphi)=\emptyset,\\
           \{\alpha\}, &\text{otherwise}
        \end{cases}$
        \item $J(\varphi_1\vee\varphi_2)=J(\varphi_1)\cup J(\varphi_2)$,
        \item $J(\varphi_1\wedge\varphi_2)=J(\varphi_1)\cap J(\varphi_2)$.
    \end{enumerate}
\end{definition}

\begin{lemma}\label{lem:J(phi)-property}
    For every $\varphi\in\mathL_{CS}$, $\varphi$ is unsatisfiable iff $J(\varphi)=\emptyset$.
\end{lemma}
\begin{proof}
    We prove the lemma by proving the following claims simultaneously by induction on the structure of $\varphi$.
    \begin{description}
        \item[Claim 1.] $\varphi\equiv\zero$ iff $J(\varphi)=\{\emptyset\}$.
        \item[Claim 2.] $\varphi\equiv\ff$ iff $J(\varphi)=\emptyset$.
        \item[Claim 3.] $\varphi$ is satisfiable and $\zero\not\models\varphi$ iff $J(\varphi)=\{\alpha\}$.
        \item[Claim 4.] $\zero\models\varphi$ and $\varphi\not\models\zero$ iff $J(\varphi)=\{\emptyset,\alpha\}$.
    \end{description}
\noindent\textbf{\textcolor{darkgray}{Proof of Claim 1.}} Note that $\varphi\equiv\zero$ iff $\varphi$ is satisfied exactly by $\mathtt{0}$.\\
$(\Rightarrow)$ Let $\varphi$ be satisfied exactly by $\mathtt{0}$. Then $\varphi$ can have one of the following forms.
\begin{description}
    \item[$\varphi=\zero$.] Then $J(\varphi)=\{\emptyset\}$.
    \item[$\varphi=\varphi_1\vee\varphi_2$.] Then either both $\varphi_i$, $i=1,2$, are satisfied exactly by $\mathtt{0}$, in which case, from inductive hypothesis, $J(\varphi_1)= J(\varphi_2) =\{\emptyset\}$, and so $J(\varphi)=J(\varphi_1)\cup J(\varphi_2)=\{\emptyset\}$, or one of them, w.l.o.g.\ assume that this is $\varphi_1$, is satisfied exactly by $\mathtt{0}$ and $\varphi_2$ is unsatisfiable, in which case $J(\varphi_2)=\emptyset$ from inductive hypothesis of Claim 2, and $J(\varphi)=J(\varphi_1)\cup J(\varphi_2)=\{\emptyset\}\cup\emptyset=\{\emptyset\}$.
    \item[$\varphi=\varphi_1\wedge\varphi_2$.] Then, one of three cases is true. 
    \begin{itemize}
        \item  Both $\varphi_i$, $i=1,2$, are satisfied exactly by $\mathtt{0}$. Then, $J(\varphi)=J(\varphi_1)\cap J(\varphi_2)=\{\emptyset\}\cap\{\emptyset\}=\{\emptyset\}$.
        \item W.l.o.g.\ $\varphi_1$ is satisfied exactly by $\mathtt{0}$ and $\varphi_2\equiv\true$. Then, $J(\varphi)=J(\varphi_1)\cap J(\varphi_2)=\{\emptyset\}\cap \{\emptyset,\alpha\}=\{\emptyset\}$.
        \item W.l.o.g.\ $\varphi_1$ is satisfied exactly by $\mathtt{0}$ and $\varphi_2$ is satisfied in $\mathtt{0}$ and some other processes. From inductive hypothesis of Claim 4, $J(\varphi_2)=\{\emptyset,\alpha\}$. Then, $J(\varphi)=J(\varphi_1)\cap J(\varphi_2)=\{\emptyset\}\cap \{\emptyset,\alpha\}=\{\emptyset\}$.
    \end{itemize} 
\end{description}
$(\Leftarrow)$ Let $J(\varphi)=\{\emptyset\}$. Then,  one of the following holds.
\begin{description}
    \item[$\varphi=\zero$.] Trivial.
    \item[$\varphi_1\vee\varphi_2$.] Then, since $J(\varphi)=J(\varphi_1)\cup J(\varphi_2)$, it must be the case that w.l.o.g.\ $J(\varphi_1)=\{\emptyset\}$ and either $J(\varphi_2)=\{\emptyset\}$ or $J(\varphi_2\}=\emptyset$. Hence, either both $\varphi_i$, $i=1,2$, are satisfied exactly by $\mathtt{0}$, or $\varphi_1$ is satisfied exactly by $\mathtt{0}$ and $\varphi_2$ is unsatisfiable. In both cases, $\varphi$ is satisfied exactly by $\mathtt{0}$.
    \item[$\varphi_1\wedge\varphi_2$.] 
    Since $J(\varphi)=J(\varphi_1)\cap J(\varphi_2)$, it must be the case that either $J(\varphi_1)=J(\varphi_2)=\{\emptyset\}$, or w.l.o.g.\ $J(\varphi_1)=\{\emptyset\}$ and $J(\varphi_2)=\{\emptyset,\alpha\}$. In the former case, $\varphi_1\equiv\varphi_2\equiv\zero$, and so $\varphi_1\wedge\varphi_2\equiv\zero$. In the latter case,  $\varphi_1\equiv\zero$, $\zero\models\varphi_2$ and $\varphi_2\not\models\zero$, and so $\varphi_1\wedge\varphi_2$ is satisfied exactly by $\mathtt{0}$.
\end{description}
\noindent\textbf{\textcolor{darkgray}{Proof of Claim 2.}}   Note that $\varphi\equiv\ff$ iff $\varphi$ is unsatisfiable.\\
$(\Rightarrow)$ Let $\varphi$ be unsatisfiable. Then, $\varphi$ can have one of the following forms.
\begin{description}
    \item[$\varphi=\ff$.] Trivial.
    \item[$\varphi=\langle a\rangle \varphi'$.] Then, $\varphi'$ is unsatisfiable, so from inductive hypothesis $J(\varphi')=\emptyset$ and $J(\varphi)=\emptyset$.
    \item[$\varphi=\varphi_1\vee\varphi_2$.] Then, both $\varphi_1$ and $\varphi_2$ are unsatisfiable, and $J(\varphi)=J(\varphi_1)\cup J(\varphi_2)=\emptyset\cup\emptyset=\emptyset$.
    \item[$\varphi=\varphi_1\wedge\varphi_2$.] Then, one of the following cases holds.
        \begin{itemize}
        \item W.l.o.g.\ $\varphi_1$ is unsatisfiable. Then, $J(\varphi)=J(\varphi_1)\cap J(\varphi_2)=\emptyset\cap J(\varphi_2)=\emptyset$.
        \item W.l.o.g.\ $\varphi_1$ is satisfied exactly by $\mathtt{0}$ and $\varphi_2$ is satisfied only by processes that are not $\mathtt{0}$. From inductive hypothesis of Claims 1 and 3, $J(\varphi_1)=\{\emptyset\}$, and $J(\varphi_2)=\{\alpha\}$, respectively. Thus, $J(\varphi)=J(\varphi_1)\cap J(\varphi_2)=\emptyset$.
         \end{itemize}  
\end{description}
$(\Leftarrow)$ Assume that $J(\varphi)=\emptyset$. Then, one of the following holds.
\begin{description}
    \item[$\varphi=\ff$.] Trivial.
    \item[$\varphi=\langle a\rangle \varphi'$.] Then, $J(\varphi')=\emptyset$, which implies that $\varphi'$ is unsatisfiable, and so $\varphi$ is unsatisfiable.
    \item[$\varphi=\varphi_1\vee\varphi_2$.] Then for both $i=1,2$, $J(\varphi_i)=\emptyset$, and from inductive hypothesis, both $\varphi_i$ are unsatisfiable, which implies that $\varphi$ is unsatisfiable as well.
    \item[$\varphi=\varphi_1\wedge\varphi_2$.] We distinguish between the following cases.
    \begin{itemize}
        \item W.l.o.g.\ $J(\varphi_1)=\emptyset$. From inductive hypothesis, $\varphi_1$ is unsatisfiable, and so $\varphi$ is also unsatisfiable.
        \item W.l.o.g.\ $J(\varphi_1)=\{\emptyset\}$ and $J(\varphi_2)=\{\alpha\}$. Then, from inductive hypothesis of Claims 1 and 3, $\varphi_1$ is satisfied exactly by $\mathtt{0}$  and $\varphi_2$ is not satisfied in $\mathtt{0}$, respectively, which implies that $\varphi$ is unsatisfiable.
    \end{itemize}
\end{description}
\noindent\textbf{\textcolor{darkgray}{Proof of Claim 3.}} 
$(\Rightarrow)$ Let $\varphi$ be satisfiable and $\zero\not\models\varphi$. Then, $\varphi$ can have one of the following forms.
\begin{description}
    \item[$\varphi=\langle a\rangle \varphi'$.] Then, $\varphi'$ is satisfiable, and so $J(\varphi)=\{\alpha\}$.
    \item[$\varphi_1\wedge\varphi_2$.] One of the following is true.
    \begin{itemize}
        \item For both $i=1,2$, $\varphi_i$ is satisfiable and $\zero\not\models\varphi_i$. From inductive hypothesis, for both $i=1,2$, $J(\varphi_i)=\{\alpha\}$, and $J(\varphi)=J(\varphi_1)\cap J(\varphi_2)=\{\alpha\}$ as well.
        \item For both $i=1,2$, $\varphi_i$ is satisfiable and w.l.o.g.\ $\zero\models\varphi_1$ and $\zero\not\models\varphi_2$. Suppose that $\varphi_1\equiv \zero$. Then, $\varphi_1\wedge\varphi_2$ is not satisfiable, contradiction. So, $\zero\models\varphi_1$ and $\varphi_1\not\models \zero$. From inductive hypothesis of Claims 3 and 4, $J(\varphi_2)=\{\alpha\}$ and $J(\varphi_1)=\{\emptyset,\alpha\}$, respectively. Thus, $J(\varphi)=\{\alpha\}$.
    \end{itemize}
    \item[$\varphi_1\vee\varphi_2$.]  Then, for both $i=1,2$, $\varphi_i$ is  satisfiable and $\zero\not\models\varphi_i$, or w.l.o.g.\ $\varphi_1$ is  satisfiable, $\zero\not\models\varphi_1$ and $\varphi_2$ is unsatisfiable. In both cases, $J(\varphi)=\{\alpha\}$.
\end{description}
$(\Leftarrow)$ Let $J(\varphi)=\{\alpha\}$. Then,
\begin{description}
    \item[$\varphi=\langle a\rangle \varphi'$.] Since $J(\langle a\rangle \varphi')\neq\emptyset$, $J(\varphi')\neq\emptyset$, which implies that $\varphi'$ is satisfiable. So, $\langle a\rangle \varphi'$ is satisfiable and $\zero\not\models\langle a\rangle \varphi'$.
    \item[$\varphi_1\wedge\varphi_2$.] One of the following is true.
    \begin{itemize}
        \item $J(\varphi_1)=J(\varphi_2)=\{\alpha\}$, and so $\varphi_i$ is satisfiable and $\zero\not\models\varphi_i$ for both $i=1,2$. In this case, there are processes $p_1,p_2\neq\mathtt{0}$ such that $p_1\models\varphi_1$ and $p_2\models\varphi_2$. Since $p_1,p_2\neq\mathtt{0}$, it holds that $p_i\curle_{CS} p_1+p_2$ for both $i=1,2$. From Proposition~\ref{logical_characterizations}, $p_1+p_2\models\varphi_1\wedge\varphi_2$, and so $\varphi$ is satisfiable. It is immediate from  $\zero\not\models\varphi_i$ that $\zero\not\models\varphi_1\wedge\varphi_2$.
        \item W.l.o.g.\ $J(\varphi_1)=\{\emptyset,\alpha\}$ and $J(\varphi_2)=\{\alpha\}$.  Then, $\varphi_2$ is satisfiable, $\zero\not\models\varphi_2$, $\zero\models\varphi_1$, and $\varphi_1\not \models\zero$. So, there are processes $p_1,p_2\neq\mathtt{0}$ such that $p_1\models\varphi_1$ and $p_2\models\varphi_2$. As in the previous case, $p_1+p_2\models\varphi_1\wedge\varphi_2$, which implies that $\varphi$ is satisfiable, and $\zero\not\models\varphi_1\wedge\varphi_2$ because of $\zero\not\models\varphi_2$. 
    \end{itemize}
    \item[$\varphi_1\vee\varphi_2$.]  We distinguish between two cases. $J(\varphi_i)=\{\alpha\}$ for both $i=1,2$. Then, from inductive hypothesis, for both $i=1,2$, $\varphi_i$ is satisfiable and $\zero\not\models\varphi_i$, which also holds for $\varphi_1\vee\varphi_2$. Otherwise, w.l.o.g.\ $J(\varphi_1)=\{\alpha\}$ and $J(\varphi_2)=\emptyset$, which implies that $\varphi_1$ is satisfiable, $\zero\not\models\varphi_1$, and $\varphi_2$ is unsatisfiable. So, $\varphi_1\vee\varphi_2$ is satisfiable and $\zero\not\models\varphi_1\vee\varphi_2$.
\end{description}
\noindent\textbf{\textcolor{darkgray}{Proof of Claim 4.}} $(\Rightarrow)$ Let  $\zero\models\varphi$ and $\varphi\not\models\zero$ be both true.
\begin{description}
    \item[$\varphi=\true$.] Trivial.
    \item[$\varphi=\varphi_1\wedge\varphi_2$.] Then, $\zero\models\varphi$ implies that for both $i=1,2$, $\zero\models\varphi_i$, and  $\varphi\not\models\zero$  means that there is $p\neq\mathtt{0}$ such that $p\models\varphi_1\wedge\varphi_2$, or equivalently, $p\models\varphi_1$ and $p\models\varphi_2$. Thus, $J(\varphi)=J(\varphi_1)\cap J(\varphi_2)=\{\emptyset,\alpha\}\cap\{\emptyset,\alpha\}=\{\emptyset,\alpha\}$. 
    \item[$\varphi=\varphi_1\vee\varphi_2$.] $\zero\models\varphi_1\vee\varphi_2$ and $\varphi_1\vee\varphi_2\not\models\zero$ implies one of the following cases.
    \begin{itemize}
        \item $\zero\models\varphi_i$ and $\varphi_i\not\models\zero$ for some $i=1,2$. From inductive hypothesis and the fact that $J(\varphi)=J(\varphi_1)\cup J(\varphi_2)$, we have that $J(\varphi)=\{\emptyset,\alpha\}$.
        \item $\varphi_i\equiv\zero$ for some $i=1,2$, assume w.l.o.g that $\varphi_1\equiv\zero$, $\zero\not\models\varphi_2$, and $\varphi_2\not\models\zero$. From $\varphi_2\not\models\zero$, we have that $\varphi_2$ is satisfiable. From inductive hypothesis of Claims 2 and 3, $J(\varphi_1)=\{\emptyset\}$, and $J(\varphi_2)=\{\alpha\}$, respectively. Thus, $J(\varphi_1\vee\varphi_2)=\{\emptyset,\alpha\}$.
    \end{itemize}
\end{description}
$(\Leftarrow)$ Let $J(\varphi)=\{\emptyset,\alpha\}$.
\begin{description}
    \item[$\varphi=\true$.] Trivial.
    \item[$\varphi=\varphi_1\wedge\varphi_2$.] For both $i=1,2$, $J(\varphi_i)=\{\emptyset,\alpha\}$. So, for both $i=1,2$, $\zero\models\varphi_i$ and $\varphi_i\not\models\zero$, which implies that $\zero\models\varphi_1\wedge\varphi_2$ and for both $i=1,2$, there is $p_i\neq\mathtt{0}$ such that $p_i\models\varphi_i$. As shown above, $p_1+p_2\models\varphi_1\wedge\varphi_2$, and so $\varphi\not\models\zero$.
     \item[$\varphi=\varphi_1\vee\varphi_2$.] One of the following cases is true.
     \begin{itemize}
         \item W.l.o.g.\ $J(\varphi_1)=\emptyset$ and $J(\varphi_2)=\{\emptyset,\alpha\}$. Then, $\varphi_1$ is unsatisfiable, $\zero\models\varphi_2$, and $\varphi_2\not\models\zero$. Then, $\zero\models\varphi$ and $\varphi\not\models\zero$.
         \item W.l.o.g.\ $J(\varphi_1)=\{\emptyset\}$ and $J(\varphi_2)=\{\alpha\}$. Then $\varphi_1\equiv\zero$, $\varphi_2$ is satisfiable, and $\zero\not\models\varphi_2$. As a result, $\zero\models\varphi_1\vee\varphi_2$. Suppose that $\varphi_1\vee\varphi_2\models\zero$. Then, from Lemma~\ref{lem:disjunction_lemma}, $\varphi_1\models\zero$ and $\varphi_2\models\zero$. Since $\varphi_2$ is satisfiable,  $\varphi_2\models\zero$ implies that $\varphi_2\equiv\zero$, which contradicts with $\zero\not\models\varphi_2$. So, $\varphi_1\vee\varphi_2\not\models\zero$.
         \item W.l.o.g.\ $J(\varphi_1)=\{\emptyset\}$ and $J(\varphi_2)=\{\emptyset,\alpha\}$. This can be proven similarly to the previous case.
         \item W.l.o.g.\ $J(\varphi_1)=\{\alpha\}$ and $J(\varphi_2)=\{\emptyset,\alpha\}$. This can be proven similarly to the previous two cases.
         \item For both $i=1,2$, $J(\varphi_i)=\{\emptyset,\alpha\}$. Then, $\zero\models\varphi_i$ and $\varphi_i\not\models\zero$ for both $i=1,2$. Consequently, $\zero\models\varphi_1\vee\varphi_2$ and from Lemma~\ref{lem:disjunction_lemma}, $\varphi_1\vee\varphi_2\not\models\zero$.\qedhere
     \end{itemize}
\end{description}
\end{proof}

\begin{proposition}\label{cor:cs-satisfiability}
    Satisfiability of formulae in $\mathL_{CS}$ is linear-time decidable.
\end{proposition}
\begin{proof}
    Let $\varphi\in\mathL_{CS}$. Consider algorithm \conscs that recursively computes $J(\varphi)$. We can easily prove by induction on the structure of $\varphi$, that \conscs requires linear time in $|\varphi|$. \conscs accepts $\varphi$ iff $J(\varphi)\neq\emptyset$. The correctness of the algorithm is immediate from Lemma~\ref{lem:J(phi)-property}.
\end{proof}

\begin{proposition}\label{prop:cs-satisfiability-satisfiable}
There is a polynomial-time algorithm that on input a satisfiable $\varphi\in\mathL_{CS}$, it returns $\varphi'$ such that (a) $\varphi\equiv\varphi'$, and (b) every $\psi\in\sub(\varphi')$ is satisfiable.
\end{proposition}
\begin{proof}
   The proof is completely analogous to the proof of Proposition~\ref{prop:rs-satisfiability-satisfiable}. Let $\varphi\in\mathL_{CS}$ be a satisfiable formula. Consider algorithm \subcs that computes $J(\psi)$ for every $\psi\in\sub(\varphi)$ and stores  $J(\psi)$ in memory. For every $\psi\in\sub(\varphi)$ such that $J(\psi)=\emptyset$, \subcs substitutes $\psi$ with $\ff$ in $\varphi$. We denote by $\varphi^{\mathrm{ff}}$ the obtained formula. Then, \subcs repeatedly applies the rules $\langle a\rangle\ff\ruleff \ff$, $\ff\vee\psi\ruleff\psi$, $\psi\vee\ff\ruleff\psi$, $\ff\wedge\psi\ruleff \ff$, and $\psi\wedge\ff\ruleff \ff$ on $\varphi^{\mathrm{ff}}$ until no rule can be applied, and returns the resulting formula, which we denote by $\varphi'$. Since every substitution has replaced a formula $\psi$ with some $\psi'\equiv\psi$, from Lemma~\ref{lem:substitution_bs}, $\varphi\equiv\varphi'$. Suppose that there is some $\psi\in\sub(\varphi')$ such that $\psi$ is unsatisfiable. Suppose that $\ff\not\in\sub(\psi)$. Then, either $\psi\in\sub(\varphi)$ or $\psi$ is the result of some substitution. In the latter case, since $\ff\not\in\sub(\psi)$, $\psi$ can only be the result of rules of the form $\ff\vee\psi\ruleff\psi$ (and $\psi\vee\ff\ruleff\psi$), and so again $\psi\in\sub(\varphi)$. However, if $\psi\in\sub(\varphi)$, then since $J(\psi)=\emptyset$, \subcs would have substituted $\psi$ with $\ff$ in $\varphi$ and $\psi\not\in\sub(\varphi')$, contradiction. So, $\ff\in\sub(\psi)$, which implies that some rule can be applied on $\varphi'$, contradiction. So, $\psi$ cannot be unsatisfiable.
\end{proof}

\subsection{Satisfiability in $\mathL_{S}$}\label{subsection:s-satisfiability}

To decide satisfiability in $\mathL_S$, we need even less information than in the case of complete simulation. We can define and use a variant of $J(\varphi)$, namely $K(\varphi)$, as follows.

\begin{definition}\label{def:s-K(phi)}
Let $\varphi\in\mathL_{S}$. We define $K(\varphi)$ inductively as follows:
    \begin{enumerate}[(a)]
        \item $K(\true)=\{\alpha\}$,
        \item $K(\ff)=\emptyset$,
        \item $K(\langle a\rangle \varphi)=\begin{cases}
           \emptyset, &\text{if } K(\varphi)=\emptyset,\\
           \{\alpha\}, &\text{otherwise}
        \end{cases}$
        \item $K(\varphi_1\vee\varphi_2)=K(\varphi_1)\cup K(\varphi_2)$,
        \item $K(\varphi_1\wedge\varphi_2)=K(\varphi_1)\cap K(\varphi_2)$.
    \end{enumerate}
\end{definition}

The following statements can be proven similarly to the case of complete simulation.

\begin{proposition}\label{prop:s-satisfiability}
    Satisfiability of formulae in $\mathL_{S}$ is linear-time decidable.
\end{proposition}

\begin{proposition}\label{prop:s-satisfiability-satisfiable}
There is a polynomial-time algorithm that on input a satisfiable $\varphi\in\mathL_{S}$, it returns $\varphi'$ such that (a) $\varphi\equiv\varphi'$, and (b) every $\psi\in\sub(\varphi')$ is satisfiable.
\end{proposition}

Alternatively, in this case, we can characterize unsatisfiable formulae as follows.

\begin{lemma}
   Every unsatisfiable formula $\varphi\in\mathL_S$ is given by the following grammar:
   $$\mathrm{unsat}_{S} ::= ~ \ff ~ \mid ~ \langle a \rangle \mathrm{unsat}_{S} ~ \mid ~ \mathrm{unsat}_{S} \wedge \varphi_S ~ \mid ~  \varphi_S \wedge  \mathrm{unsat}_{S}~ \mid ~ \mathrm{unsat}_{S}\vee \mathrm{unsat}_{S},$$
   where $\varphi_S\in \mathL_S$.
\end{lemma}

As a result, given $\varphi\in\mathL_S$, it suffices to apply the rules $\langle a\rangle\ff \ruleff \ff$,
$\ff\wedge\varphi \ruleff \ff$,
$\varphi\wedge\ff \ruleff \ff$,
$\ff\vee\varphi \ruleff \varphi$,
$\varphi\vee\ff \ruleff \varphi$, on $\varphi$. Then, the resulting formula $\varphi'$ is $\ff$ iff $\varphi$ is unsatisfiable, contains no unsatisfiable subformulae if $\varphi$ is satisfiable, and is logically equivalent to $\varphi$.

\polysat*
\begin{proof}
    Immediate from Propsitions~\ref{prop:rs-satisfiability}, \ref{cor:cs-satisfiability}, and~\ref{prop:s-satisfiability}.
\end{proof}

\subsection{Satisfiability in $\mathL_{TS}$}\label{subsection:ts-satisfiability}

\subsubsection{The case of $|A|>1$}

\begin{proposition}\label{prop:sat-TS}
 Let $|A|>1$. Satisfiability of formulae in $\mathL_{TS}$  is $\NP$-hard. 
\end{proposition}
\begin{proof} Assume that $A=\{0,1\}$. Consider a propositional formula $\psi$ in CNF with variables $x_1,\dots, x_n$. We use $\mathL_{TS}$ formulae of logarithmic size to encode literals of $\varphi$. 
We associate a positive literal $x_i$, $i=1,\dots,n$, with the binary representation of $i$, i.e.\ ${b_i}_1\dots{b_i}_k$, where every ${b_i}_j\in\{0,1\}$ and $k=\lceil\log n\rceil$. The binary string ${b_i}_1\dots {b_i}_k$ can now be mapped to formula 
$\enc(x_i)= \langle {b_i}_1\rangle \langle {b_i}_2\rangle\dots \langle {b_i}_k\rangle\true$.
We map a negative literal $\neg x_i$ to $\enc(\neg x_i)=[{b_i}_1] [{b_i}_2]\dots [{b_i}_k]\ff$. We construct formula $\varphi\in\mathL_{TS}$ by starting with $\psi$ and substituting every literal $l$ with $\enc(l)$ in $\psi$. It is not hard to see that $\varphi$ is satisfiable iff $\psi$ is satisfiable.
\end{proof}

To prove that the problem is in \NP, we first prove that it can be solved efficiently when the input is a formula that does not contain disjunctions. We associate two sets to a formula $\varphi$, namely the set of traces required by $\varphi$, denoted by $\mathrm{traces}(\varphi)$, and the set of traces forbidden by $\varphi$, denoted by $\mathrm{forbidden}(\varphi)$. The sets are defined as follows.

\begin{definition}\label{def:sets-traces-forbidden}
    Let $\varphi\in\mathL_{TS}$ be a formula that does not contain disjunctions. Sets $\mathrm{traces}(\varphi)$ and $\mathrm{forbidden}(\varphi)$ are defined inductively as follows:
    \begin{itemize}
        \item $\mathrm{traces}(\true)=\{\varepsilon\}$ and $\mathrm{forbidden}(\true)=\emptyset$,
        \item $\mathrm{traces}(\ff)=\emptyset$ and $\mathrm{forbidden}(\ff)=\{\varepsilon\}$, 
        \item $\mathrm{traces}([a]\varphi')=\{\varepsilon\}$ and $\mathrm{forbidden}([a]\varphi')=\{at | t\in\mathrm{forbidden}(\varphi')\}$,
        \item $\mathrm{traces}(\varphi_1\wedge\varphi_2)=\mathrm{traces}(\varphi_1)\cup\mathrm{traces}(\varphi_2)$ and \\$\mathrm{forbidden}(\varphi_1\wedge\varphi_2)=\mathrm{forbidden}(\varphi_1)\cup\mathrm{forbidden}(\varphi_2)$,
        \item $\mathrm{traces}(\langle a\rangle\varphi')=\{at | t\in\mathrm{traces}(\varphi')\}\cup\{\varepsilon\}$ and $\mathrm{forbidden}(\langle a\rangle\varphi')=\emptyset$.
    \end{itemize}
\end{definition}

\begin{lemma}\label{lem:sets-traces-forbidden}
 Let $\varphi\in\mathL_{TS}$ be a formula that does not contain disjunctions such that if $\ff\in\sub(\varphi)$, then $\ff$ occurs in $\varphi$ only in the scope of some $[a]$. Then, $\varphi$ is satisfiable iff for every $\varphi_1\wedge\varphi_2\in\sub(\varphi)$, $\mathrm{traces}(\varphi_1\wedge\varphi_2)\cap\mathrm{forbidden}(\varphi_1\wedge\varphi_2)=\emptyset$.
\end{lemma}

\begin{lemma}\label{lem:TS-poly-sat-no-disj}
 Let $|A|>1$ and $\varphi$ be a formula in $\mathL_{TS}$ that does not contain disjunctions. Then, the satisfiability of $\varphi$ can be decided in polynomial time.
\end{lemma}
\begin{proof}
Consider the algorithm that first repeatedly applies the rules $\langle a\rangle\ff\ruleff \ff$, $\ff\wedge\psi\ruleff \ff$, and $\psi\wedge\ff\ruleff \ff$ on $\varphi$ until no rule can be applied. If the resulting formula $\varphi'$ is $\ff$, it rejects. Otherwise, for every $\varphi_1\wedge\varphi_2\in\sub(\varphi)$, the algorithm checks whether $\mathrm{traces}(\varphi_1\wedge\varphi_2)\cap\mathrm{forbidden}(\varphi_1\wedge\varphi_2)=\emptyset$. From Lemma~\ref{lem:sets-traces-forbidden}, this is a necessary and sufficient condition for acceptance. Since the sets $\mathrm{traces}(\varphi')$ and $\mathrm{forbidden}(\varphi')$ can be efficiently computed for every $\varphi'\in\sub(\varphi)$, the algorithm requires polynomial time.
\end{proof}

\begin{corollary}\label{cor:TS-sat-NP}
Let $|A|>1$. Satisfiability of formulae in $\mathL_{TS}$  is $\NP$-complete.
\end{corollary}
\begin{proof}
The proof is analogous to the proof of Corollary~\ref{cor:RS-sat-NP}. For the membership in \NP, Lemma~\ref{lem:TS-poly-sat-no-disj} can be applied here instead of Lemma~\ref{lem:RS-poly-sat-no-disj}. \NP-hardness of the problem follows from Proposition~\ref{prop:sat-TS}.
\end{proof}

\subsection{Satisfiability in $\mathL_{2S}$}\label{subsection:2s-satisfiability}

\subsubsection{The case of $|A|>1$}

\begin{proposition}\label{prop:tws-sat}
    Let $|A|>1$. Satisfiability of formulae in $\mathL_{2S}$ is \NP-complete.
\end{proposition}
\begin{proof}
    \NP-hardness is a corollary of Proposition~\ref{prop:sat-TS} and the fact that $\mathL_{TS}\subseteq\mathL_{2S}$. We can decide whether $\varphi\in\mathL_{2S}$ is satisfiable in a standard way by constructing a tableau for $\varphi$ and checking whether its root is satisfiable as described in~\cite{HalpernM92}. To avoid giving here the details of this construction, we refer the reader to~\cite[Section 6.3]{HalpernM92}. Consider the non-deterministic polynomial-time Turning machine $M$ such that each path of $M$ constructs a branch of a tableau for $\varphi$ as follows. It starts from the node $r$ of the tableau that corresponds to $\varphi$ labelled with $L(r)=\{\varphi\}$. Let $s$ be a node of the tableau that has been already created by $M$ labelled with $L(s)$. If one successor $s'$ of a node $s$ must be created because of a formula $\psi\wedge\psi'\in L(s)$, then $M$ creates $s'$. If for every $\langle a_i\rangle\varphi'\in L(s)$, an $i$-successor of $s$ must be created, then $M$ creates all these successors. If two different successors $s'$ and $s''$ of $s$ that are not $i$ successors of $s$ must be created, then $M$ non-deterministically chooses to create one of them. In this last case, $s'$ and $s''$ correspond either to some $\psi\vee\psi'\in L(s)$ or some $\psi\in\sub(\varphi')$, where $\varphi'\in L(s)$, such that $\psi\not\in L(s)$ and $\neg\psi\in L(s)$. To decide if the root $r$ is marked satisfiable, we need to know whether at least one of $s$, $s''$ is marked satisfiable and this will be done by $M$ using non-determinism. After constructing this branch of the tableau, $M$ propagates information from the leaves to the root and decides whether the root must be marked ``satisfiable''. It accepts iff $r$ is marked ``satisfiable''. It holds that $\varphi$ is satisfiable iff there is some branch that makes $r$ satisfiable iff $M$ has an accepting path. Since $\varphi\in\mathL_{2S}$, the formula has no diamond operators $\langle a_i\rangle$ in the scope of box operators $[a_i]$. So, the longest path of a branch is polynomial in the number of nested diamond operators occurring in $\varphi$ and a branch can be computed in polynomial time.
\end{proof}

\twostssat*
\begin{proof}
\NP-hardness is immediate from Propositions~\ref{prop:sat-validity-RS-modal-depth-one} and \ref{prop:sat-TS} and the fact that $\mathL_{TS}\subseteq\mathL_{2S}$. Membership in \NP follows from Propositions~\ref{cor:RS-sat-NP}, \ref{cor:TS-sat-NP}, and~\ref{prop:tws-sat}. In fact Proposition~\ref{prop:tws-sat} suffices to prove membership of the satisfiability problem in \NP for all three logics.
\end{proof}

\subsection{Satisfiability in $\mathL_{3S}$}\label{subsection:3s-satisfiability}

\subsubsection{The case of $|A|>1$}

Let $|A|>1$. We show that $\mathL_{2S}$-validity is \pspace-complete. In fact, this is done by proving that $\compL_{2S}$-satisfiability is \pspace-complete. 

In what follows $\mathL_{\Box\Diamond}$ denotes the dual fragment of $\mathL_{2S}$, which consists of all \hml formulae that have no box subformulae in the scope of a diamond operator.

\begin{theorem}\label{thm:2s-validity}
Let $|A|>1$. Validity of formulae in $\mathL_{2S}$ is \pspace-complete.
\end{theorem}
\begin{proof}
    That $\mathL_{2S}$-validity is in \pspace\ is a direct result of the fact that validity for Modal Logic \textbf{K} and \hml\ is in \pspace~\cite{ladner1977computational,HalpernM92}.
    To prove the \pspace-hardness of $\mathL_{2S}$-validity, we consider 
    $\mathL_{\Box\Diamond x}$ to be the extension of $\mathL_{\Box\Diamond}$ with literals, i.e.\ with propositional variables and their negation.
    We can interpret variables and their negation in the usual way by introducing a labelling of each process in an LTS with a set of propositional variables (see \cite{HalpernM92} for instance).    As $\mathL_{\Box\Diamond}$ is the dual fragment of $\mathL_{2S}$, it suffices to prove that $\mathL_{\Box\Diamond}$-satisfiability is \pspace-hard.
    We observe that Ladner's reduction in the proof for the \pspace-hardness of \textbf{K}-satisfiability from \cite{ladner1977computational} constructs a one-action formula in $\mathL_{\Box\Diamond x}$, and therefore $\mathL_{\Box\Diamond x}$-satisfiability is \pspace-hard, even with only one action.

    We now give a reduction from $\mathL_{\Box\Diamond x}$-satisfiability to $\mathL_{\Box\Diamond}$-satisfiability by encoding literals with formulae that have no box modalities.
    Let $\mathL_{[a]\langle a \rangle^d x}^k$ be the fragment of $\mathL_{\Box\Diamond x}$ that includes the formulae of modal depth up to $d$ that use only action $a$ and $k$ propositional variables, $x_1,x_2,\ldots ,x_k$. 
    We write the negation of variable $x$ as $\Bar{x}$; let $b \neq a$ be an action.
    We now describe how to encode each $x_i$ and $\Bar{x_i}$.
    For each $0 \leq i \leq k$ and $0\leq j <  \lceil\log k \rceil$, let $\alpha(i,j) = a$, if position $j$ in  the binary representation of $i$ (using $\lceil\log k \rceil$ bits) has bit $1$, and $\alpha(i,j) = b$ otherwise.
    Let $e (x_i)  = \langle b \rangle\langle \alpha(i,0) \rangle\langle \alpha(i,1) \rangle\cdots \langle \alpha(i,\lceil\log k \rceil) \rangle\langle a \rangle \true$, and 
    $e (\Bar{x_i})  = \langle b \rangle\langle \alpha(i,0) \rangle\langle \alpha(i,1) \rangle\cdots \langle \alpha(i,\lceil\log k \rceil) \rangle\langle b \rangle \true$.
    The negations of these are defined as
    $\neg e( x_i ) = [ b ][ \alpha(i,0) ][ \alpha(i,1) ]\cdots [ \alpha(i,\lceil\log k \rceil) ] [a]  \ff$, and 
    $\neg e( \Bar{x_i} ) = [ b ][ \alpha(i,0) ][ \alpha(i,1) ]\cdots [ \alpha(i,\lceil\log k \rceil) ] [b]  \ff$.
    The formula $e(x_i)$ asserts that a process has a trace of the form $bt_ia$, where $t_i$ encodes $i$ in binary, where $a$ stands for 1 and $b$ for 0; and $e(\Bar{x_i})$ asserts that a process has a trace of the form $bt_ib$.
    Notice that for $i \neq j$, $t_i \neq t_j$, and  therefore every conjunction than can be formed from the $e(x_i)$'s and $e(\Bar{x_i})$'s is satisfiable.
    For each $d,k\geq 0$, 
    let $\varphi_d^k = \bigwedge_{j=0}^d [a]^j \bigwedge_{i=1}^k \left(
    (e(x_i) \land \neg e(\Bar{x_i}))
    \lor 
    (\neg e(x_i) \land  e(\Bar{x_i}))
    \right)$. The formula $\varphi_d^k$ asserts that for every process that can be reached with up to $k$ $a$-transitions, for each $1 \leq i \leq k$, exactly one of $e(x_i)$ and $e(\Bar{x_i})$ must be true.

    For each formula $\varphi \in \mathL_{[a]\langle a \rangle x}$ of modal depth $d$ and on variables $\{x_1,x_2,\ldots, x_k\}$, let $\varphi_{-x} = \varphi' \land \varphi_d^k$, where $\varphi'$ is the result of replacing each positive occurrence of $x_i$ in $\varphi$ with $e(x_i)$ and each occurrence of $\Bar{x_i}$ in $\varphi$ with $e(\Bar{x_i})$.
    For each $1 \leq i \leq k$, let $p_i$ be the process that only has $t_i a$ and its prefixes as traces; and let $\neg p_i$ be the process that only has $t_i b$ and its prefixes as traces.
    For each labelled LTS $\mathS$ with only $a$-transitions, we  can define the LTS $\mathS_e$
    that additionally includes the processes $p_i$ and $\neg p_i$ and for every $p$ in $\mathS$ and $1 \leq i \leq k$, it includes an $a$-transition to $p_i$, if $x_i$ is in the labelling of $p$, and an $a$-transition to $\neg p_i$, otherwise.
    It is easy to see that for every $p$ in $\mathS$, $p$ satisfies $\varphi_d^k$; also by straightforward induction on $\varphi$, $p$ satisfies $\varphi$ in $\mathS$ if and only if 
    $p$ satisfies $\varphi'$ in $\mathS_e$.
    Therefore, if $\varphi$ is $\mathL_{\Box\Diamond x}$-satisfiable, then $\varphi_{-x}$ is $\mathL_{\Box\Diamond }$-satisfiable.

    Let $\mathS$ be an LTS; we define $\mathS_x$ to be the labelled LTS that results from labelling each $p$ in $\mathS$ with $\{ x_i \mid b t_i a \in \mathrm{traces}(p) \}$.
    It is then not hard to use induction on $\varphi$ to prove that for every process $p$ in $\mathL$, if $p$ satisfies $\varphi_{-x}$ in $\mathL$, then $p$ satisfies $\varphi$ in $\mathL_x$.
    Therefore, if $\varphi_{-x}$ is $\mathL_{\Box\Diamond }$-satisfiable, then $\varphi$ is $\mathL_{\Box\Diamond x}$-satisfiable.
\end{proof}

\threessat*
\begin{proof}
    It holds that $\mathL_{\Box\Diamond}\subseteq\mathL_{3S}$, and so from Theorem~\ref{thm:2s-validity}, $\mathL_{3S}$-satisfiability is \pspace-hard. Again, \pspace-membership is an immediate implication of the fact that satisfiability for Modal Logic \textbf{K} and \hml is in \pspace \cite{ladner1977computational,HalpernM92}.
\end{proof}

\section{The complexity of primality}\label{section:deciding-primality-appendix}

Proofs of statements in Subection~\ref{section:deciding-primality} are given in detail below.
Let $\mathL$ be $\mathL_S$, $\mathL_{CS}$, or $\mathL_{RS}$ with a bounded action set. From Subsection~\ref{section:deciding-satisfiability} and Appendix~\ref{section:deciding-satisfiability-appendix} satisfiability in $\mathL$ is linear-time decidable. Moreover, in case that the input formula is satisfiable, we can efficiently compute a logically equivalent formula that contains no unsatisfiable formulae. Therefore, we examine formulae of this form when we examine the complexity of primality.

\subsection{Primality in $\mathL_S$}\label{subsection:sim-primality-appendix}

We first provide some simple lemmas. 

\begin{lemma}\label{lem:conjunction_lemma_simulation}
 For every $\varphi_1,\varphi_2,\psi\in \mathL_{S}$, $\varphi_1\wedge\varphi_2\models \langle a\rangle \psi$ iff $\varphi_1\models \langle a\rangle \psi   \text{ or }\varphi_2\models \langle a\rangle \psi$.
\end{lemma}
\begin{proof}
($\Leftarrow$) If $\varphi_1\models \langle a\rangle \psi   \text{ or }\varphi_2\models \langle a\rangle \psi$, then $\varphi_1\wedge\varphi_2\models \langle a\rangle \psi$ immediately holds. \\
($\Rightarrow$) Assume that $\varphi_1\wedge\varphi_2\models \langle a\rangle \psi$ and w.l.o.g.\ $\varphi_1\not\models \langle a\rangle \psi$. We  show that $\varphi_2\models \langle a\rangle \psi$. Since  $\varphi_1\not\models \langle a\rangle \psi$, there is some $p_1$ such that $p_1\models\varphi_1$ and $p_1\not\models \langle a\rangle \psi$. If $\varphi_2$ is not satisfiable, then $\varphi_2\models\langle a\rangle \psi$ trivially holds. Assume now that $\varphi_2$ is satisfiable and let $p_2\models\varphi_2$. Observe that 
$p_1 \curle_S p_1+p_2$ and $p_2 \curle_S p_1+p_2$.  Then, $p_1+p_2\models \varphi_1\wedge\varphi_2$ by Proposition~\ref{logical_characterizations}, and therefore $p_1+p_2\models\langle a\rangle \psi$. This means that either $p_1\myarrowa p'$ or $p_2\myarrowa p'$ for some $p'$ such that $p'\models \psi$. Since $p_1\not\models \langle a\rangle \psi$, we have that $p_2\myarrowa p'$, and so $p_2\models \langle a\rangle \psi$, which was to be shown.
\end{proof}

Lemma~\ref{lem:prime-no-ff-disj} states that all formulae in $\mathL_S$ that do not contain $\ff$ and disjunctions are prime. Similarly to the previous subsection, we associate a process $p_\varphi$ to a formula $\varphi$ of this form and prove that $\varphi$ is characteristic within $\mathL_S$ for $p_\varphi$.

\begin{definition}\label{def:simulation-associated-process}
   Let $\varphi\in\mathL_{S}$ given by the grammar $\varphi::=\true ~|~ \langle a \rangle \varphi ~|~ \varphi\wedge\varphi$. We define process $p_\varphi$ inductively as follows.
\begin{itemize}
    \item If $\varphi=\true$, then $p_\varphi=\mathtt{0}$.
     \item If $\varphi=\langle a\rangle \varphi'$, then $p_\varphi=a. p_{\varphi'}$.
     \item If $\varphi=\varphi_1\wedge \varphi_2$, then $p_\varphi=p_{\varphi_1}+p_{\varphi_2}$.
\end{itemize}
\end{definition}

\begin{lemma}\label{lem:prime-no-ff-disj}
Let $\varphi$ be a formula given by the  grammar $\varphi::=\true~|~ \langle a\rangle \varphi ~|~ \varphi\wedge\varphi$. Then, $\varphi$ is prime. In particular, $\varphi$ is characteristic within $\mathL_S$ for $p_\varphi$. 
\end{lemma}
\begin{proof}  We prove that $\varphi$ is characteristic within $\mathL_S$ for $p_\varphi$. Then, from Proposition~\ref{prop:charact-via-primality}, $\varphi$ is also prime.
Let $p\models\varphi$.  From Corollary~\ref{cor:characteristic}, it suffices to show that $p_\varphi\curle_S p$. We show that by induction on the structure of $\varphi$.
\begin{description}
    \item[Case $\varphi=\true$.] It holds $p_\varphi=\mathtt{0}\curle_S p$, since $\mathtt{0}\curle_S p$ holds for every process $p$.
    \item[Case $\varphi=\langle a\rangle \varphi'$.] As $p\models \langle a\rangle\varphi'$, there is $p\myarrowa p'$ such that $p'\models\varphi'$. By inductive hypothesis, $p_{\varphi'}\curle_S p'$. Thus, $a.p_{\varphi'}\curle_S p$, or $p_\varphi\curle_S p$. 
    \item[Case $\varphi=\varphi_1\wedge \varphi_2$.] As $p\models\varphi_1\wedge\varphi_2$, then $p\models\varphi_1$ and $p\models\varphi_2$. By inductive hypothesis, $p_{\varphi_1}\curle_S p$ and $p_{\varphi_2}\curle_S p$. As noted in Remark~\ref{rem:preorders-preserved-under-plus}, $\curle_S$ is preserved under $+$, and so $p_{\varphi_1}+p_{\varphi_2}\curle_S p+p\sim p$. Hence, $p_\varphi\curle_S p$. 
\end{description}
The inverse implication, namely $p_\varphi\curle_{CS} p\implies p\models\varphi$, can be easily proven by induction on the structure of $\varphi$.
\end{proof}

The next more general statement follows immediately from Lemma~\ref{lem:prime-no-ff-disj} and the fact that unsatisfiable formulae are prime.

\begin{corollary}\label{cor:prime-no-disj}
   Let $\varphi\in\mathL_S$ be a formula given by the  grammar $\varphi::=\true~|~ \ff~|~ \langle a\rangle \varphi ~|~ \varphi\wedge\varphi$. Then, $\varphi$ is prime. 
\end{corollary}

Proposition~\ref{prop:primality-LS} provides a necessary and sufficient condition for primality in $\mathL_S$. 

\begin{proposition}\label{prop:primality-LS}
Let $\varphi\in \mathL_{S}$ and $\bigvee_{i=1}^k \varphi_i$ be $\varphi$ in DNF. Then, $\varphi$ is prime iff $\varphi\models\varphi_j$ for some $1\leq j\leq k$.
\end{proposition}
\begin{proof}
($\Rightarrow$) Assume that $\varphi$ is prime. From Lemma~\ref{lem:DNF-equiv}, $\varphi\equiv\bigvee_{i=1}^k \varphi_i$. By the definition of primality, $\varphi\models\varphi_j$, for some $1\leq j\leq k$. \\
($\Leftarrow$) From Lemma~\ref{lem:DNF-equiv}, it suffices to  show the claim for $\bigvee_{i=1}^k\varphi_i$. Assume that $\bigvee_{i=1}^k\varphi_i\models \varphi_j$, for some $1\leq j\leq k$.  Let $\bigvee_{i=1}^k\varphi_i\models \bigvee_{l=1}^m \phi_l$. From Lemma~\ref{lem:disjunction_lemma}, $\varphi_i\models\bigvee_{l=1}^m \phi_l$, for every $1\leq i\leq k$, and in particular $\varphi_j\models\bigvee_{l=1}^m \phi_l$. But $\varphi_j$ does not contain disjunctions and  Corollary~\ref{cor:prime-no-disj} guarantees that it is prime. Thus, $\varphi_j\models \phi_s$, for some $1\leq s\leq m$, and since $\bigvee_{i=1}^k\varphi_i\models \varphi_j$, it holds that $\bigvee_{i=1}^k\varphi_i\models \phi_s$. 
\end{proof}

To give necessary and sufficient conditions for the primality of formulae in $\mathL_S$ without $\ff$, we examine and state results regarding the DNF of such formulae.

\begin{lemma}\label{lem:disjuncts-of-DNF-no-ff}
    Let $\varphi$ be given by the grammar $\varphi::= \true ~|~ \langle a\rangle \varphi ~|~ \varphi\wedge\varphi  ~|~ \varphi\vee\varphi$; let also $\bigvee_{i=1}^k \varphi_i$ be $\varphi$ in DNF. Then, for every $1\leq i\leq k$, $\varphi_i$ is characteristic for $p_{\varphi_i}$.
\end{lemma}
\begin{proof}
    Every $\varphi_i$ does not contain disjunctions and so it is characteristic for $p_{\varphi_i}$ from Lemma~\ref{lem:prime-no-ff-disj}.
\end{proof}

\begin{lemma}\label{lem:common-divisor-pairs}
    Let $\varphi$ be given by the grammar $\varphi::= \true ~|~ \langle a\rangle \varphi ~|~ \varphi\wedge\varphi  ~|~ \varphi\vee\varphi$;
    let also $\bigvee_{i=1}^k \varphi_i$ be $\varphi$ in DNF. If for every pair $p_{\varphi_i},p_{\varphi_j}$, $1\leq i,j\leq k$, there is some process $q$ such that $q\curle_S p_{\varphi_i}$, $q\curle_S p_{\varphi_j}$, and $q\models\varphi$, then there is some process $q$ such that $q\curle_S p_{\varphi_i}$ for every $1\leq i\leq k$, and $q\models\varphi$. 
\end{lemma}
\begin{proof}  We prove that for every $2\leq m\leq k$ processes $p_{\varphi_{i_1}}, \dots,p_{\varphi_{i_m}}$ there is some process $q$ such that $q\curle_S p_{\varphi_{i_1}},\dots, q\curle_S p_{\varphi_{i_m}}$ and $q\models\varphi$. The proof is by strong induction on $m$.
\begin{description}
    \item[Base case.] Let $m=2$. This is true from the hypothesis of the lemma.
    \item[Inductive step.] Let the argument be true for every $m\leq n-1$. We show that it is true for $m=n$. Let $p_{\varphi_{i_1}}, \dots,p_{\varphi_{i_n}}$ be processes associated to $\varphi_{i_1}, \dots,\varphi_{i_n}$, respectively. Consider the pairs $(p_{\varphi_{i_1}},p_{\varphi_{i_2}})$, $(p_{\varphi_{i_3}},p_{\varphi_{i_4}}),$ $\dots,$
    $(p_{\varphi_{i_{n-1}}},p_{\varphi_{i_n}})$. By inductive hypothesis, there are $q_1,\dots, q_{n/2}$ such that $q_1\curle_S p_{\varphi_{i_1}}, p_{\varphi_{i_2}}$, $q_2\curle_S p_{\varphi_{i_3}}, p_{\varphi_{i_4}}$, $\dots$, $q_{n/2}\curle_S p_{\varphi_{i_{n-1}}},p_{\varphi_{i_n}}$, and $q_i\models\varphi$ for every $1\leq i\leq n/2$. Thus, there are $\varphi_{j_i}$, $1\leq j_i\leq k$, $1\leq i\leq n/2$, such that $q_i\models \varphi_{j_i}$. From Lemma~\ref{lem:disjuncts-of-DNF-no-ff}, every $\varphi_{j_i}$ is characteristic for $p_{\varphi_{j_i}}$ and from Corollary~\ref{cor:characteristic}, $p_{\varphi_{j_i}}\curle_S q_i$ for every  $1\leq j_i\leq k$ and $1\leq i\leq n/2$. By inductive hypothesis, there is $q$ such that $q\curle_S p_{\varphi_{j_1}},\dots,p_{\varphi_{j_{n/2}}}$ and $q\models\varphi$. By transitivity of $\curle_S$, $q\curle_S p_{\varphi_{i_1}}, \dots,p_{\varphi_{i_n}}$.  \qedhere
\end{description}
\end{proof}

\begin{remark}
    Note that if we consider $k$ processes $p_1,\dots, p_k$ and for every pair $p_i$, $p_j$, there is $q\neq\mathtt{0}$ such that $q\curle_S p_i$ and $q\curle_S p_j$, then it may be the case that there is no $q\neq\mathtt{0}$ such that $q\curle_S p_i$ for every $1\leq i\leq k$. For example, this is true for processes $p_1$, $p_2$, and $p_3$, where $p_1\myarrowa \mathtt{0}$, $p_1\myarrowb \mathtt{0}$, $p_1\myarrowc \mathtt{0}$, $p_2\myarrowc \mathtt{0}$, $p_2\myarrowd \mathtt{0}$, $p_2\myarrowe \mathtt{0}$, $p_3\myarrowe \mathtt{0}$, $p_3\myarrowf \mathtt{0}$, and $p_3\myarrowa \mathtt{0}$.
\end{remark}

\begin{corollary}\label{cor:common-divisor-pairs}
    Let $\varphi$ be given by the grammar $\varphi::= \true ~|~ \langle a\rangle \varphi ~|~ \varphi\wedge\varphi  ~|~ \varphi\vee\varphi$;
    let also $\bigvee_{i=1}^k \varphi_i$ be $\varphi$ in DNF. If for every pair $p_{\varphi_i},p_{\varphi_j}$, $1\leq i,j\leq k$, there is some process $q$ such that $q\curle_S p_{\varphi_i}$, $q\curle_S p_{\varphi_j}$, and $q\models\varphi$, then there is some $1\leq m\leq k$, such that $p_{\varphi_m}\curle_S p_{\varphi_i}$ for every $1\leq i\leq k$. 
\end{corollary}
\begin{proof}
    From Lemma~\ref{lem:common-divisor-pairs}, there is $q$ such that $q\curle_S p_{\varphi_i}$ for every $1\leq i\leq k$, and $q\models\varphi$. Consequently, $q\models\varphi_m$ for some $1\leq m\leq k$, so $p_{\varphi_m}\curle_S q$. By transitivity of $\curle_S$, $p_{\varphi_m}\curle_S p_{\varphi_i}$ for every $1\leq i\leq k$. Since, $p_{\varphi_m}\models\varphi_m$, we have that $p_{\varphi_m}\models\varphi$ as well.
\end{proof}

Proposition~\ref{prop:primality-LS-2} provides a necessary and sufficient condition for primality in $\mathL_S$ given that the formulae do not contain $\ff$.

\begin{proposition}\label{prop:primality-LS-2}
    Let $\varphi$ be given by the grammar $\varphi::= \true ~|~ \langle a\rangle \varphi ~|~ \varphi\wedge\varphi  ~|~ \varphi\vee\varphi$;
    let also $\bigvee_{i=1}^k \varphi_i$ be $\varphi$ in DNF. Then, $\varphi$ is prime iff for every pair $p_{\varphi_i},p_{\varphi_j}$, $1\leq i,j\leq k$, there is some process $q$ such that $q\curle_S p_{\varphi_i}$, $q\curle_S p_{\varphi_j}$, and $q\models \varphi$.
\end{proposition}
\begin{proof}
    $(\Leftarrow)$ From Corollary~\ref{cor:common-divisor-pairs}, there is some $1\leq m\leq k$, such that $p_{\varphi_m}\curle_S p_{\varphi_i}$ for every $1\leq i\leq k$. From the fact that $\varphi_i$ is characteristic for $p_{\varphi_i}$ and Corollary~\ref{cor:logical-char-implication}, $\varphi_i\models \varphi_m$ for every $1\leq i\leq m$. From Proposition~\ref{prop:primality-LS}, $\varphi$ is prime. 
    $(\Rightarrow)$ Let $\varphi$ be prime. From Lemma~\ref{lem:prime-no-ff-disj}, there is some $1\leq m
    \leq k$ such that $\varphi\models\varphi_m$. From Lemmas~\ref{lem:DNF-equiv} and~\ref{lem:disjunction_lemma}, $\varphi_i\models\varphi_m$ for all $1\leq i\leq k$. From Corollary~\ref{cor:logical-char-implication}, $p_{\varphi_m}\curle_S p_{\varphi_i}$ for every $1\leq i\leq k$. As a result, for every pair $p_{\varphi_i},p_{\varphi_j}$, it holds that $p_{\varphi_m}\curle_S p_{\varphi_i}$, $p_{\varphi_m}\curle_S p_{\varphi_j}$, and $p_{\varphi_m}\models\varphi_m$, which implies that $p_{\varphi_m}\models\varphi$.
\end{proof}

\begin{corollary}\label{cor:primality-LS-3}
    Let $\varphi$ be given by the grammar $\varphi::= \true ~|~ \langle a\rangle \varphi ~|~ \varphi\wedge\varphi  ~|~ \varphi\vee\varphi$;
    let also $\bigvee_{i=1}^k \varphi_i$ be $\varphi$ in DNF. Then, $\varphi$ is prime iff for every pair $\varphi_i$, $\varphi_j$ there is some $1\leq m \leq k$ such that $\varphi_i\models \varphi_m$ and $\varphi_j\models\varphi_m$.
\end{corollary}
\begin{proof}
  $(\Rightarrow)$ From Proposition~\ref{prop:primality-LS}, if $\varphi$ is prime, then there is some $1\leq m \leq k$ such that $\varphi_i\models \varphi_m$ for every $1\leq i \leq k$. $(\Leftarrow)$ If for every $\varphi_i$, $\varphi_j$ there is some $1\leq m \leq k$ such that $\varphi_i\models \varphi_m$ and $\varphi_j\models\varphi_m$, then from the fact that $\varphi_i,\varphi_j$ are characteristic for $p_{\varphi_i}$ and $p_{\varphi_j}$, and Corollary~\ref{cor:logical-char-implication}, $p_{\varphi_m}\curle_S p_{\varphi_i}$ and $p_{\varphi_m}\curle_S p_{\varphi_j}$. It also holds that  $p_{\varphi_m}\models\varphi$, so from Proposition~\ref{prop:primality-LS-2}, $\varphi$ is prime.
\end{proof}


%

To prove correctness of algorithm $\algos$, we need the following lemmas.

\begin{lemma}
    Let $\varphi\in\mathL_S$ such that $\ff\not\in\sub(\varphi)$. For every formula $\psi$ that appears in vertices of $G_\varphi$, it holds that $\ff\not\in\sub(\psi)$.
\end{lemma}

\begin{lemma}\label{lem:sim-algo-correct-1}
Let $\varphi\in\mathL_S$ such that $\ff\not\in\sub(\varphi)$. If there is an alternating path in $G_\varphi$ from $(\varphi,\varphi\Rightarrow\varphi)$ to \textsc{True}, then $\varphi$ is prime.
\end{lemma}
\begin{proof}
Let $\varphi_1,\varphi_2\Rightarrow \psi$ be a vertex in an alternating path from $(\varphi,\varphi\Rightarrow\varphi)$ to \textsc{True} and $\bigvee_{i=1}^{k_1}\varphi_{1}^i$, $\bigvee_{i=1}^{k_2}\varphi_2^i$, and $\bigvee_{i=1}^{k_3}\psi^{i}$ be $\varphi_1$, $\varphi_2$, and $\psi$ in DNF, respectively. We show that $\varphi_1,\varphi_2 \text{ and } \psi$ satisfy the following property $P_1$: 
\begin{center}
    ``For every $\varphi_1^i,\varphi_2^j$ there is $\psi^k$ such that $\varphi_1^i\models\psi^k$ and $\varphi_2^j\models\psi^k$.''
\end{center}
Edges of $G_\varphi$ correspond to the application of some rule of Table~\ref{tab:S-rules}. In an alternating path from $(\varphi,\varphi\Rightarrow\varphi)$ to \textsc{True}, the edges connect the top vertex of a rule to one or two vertices that correspond to the bottom of the same rule: if a rule is universal, the top vertex connects to two children. Otherwise, it connects to one child. We show that $P_1$ is true for every vertex in the alternating path from $s$ to $t$ by induction on the type of rules, read from their conclusions to their premise. 
\begin{description}
    \item[Case (tt).] $P_1$ is trivial for $\varphi_1$, $\varphi_2$, and $\true$, since for every $\varphi_1^i$, $\varphi_2^j$, it holds that $\varphi_1^i\models\true$ and $\varphi_2^j\models\true$.
    \item[Case (L$\vee_1$).] Since rule (L$\vee_1$) is universal, assume that $P_1$ is true for both $\varphi_1,\varphi,\psi$ and $\varphi_2,\varphi,\psi$. Let $\bigvee_{i=1}^{k_{12}}\varphi_{12}^i$  be $\varphi_1\vee\varphi_2$  in DNF. Then, $\bigvee_{i=1}^{k_{12}}\varphi_{12}^i=\bigvee_{i=1}^{k_1}\varphi_1^i\vee\bigvee_{i=1}^{k_2}\varphi_2^i$. So, $P_1$ holds for $\varphi_1\vee\varphi_2,\varphi,\psi$ as well. Case (L$\vee_2$) is completely analogous.
    \item[Case (L$\wedge_1$).] Since rule (L$\wedge_1$) is existential, assume that $P_1$ is true for either $\varphi_1,\varphi,\langle a\rangle\psi$ or $\varphi_2,\varphi,\langle a\rangle\psi$. Let $\bigvee_{i=1}^{k_{12}}\varphi_{12}^i$ and $\bigvee_{i=1}^k\varphi^i$ be $\varphi_1\wedge\varphi_2$ and $\varphi$ in DNF, respectively. Then, every $\varphi_{12}^i$ is $\varphi_1^j\wedge\varphi_2^k$ for some $1\leq j\leq k_1$ and $1\leq k\leq k_2$. Property $P_1$ for $\varphi_1\wedge\varphi_2,\varphi,\langle a\rangle\psi$ is as follows: for every $\varphi_1^j\wedge\varphi_2^k, \varphi^i$ there is $\langle a\rangle \psi^m$ such that $\varphi_1^j\wedge\varphi_2^k\models\langle a\rangle\psi^m$ and $\varphi^i\models\langle a \rangle\psi^m$. This is true since $\varphi_1^j\wedge\varphi_2^k\models\langle a \rangle\psi^m$ is equivalent to $\varphi_1^j\models\langle a\rangle \psi^m$ or $\varphi_2^k\models\langle a \rangle\psi^m$ from Lemma~\ref{lem:conjunction_lemma_simulation}. Case (L$\wedge_2$) is completely analogous.
    \item[Case (R$\wedge$).] Let $\bigvee_{i=1}^{m_1}\psi_1^i$ and $\bigvee_{i=1}^{m_2}\psi_2^i$, and $\bigvee_{i=1}^{m}\psi_{12}^i$ be $\psi_1$, $\psi_2$, and $\psi_1\wedge\psi_2$ in DNF, respectively. Assume $P_1$ is true for $\varphi_1,\varphi_2,\psi_1$ and $\varphi_1,\varphi_2,\psi_2$, which means that for every  $\varphi_1^i,\varphi_2^j$ there are $\psi_1^{k_1},\psi_2^{k_2}$ such that $\varphi_1^i\models\psi_1^{k_1},\psi_2^{k_2}$ and $\varphi_2^j\models\psi_1^{k_1},\psi_2^{k_2}$. So,  $\varphi_1^i\models\psi_1^{k_1}\wedge\psi_2^{k_2}$ and $\varphi_2^j\models\psi_1^{k_1}\wedge\psi_2^{k_2}$. Since every $\psi_{12}^i$ in the DNF of $\psi_1\wedge\psi_2$ is $\psi_1^j\wedge\psi_2^k$ for some $1\leq j\leq m_1$ and $1\leq k\leq m_2$, $P_1$ is also true for $\varphi_1,\varphi_2,\psi_1\wedge\psi_2$.
    \item[Case (R$\vee$).] Let $\bigvee_{i=1}^{m_1}\psi_1^i$ and $\bigvee_{i=1}^{m_2}\psi_2^i$, and $\bigvee_{i=1}^{m}\psi_{12}^i$ be $\psi_1$, $\psi_2$, and $\psi_1\vee\psi_2$ in DNF, respectively. Assume $P_1$ is true for $\varphi_1,\varphi_2,\psi_1$ or $\varphi_1,\varphi_2,\psi_2$, which means that for every  $\varphi_1^i,\varphi_2^j$ there is $\psi_n^{k}$  such that $\varphi_1^i\models\psi_n^{k}$ and $\varphi_2^j\models\psi_n^{k}$, where $n=1$ or $n=2$. Since every $\psi_{12}^i$ in the DNF of $\psi_1\vee\psi_2$ is either some $\psi_1^j$, $1\leq j\leq m_1$, or $\psi_2^k$, $1\leq k\leq m_2$, $P_1$ is also true for $\varphi_1,\varphi_2,\psi_1\vee\psi_2$.
    \item[Case ($\diamond$).] Assume that $P_1$ is true for $\varphi_1,\varphi_2,\psi$, so for every $\varphi_1^i,\varphi_2^j$, there is some $\psi^k$ such that $\varphi_1^i\models\psi^k$ and $\varphi_2^j\models\psi^k$. The DNFs of $\langle a\rangle\varphi_1$, $\langle a\rangle\varphi_2$, and $\langle a\rangle\psi$ are $\bigvee_{i=1}^{k_1}\langle a\rangle \varphi_1^i$, $\bigvee_{i=1}^{k_2}\langle a\rangle \varphi_2^i$, and $\bigvee_{i=1}^{k_3}\langle a\rangle \psi^i$, respectively.  From Lemma~\ref{lem:diamond_lemma}, $\langle a\rangle\varphi_1^i\models\langle a\rangle\psi^k$ and $\langle a\rangle\varphi_2^j\models\langle a\rangle\psi^k$ hold and 
    $P_1$ is also true for $\langle a\rangle\varphi_1,\langle a\rangle \varphi_2,\langle a\rangle\psi$.
\end{description}
Consequently, if  $(\varphi_1,\varphi_2\Rightarrow\psi)$ is a vertex in $G_\varphi$, $P_1$ is true for $\varphi_1,\varphi_2,\psi$. In particular, $(\varphi,\varphi\Rightarrow\varphi)$ is a vertex in $G_\varphi$. Thus, for every $\varphi^i,\varphi^j$ there is $\varphi^k$ such that $\varphi^i\models\varphi^k$ and $\varphi^j\models\varphi^k$. From Corollary~\ref{cor:primality-LS-3}, $\varphi$ is prime.
\end{proof}

\begin{lemma}\label{lem:sim-algo-correct-2}
 Let $\varphi\in\mathL_S$ such that $\ff\not\in\sub(\varphi)$. If $\varphi$ is prime, then there is an alternating path in $G_\varphi$ from $(\varphi,\varphi\Rightarrow\varphi)$ to \textsc{True}.
\end{lemma}
\begin{proof}
   Assume that $\varphi$ is prime. Let $\varphi_1,\varphi_2,\psi\in\mathL_S$, such that they do not contain $\ff$, and $\bigvee_{i=1}^{k_1}\varphi_{1}^i$, $\bigvee_{i=1}^{k_2}\varphi_2^i$, and $\bigvee_{i=1}^{k_3}\psi^{i}$ be their DNFs, respectively. We say that $\varphi_1,\varphi_2,\psi$ satisfy property $P_2$ if there is $\psi^k$ such that $\varphi_1\models\psi^k$ and $\varphi_2\models\psi^k$. We prove Claims 1 and 2.
   \begin{description}
        \item[Claim 1.]  For every vertex $x=(\varphi_1,\varphi_2\Rightarrow \psi)$  in $G_\varphi$ such that $\psi\neq\true$ and $\varphi_1,\varphi_2,\psi$ satisfy $P_2$, the following are true: 
        \begin{enumerate}[(a)]
            \item One of the rules from Table~\ref{tab:S-rules} can be applied on $x$.
            \item If an existential rule is applied on $x$, then there is some $z=(\varphi_1',\varphi_2'\Rightarrow \psi')$ such that $(x,z)\in E$ and $\varphi_1',\varphi_2',\psi'$ satisfy $P_2$. 
            \item If a universal rule is applied on $x$, then for all $z=(\varphi_1',\varphi_2'\Rightarrow \psi')$ such that $(x,z)\in E$, $\varphi_1',\varphi_2',\psi'$ satisfy $P_2$.
        \end{enumerate}
        \item[Claim 2.] If $x$ is a vertex $(\varphi_1,\varphi_2\Rightarrow\psi)$ in $G_\varphi$ such that $\varphi_1,\varphi_2,\psi$ satisfy $P_2$, then there is an alternating path from $x$ to \textsc{True}.
   \end{description}
\textbf{\textcolor{darkgray}{Proof of Claim 1.}} We first prove statement (a). To this end, suppose that $(\varphi_1,\varphi_2\Rightarrow \psi)$ is a vertex such that no rule from Table~\ref{tab:S-rules} can be applied and  $\varphi_1,\varphi_2,\psi$ satisfy $P_2$. Then, it must be the case that $\varphi_1=\langle a\rangle \varphi_1'$, $\varphi_2=\langle b\rangle \varphi_2'$, and $\psi=\langle c\rangle \psi'$, where $a=b=c$ is not true. Assume that $a=c\neq b$. Hence, there is $\psi^k$ such that $\langle a\rangle \varphi_1'\models\psi^k$ and $\langle b\rangle \varphi_2'\models\psi^k$. However, $\psi^k=\langle a\rangle \psi'$ for some $\psi'$, and $\langle b\rangle \varphi_2'\not\models\langle a\rangle \psi'$, contradiction. The other cases that make $a=b=c$ false can be proven analogously.

We prove parts (b) and (c) of the claim by induction on the type of the rules. 
\begin{description}
    \item[Case (L$\vee_1$).] Assume there is $\psi^k$ such that $\varphi_1\vee\varphi_2\models\psi^k$ and $\varphi\models\psi^k$. From Lemma~\ref{lem:disjunction_lemma}, it holds that $\varphi_1\models\psi^k$ and $\varphi_2\models\psi^k$, and so $P_2$ is true for both $\varphi_1,\varphi,\psi$ and $\varphi_2,\varphi,\psi$. Case (L$\vee_2$) is similar.
    \item[Case (L$\wedge_1$).] Assume there is $\langle a\rangle \psi^k$ such that $\varphi_1\wedge\varphi_2\models\langle a\rangle\psi^k$ and $\varphi\models\langle a\rangle\psi^k$. From Lemma~\ref{lem:conjunction_lemma_simulation}, it holds that $\varphi_1\models\langle a\rangle\psi^k$ or $\varphi_2\models\langle a\rangle\psi^k$, and so $P_2$ is true for either $\varphi_1,\varphi,\langle a\rangle\psi$ or $\varphi_2,\varphi,\langle a\rangle\psi$. Case (L$\wedge_2$) is similar.
    \item[Case (R$\wedge$).] Let $\bigvee_{i=1}^m \psi_{12}^i$ be the DNF of $\psi_1\wedge\psi_2$. Assume there is $\psi_{12}^k$ such that $\varphi_1\models\psi_{12}^k$ and $\varphi_2\models\psi_{12}^k$. Since every $\psi_{12}^k$ is $\psi_1^i\wedge\psi_2^j$ for some $1\leq i\leq k_1$ and $1\leq j\leq k_2$, it holds that $P_2$ is true for both $\varphi_1,\varphi_2,\psi_1$ and $\varphi_1,\varphi_2,\psi_2$.
    \item[Case (R$\vee$).] Let  $\bigvee_{i=1}^m \psi_{12}^i$ denote the DNF of $\psi_1\vee\psi_2$. Then, $\bigvee_{i=1}^m \psi_{12}^i=\bigvee_{i=1}^{k_1}\varphi_1^i\vee\bigvee_{i=1}^{k_2}\varphi_2^i$. This immediately implies that if $P_2$ is true for $\varphi_1,\varphi_2, \psi_1\vee\psi_2$, then $P_2$ is true for $\varphi_1,\varphi_2, \psi_1$ or $\varphi_1,\varphi_2, \psi_2$.
    \item[Case ($\diamond$).] If there is $\langle a \rangle \psi^k$ such that $\langle a\rangle \varphi_1\models \langle a \rangle \psi^k$ and $\langle a\rangle \varphi_2\models \langle a \rangle \psi^k$, then from Lemma~\ref{lem:diamond_lemma}, $\varphi_1\models\psi^k$ and $\varphi_2\models\psi^k$. From the fact that the DNFs of $\langle a\rangle\varphi_1$, $\langle a\rangle\varphi_2$, and $\langle a\rangle\psi$ are $\bigvee_{i=1}^{k_1}\langle a\rangle \varphi_1^i$, $\bigvee_{i=1}^{k_2}\langle a\rangle \varphi_2^i$, and $\bigvee_{i=1}^{k_3}\langle a\rangle \psi^i$, respectively, we have that $P_2$ is true for $\varphi_1, \varphi_2, \psi$.
\end{description}
\textbf{\textcolor{darkgray}{Proof of Claim 2.}} Let $x=(\varphi_1,\varphi_2\Rightarrow\psi)$ be a vertex in $G_\varphi$ such that $\varphi_1,\varphi_2,\psi$ satisfy $P_2$. We prove the claim by induction on the form of $x$.
\begin{description}
   \item[Case $x=(\varphi_1,\varphi_2\Rightarrow\true)$.] In this case, $\varphi_1,\varphi_2,\psi$ satisfy $P_2$ and $P^G(x,\textsc{True})$ trivially holds. 
   \item[Case $x=(\varphi_1\vee\varphi_2\Rightarrow\psi)$.] In this case, $x$ is universal and from Claim 1(c), all $z$ such that $(x,z)\in E$, i.e.\ $z_1=(\varphi_1,\varphi\Rightarrow\psi)$ and $z_2=(\varphi_2,\varphi\Rightarrow\psi)$, are vertices such that  $\varphi_1,\varphi,\psi$ and $\varphi_2,\varphi,\psi$ satisfy $P_2$, respectively. By  inductive hypothesis, $P^G(z_1,\textsc{True})$ and $P^G(z_2,\textsc{True})$. As a result, $P^G(x,\textsc{True})$.
\end{description}
Similarly to the case $x=(\varphi_1\vee\varphi_2\Rightarrow\psi)$ and using Claims 1(b) and (c), Claim 2 can be proven for all the other cases that correspond to the top of the rules in Table~\ref{tab:S-rules}. From Claim 1(a), we know that these are the only forms that vertex $x$ can have. Finally, primality of $\varphi$ and Proposition~\ref{prop:primality-LCS} imply that $P_2$ is true for $\varphi,\varphi,\varphi$, and so there is an alternating path from $(\varphi,\varphi\Rightarrow\varphi)$ to \textsc{True}. 
\end{proof}

Polynomial-time complexity of algorithm $\algos$ on $\varphi$ derives from the polynomial size of $G_\varphi$ and linear-time complexity of \reacha.

\begin{lemma}\label{lem:graph-size}
    Given a formula $\varphi\in\mathL_S$ such that $\ff\not\in\sub(\varphi)$, the size of $G_\varphi$ is polynomial in $|\varphi|$.
\end{lemma}
\begin{proof}
 Let $|\varphi|=n$. To construct $G_\varphi$, we start from $(\varphi,\varphi\Rightarrow\varphi)$ and apply repeatedly rules from Table~\ref{tab:S-rules} until no rule can be applied. Let $x=(\varphi_1,\varphi_2\Rightarrow\psi)$ be a vertex in $G_\varphi$. Apart from (tt), every rule generates new vertices by replacing at least one of $\varphi_1,\varphi_2,\psi$ with one of its subformulae. 
 Thus, every vertex of the form $(\varphi_1',\varphi_2'\Rightarrow\psi')$ is such that all $\varphi_1',\varphi_2',\psi'\in\sub(\varphi)$. Since $|\sub(\varphi)|=\mathcal{O}(n)$, the number of different vertices is at most $\mathcal{O}(n^3)$.
\end{proof}

\simprime*
\begin{proof}
The algorithm \algos described in the main body of the paper (in the proof of Proposition~\ref{prop:sim-primality}) decides whether $\varphi$ is prime in polynomial time. From Lemma~\ref{lem:graph-size} and the linear-time complexity of \reacha \cite{Immerman99}, $\algos$ runs  in $\mathcal{O}(|\varphi|^3)$. 
Correctness of $\algos$ is immediate from Lemmas~\ref{lem:sim-algo-correct-1} and~\ref{lem:sim-algo-correct-2}.
\end{proof}

\begin{corollary}\label{cor:find-p-algo}
Let $\varphi\in\mathL_S$ such that $\ff\not\in\sub(\varphi)$. If $\varphi$ is prime, there is a polynomial-time algorithm that constructs a process for which  $\varphi$ is characteristic within $\mathL_S$.
\end{corollary}
\begin{proof}
    Let $\varphi$ be satisfiable and prime and $p_\varphi$ be a process for which $\varphi$ is characteristic within $\mathL_S$. From Proposition~\ref{prop:primality-LS}, there is $1\leq j\leq k$, such that $\varphi\models\varphi_j$. If $p_j$ denotes a process for which $\varphi_j$ is characteristic within $\mathL_S$, then from Corollary~\ref{cor:logical-char-implication}, $p_j\curle_S p_{\varphi}$ and so $p_j\equiv_S p_\varphi$. Consider now algorithm \algos described in the proof of Proposition~\ref{prop:sim-primality} (in the main body of the paper). When \algos checks whether there is an alternating path in $G_\varphi$ from $s$ to $t$, it can also find an alternating path, denoted here by $\mathP_a$. As we move from the starting vertex $s=(\varphi,\varphi\Rightarrow\varphi)$ to the descendants of $s$ along $\mathP_a$, formula $\varphi$ on the right-hand side of $\Rightarrow$ gets deconstructed to give some $\varphi_j$ such that $\varphi\models\varphi_j$. We construct a process $p_j$ for which $\varphi_j$ is characteristic within $\mathL_S$, by following $\mathP_a$ bottom-up, i.e.\ from $t$ to $s$, and associating a process $p$ to every vertex $x$ in $\mathP_a$. Process $p$ depends only on the right-hand side of $\Rightarrow$ that appears in vertex $x$.  At the end, the process corresponding to $s$ is $p_j$.
    \begin{itemize}
        \item If $x=(\varphi_1,\varphi_2\Rightarrow \true)$ belongs to $\mathP_a$, then  $p=\mathtt{0}$ corresponds to $x$.
        \item If $p$ corresponds to $x=(\varphi_1,\varphi_2\Rightarrow\psi)$ and $y=(\langle a\rangle\varphi_1,\langle a\rangle\varphi_2\Rightarrow \langle a\rangle\psi)$ is the parent of $x$ in $\mathP_a$, then $q=a.p$ corresponds to $y$.
        \item If $p_1$ corresponds to $x_1=(\varphi_1,\varphi_2\Rightarrow \psi_1)$, $p_2$ corresponds to $x_2=(\varphi_1,\varphi_2\Rightarrow \psi_2)$, and $y=(\varphi_1,\varphi_2\Rightarrow \psi_1\wedge\psi_2)$ is the parent of $x_1$ and $x_2$ in $\mathP_a$, then $p_1+p_2$ corresponds to $y$.
        \item If $p\in P$ corresponds to $x=(\varphi_1,\varphi_2\Rightarrow \psi_1)$ and $y=(\varphi_1,\varphi_2\Rightarrow \psi_1\vee\psi_2)$ is the parent of $x$ in $\mathP_a$, then $p$ corresponds to $y$.
        \item If $p\in P$ corresponds to $x=(\varphi_1,\varphi\Rightarrow \langle a\rangle\psi)$ and $y=(\varphi_1\wedge\varphi_2,\varphi\Rightarrow \langle a\rangle\psi)$ (or $y=(\varphi,\varphi_1\wedge\varphi_2\Rightarrow \langle a\rangle\psi)$) is the parent of $x$ in $\mathP_a$, then $p$ corresponds to $y$.
         \item If $p_1\in P$ corresponds to $x_1=(\varphi_1,\varphi\Rightarrow \psi)$, $p_2\in P$ corresponds to $x_2=(\varphi_2,\varphi\Rightarrow \psi)$ and $y=(\varphi_1\vee\varphi_2,\varphi\Rightarrow \psi)$ (or $y=(\varphi,\varphi_1\vee\varphi_2\Rightarrow \psi)$) is the parent of $x_1$ and $x_2$ in $\mathP_a$, then w.l.o.g.\ $p_1$ corresponds to $y$.\qedhere
    \end{itemize}
\end{proof}

\begin{corollary}\label{cor:simulation-decide-characteristic}
   Deciding characteristic formulae in $\mathL_S$ is polynomial-time solvable.
\end{corollary}
\begin{proof}
   Immediate from Propositions~\ref{prop:charact-via-primality}, \ref{prop:s-satisfiability}, \ref{prop:s-satisfiability-satisfiable}, and~\ref{prop:sim-primality}. In particular, deciding characteristic formulae in $\mathL_S$ requires $\mathcal{O}(n^3)$.
\end{proof}

\subsubsection{The case of $|A|=1$}

In the case that $A$ is a singleton, i.e.\ $A=\{a\}$, we prove that every formula given by the grammar
$\varphi::= \true ~|~ \langle a\rangle \varphi ~|~ \varphi\wedge\varphi  ~|~ \varphi\vee\varphi$ 
is characteristic within $\mathL_S$ for some process. In other words, primality of formulae that do not contain unsatisfiable subformulae can be decided in constant time. 

Note that in this case, if $p\curle_S q$, then $\mathrm{traces}(p)\subseteq \mathrm{traces}(q)$, and vice versa.

\begin{definition}
 Let $\varphi$ be given by the grammar $\varphi::= \true ~|~ \langle a\rangle \varphi ~|~ \varphi\wedge\varphi  ~|~ \varphi\vee\varphi$. The trace depth of $\varphi$ is inductively defined as follows.
 \begin{itemize}
     \item $\td(\true)=0$,
     \item $\td(\langle a \rangle \varphi)= 1+\td(\varphi)$,
     \item $\td(\varphi_1\wedge \varphi_2)= \max\{\td(\varphi_1),\td(\varphi_2)\}$,
     \item $\td(\varphi_1\vee \varphi_2)= \min\{\td(\varphi_1),\td(\varphi_2)\}$.
 \end{itemize}
\end{definition}

\begin{lemma}
Let $|A|=1$ and $\varphi$ be given by the grammar $\varphi::= \true ~|~ \langle a\rangle \varphi ~|~ \varphi\wedge\varphi  ~|~ \varphi\vee\varphi$. For every process $p$, $p\models \varphi$ iff there are $i$ and $p'$ such that $p\myarrowai p'$ and $i\geq \td(\varphi)$.
\end{lemma}
\begin{proof}
    We prove both implications separately by induction on the structure of $\varphi$.\\
    $(\Rightarrow)$ We proceed by a case analysis on the form of $\varphi$.
    \begin{description}
        \item [Case $\varphi=\true$.] Trivial since $p\myarrowepsilon p$ holds for every $p$ and $|\varepsilon|\geq 0=\td(\true)$.
        \item [Case $\varphi=\langle a\rangle \varphi'$.] Assume $p\models \varphi$. Then, there is some $p'$ such that $p\myarrowa p'$ and $p'\models\varphi'$. By  inductive hypothesis, $p'\myarrowai p''$ for some $i\geq \td(\varphi')$. So $p\myarrowaiplusone p''$ and $i+1\geq \td(\varphi)$.
        \item [Case $\varphi=\varphi_1\wedge \varphi_2$.] Assume $p\models \varphi$. Then $p\models \varphi_1$ and $p\models \varphi_2$. By inductive hypothesis, $p\myarrowai p_1$ and  $p\myarrowaj p_2$ for some $i,j,p_1$, and $p_2$ such that $i\geq \td(\varphi_1)$ and $j\geq \td(\varphi_2)$. Let $m=\max\{i,j\}$. Then, $p\myarrowamax p_3$ for some $p_3$, and $m\geq\td(\varphi)=\max\{\td(\varphi_1), \td(\varphi_2)\}$.
        \item [Case $\varphi=\varphi_1\vee \varphi_2$.]  Assume, without loss of generality, that $p\models\varphi$ because $p\models\varphi_1$. By  inductive hypothesis, $p\myarrowai p'$ for some $p'$ and $i\geq\td(\varphi_1)$. If $\td(\varphi)=\td(\varphi_1)$, then we are done. Otherwise, $\td(\varphi)=\td(\varphi_2)<\td(\varphi_1)\leq i$ and we are done too.
    \end{description}
    $(\Leftarrow)$ The proof proceeds along similar lines, by a case analysis on the form of $\varphi$.
    \begin{description}
        \item [Case $\varphi=\true$.] Trivial since $p\models\true$ always holds.
        \item [Case $\varphi=\langle a\rangle \varphi'$.]  Assume that  $p\myarrowai p'$ and $i\geq \td(\varphi)$. Since $\td(\varphi)=1+\td(\varphi')$, we have that $p\myarrowa p'' \myarrowaiminusone p'$ for some $p''$ and $i-1\geq\td(\varphi')$. By inductive hypothesis, $p''\models \varphi'$, and thus $p\models \varphi$.
        \item [Case $\varphi=\varphi_1\wedge \varphi_2$.] Assume $p\myarrowai p'$ and $i\geq \td(\varphi)$. Since $\td(\varphi)=\max\{\td(\varphi_1),\td(\varphi_2)\}$, by inductive hypothesis, we have that $p\models \varphi_1$ and $p\models \varphi_2$, which implies that $p\models \varphi_1\wedge \varphi_2$.
        \item [Case $\varphi=\varphi_1\vee \varphi_2$.]  Assume $p\myarrowai p'$ and $i\geq \td(\varphi)=\min\{\td(\varphi_1),\td(\varphi_2)\}$. By  inductive hypothesis, $p\models\varphi_1$ or $p\models\varphi_2$, which implies that $p\models \varphi_1\vee \varphi_2$.\qedhere
    \end{description}
\end{proof}

Our next step is to associate a process $p_\varphi$ to each formula $\varphi$ given by the grammar $\varphi::= \true ~|~ \langle a\rangle \varphi ~|~ \varphi\wedge\varphi  ~|~ \varphi\vee\varphi$. Our intention is to prove that $p_\varphi$ is a process for which $\varphi$ is characteristic.

\begin{definition}
Let $|A|=1$ and $\varphi$ be given by the grammar $\varphi::= \true ~|~ \langle a\rangle \varphi ~|~ \varphi\wedge\varphi  ~|~ \varphi\vee\varphi$. We define process $p_\varphi$ to be  $p_\varphi=a^{\td(\varphi)}$, where $a^0=\mathtt{0}$ and $a^{n+1}=a.a^n$.
\end{definition}

\begin{proposition}
 Let $|A|=1$.  Every $\varphi$ given by the grammar $\varphi::= \true ~|~ \langle a\rangle \varphi ~|~ \varphi\wedge\varphi  ~|~ \varphi\vee\varphi$ is characteristic within $\mathL_S$ for $p_\varphi$.
\end{proposition}
\begin{proof} It suffices to show that for every $q$, $q\models\varphi$ iff $p_\varphi\curle_S q$. This is true since
    $q\models\varphi \Leftrightarrow q\myarrowai q'$  for some $i, q'$ such that $i\geq\td(\varphi)$ $\Leftrightarrow p_\varphi=a^{\td(\varphi)}\curle_S q$.
\end{proof}

\begin{corollary}
    Let $|A|=1$. Every formula given by the grammar $\varphi::= \true ~|~ \langle a\rangle \varphi ~|~ \varphi\wedge\varphi  ~|~ \varphi\vee\varphi$ is characteristic within $\mathL_S$ for some $p$. Moreover, such a $p$ can be constructed in time linear in $|\varphi|$.
\end{corollary}

\subsection{Primality in $\mathL_{CS}$}\label{subsection:cs-primality-appendix}

In this subsection we consider formulae in $\mathL_{CS}$ that contain only satisfiable subformulae and examine the complexity of deciding whether such a formula is prime. We describe a preprocessing phase during which appropriate rules are applied on $\varphi$, so that the primality of the resulting formula $\varphi'$ can give information on the primality of $\varphi$. Moreover, primality of $\varphi'$ can be efficiently decided by  an appropriate variant of algorithm \algos.

First, we make some observations about formulae in $\mathL_{CS}$ that will be used throughout this section. Let  $\varphi\in\mathL_{CS}$ be a satisfiable formula that does not contain disjunctions and $\true$. We associate a process $p_\varphi$ to $\varphi$ and prove that $\varphi$ is characteristic within $\mathL_{CS}$ for $p_\varphi$.

\begin{definition}\label{def:cs-associated-process}
    Let $\varphi\in\mathL_{CS}$ be a 
    formula given by the grammar $\varphi::=\zero ~|~ \langle a \rangle \varphi ~|~ \varphi\wedge\varphi$. We define process $p_\varphi$ inductively as follows.
\begin{itemize}
    \item If $\varphi=\zero$, then $p_\varphi=\mathtt{0}$.
     \item If $\varphi=\langle a\rangle \varphi'$, then $p_\varphi=a. p_{\varphi'}$.
     \item If $\varphi=\varphi_1\wedge \varphi_2$, then $p_\varphi=p_{\varphi_1}+p_{\varphi_2}$.
\end{itemize}
\end{definition}

\begin{lemma}\label{lem:cs-associated-process}
    Let $\varphi\in\mathL_{CS}$ be a satisfiable formula given by the grammar $\varphi::=\zero ~|~ \langle a \rangle \varphi ~|~ \varphi\wedge\varphi$. Then, $\varphi$ is prime.
    In particular, $\varphi$ is characteristic within $\mathL_{CS}$ for $p_\varphi$.
\end{lemma}
 \begin{proof} 
   The proof is analogous to the proof of Lemma~\ref{lem:prime-no-ff-disj}, where we also use the fact that $\varphi$ is satisfiable.
\end{proof}

As a corollary, in the case of complete simulation, all formulae that do not contain disjunctions and $\true$ are prime.

\begin{corollary}\label{cor:grammar_prime_cs}
    Let $\varphi\in\mathL_{CS}$ be given by the grammar $\varphi::= \ff ~|~ \zero ~|~ \varphi\wedge\varphi~|~ \langle a\rangle \varphi$. Then, $\varphi$ is prime.
\end{corollary}

Let $\varphi$ be given by the grammar $\varphi::= \zero ~|~ \langle a\rangle \varphi ~|~ \varphi\wedge\varphi  ~|~ \varphi\vee\varphi$ and $\bigvee_{i=1}^k \varphi_i$ be $\varphi$ in DNF. Propositions~\ref{prop:primality-LCS} and~\ref{prop:primality-LCS-2}, Lemma~\ref{lem:cs-common-divisor-pairs} and Corollaries~\ref{cor:cs-common-divisor-pairs} and~\ref{cor:primality-LCS-3} are variants of Propositions~\ref{prop:primality-LS} and~\ref{prop:primality-LS-2}, Lemma~\ref{lem:common-divisor-pairs} and Corollaries~\ref{cor:common-divisor-pairs} and~\ref{cor:primality-LS-3}, respectively, that hold in the case of complete simulation. Their proofs are completely analogous to those for their respective statements in Subsection~\ref{subsection:sim-primality-appendix}, where we also need that every $\varphi_i$ is prime from Corollary~\ref{cor:grammar_prime_cs}, and every satisfiable $\varphi_i$ is characteristic within $\mathL_{CS}$ for $p_{\varphi_i}$ from Lemma~\ref{lem:cs-associated-process}. 


\begin{proposition}\label{prop:primality-LCS}
     Let $\varphi$ be given by the grammar $\varphi::= \zero ~|~ \langle a\rangle \varphi ~|~ \varphi\wedge\varphi  ~|~ \varphi\vee\varphi$; let also $\bigvee_{i=1}^k \varphi_i$ be $\varphi$ in DNF. Then, 
     $\varphi$ is prime iff $\varphi\models \varphi_j$ for some $1\leq j \leq k$.
\end{proposition}

\begin{lemma}\label{lem:cs-common-divisor-pairs}
    Let $\varphi$ be given by the grammar $\varphi::= \zero ~|~ \langle a\rangle \varphi ~|~ \varphi\wedge\varphi  ~|~ \varphi\vee\varphi$ such that every $\psi\in\sub(\varphi)$ is satisfiable;
    let also $\bigvee_{i=1}^k \varphi_i$ be $\varphi$ in DNF. If for every pair $p_{\varphi_i},p_{\varphi_j}$, $1\leq i,j\leq k$, there is some process $q$ such that $q\curle_{CS} p_{\varphi_i}$, $q\curle_{CS} p_{\varphi_j}$, and $q\models\varphi$, then there is some process $q$ such that $q\curle_{CS} p_{\varphi_i}$ for every $1\leq i\leq k$, and $q\models\varphi$. 
\end{lemma}

\begin{corollary}\label{cor:cs-common-divisor-pairs}
    Let $\varphi$ be given by the grammar $\varphi::= \zero ~|~ \langle a\rangle \varphi ~|~ \varphi\wedge\varphi  ~|~ \varphi\vee\varphi$ such that every $\psi\in\sub(\varphi)$ is satisfiable;
    let also $\bigvee_{i=1}^k \varphi_i$ be $\varphi$ in DNF. If for every pair $p_{\varphi_i},p_{\varphi_j}$, $1\leq i,j\leq k$, there is some process $q$ such that $q\curle_{CS} p_{\varphi_i}$, $q\curle_{CS} p_{\varphi_j}$, and $q\models\varphi$, then there is some $1\leq m\leq k$, such that $p_{\varphi_m}\curle_{CS} p_{\varphi_i}$ for every $1\leq i\leq k$. 
\end{corollary}

\begin{proposition}\label{prop:primality-LCS-2}
    Let $\varphi$ be given by the grammar $\varphi::= \zero ~|~ \langle a\rangle \varphi ~|~ \varphi\wedge\varphi  ~|~ \varphi\vee\varphi$ such that every $\psi\in\sub(\varphi)$ is satisfiable;
    let also $\bigvee_{i=1}^k \varphi_i$ be $\varphi$ in DNF. Then, $\varphi$ is prime iff for every pair $p_{\varphi_i},p_{\varphi_j}$, $1\leq i,j\leq k$, there is some process $q$ such that $q\curle_{CS} p_{\varphi_i}$, $q\curle_{CS} p_{\varphi_j}$, and $q\models \varphi$.
\end{proposition} 

\begin{corollary}\label{cor:primality-LCS-3}
    Let $\varphi$ be given by the grammar $\varphi::= \zero ~|~ \langle a\rangle \varphi ~|~ \varphi\wedge\varphi  ~|~ \varphi\vee\varphi$ such that every $\psi\in\sub(\varphi)$ is satisfiable;
    let also $\bigvee_{i=1}^k \varphi_i$ be $\varphi$ in DNF. Then, $\varphi$ is prime iff for every pair $\varphi_i$, $\varphi_j$ there is some $1\leq m \leq k$ such that $\varphi_i\models \varphi_m$ and $\varphi_j\models\varphi_m$.
\end{corollary}

Let $\varphi\in\mathL_{CS}$ be a formula such that every $\psi\in\sub(\varphi)$ is satisfiable. We transform  $\varphi$ into a formula that we denote by $\varphi^\diamond$, such that in case $\varphi$ is prime, $\varphi^\diamond\equiv\varphi$, $\varphi^\diamond$ is prime, and primality of $\varphi^\diamond$ can be efficiently checked. To this end, we apply a set of rules on $\varphi$. First, we consider the following rewriting rules: $\true\wedge\psi\rulett \psi$ and $\true\vee\psi\rulett\true$ modulo commutativity---i.e.\ we also consider the rules $\psi\wedge\true\rulett \psi$ and $\psi\vee\true\rulett\true$.  We write  $\varphi\rulettsub\varphi'$ if $\varphi'=\varphi[\psi/\psi']$, where $\psi\rulett\psi'$ for some $\psi\in\sub(\varphi)$, and $\varphi\rulettstar\varphi'$ to denote that there is a sequence $\varphi\rulettsub\varphi_1\cdots\rulettsub\varphi'$ and there is no $\varphi''$ such that $\varphi'\rulettsub\varphi''$. 

\begin{lemma}\label{lem:phitt-poly-time-complexity}
    Let $\varphi\in\mathL_{CS}$ and $\varphi\rulettstar\varphi^{tt}$. Then, $\varphi^{tt}$ is unique, can be computed in polynomial time, and $\varphi^{tt}\equiv\varphi$.
\end{lemma}

\begin{lemma}\label{lem:phitt-diamond-true}
     Let $\varphi\in\mathL_{CS}$ and $\varphi\rulettstar\varphi^{tt}$. If $\varphi$ is not a tautology and $\true\in\sub(\varphi^{tt})$, then every occurrence of $\true$ is in the scope of some $\langle a\rangle$.
\end{lemma}

Assume that we have a formula $\varphi\in\mathL_{CS}$, such that for every $\psi\in\sub(\varphi)$, $\psi$ is satisfiable and if $\psi=\true$, then $\psi$ occurs only in the scope of some $\langle a\rangle$. We consider the following rewriting rules modulo commutativity and associativity: 
\begin{enumerate}
    \item $\zero\vee\zero\rulezero \zero$,
    \item $\zero\wedge\varphi\rulezero\zero$, 
    \item $(\zero\vee\varphi_1)\wedge\varphi_2\rulezero \varphi_1\wedge\varphi_2$, where $\varphi_2\neq\zero$ and $\varphi_2\neq\zero\vee\varphi_2'$, 
    \item $(\zero\vee\varphi_1)\wedge(\zero\vee\varphi_2)\rulezero \zero\vee (\varphi_1\wedge\varphi_2)$.
\end{enumerate}  
Note that the following rules can be derived:
\begin{enumerate}\setcounter{enumi}{4}
 \item $\zero\vee (\zero\vee\varphi)\rulezero \zero\vee\varphi$ from rule 1 and associativity, 
    \item $(\zero\vee\varphi_1)\vee\varphi_2\rulezero \zero\vee (\varphi_1\vee\varphi_2)$ from associativity, and  
    \item $(\zero\vee\varphi_1)\vee(\zero\vee\varphi_2)\rulezero \zero\vee (\varphi_1\vee\varphi_2)$ from rule 1, commutativity, and associativity.
\end{enumerate}
We apply these rules on a formula $\varphi$ from the innermost to the outermost subformulae. Formally, we write  $\varphi\rulezerosub\varphi'$ if $\varphi'=\varphi[\psi/\psi']$, where $\psi\rulezero\psi'$ for some $\psi\in\sub(\varphi)$ and there is no $\psi''\in\sub(\psi)$ on which a rule can be applied. For every $\varphi\in\mathL_{CS}$, if there is no $\varphi'$ such that $\varphi\rulezerosub\varphi'$,  we say that $\varphi$ is in \emph{zero normal form}.
We write $\varphi\rulezerostar\varphi'$  if there is a (possibly empty) sequence $\varphi\rulezerosub\varphi_1\cdots\rulezerosub\varphi'$,  
and $\varphi'$ is in zero normal form. 

\begin{lemma}\label{lem:phizero-poly-time-complexity}
    Let $\varphi\in\mathL_{CS}$ and $\varphi\rulezerostar\varphi^{0}$. Then, $\varphi^{0}$ is unique and can be computed in polynomial time.
\end{lemma}

\begin{lemma}\label{lem:guarded-zeros-property}
    Let $\varphi\in\mathL_{CS}$ such that for every $\psi\in\sub(\varphi)$, $\psi$ is satisfiable and if $\psi=\true$, then $\psi$ occurs in the scope of some $\langle a\rangle$; let also  $\varphi\rulezerostar\varphi^0$. Then, (a) $\varphi\equiv\varphi^0$ and (b) $\zero\models\varphi$ iff either $\varphi^0=\zero$ or $\varphi^0=\zero\vee\varphi'$, where $\zero\not\models\varphi'$.
\end{lemma}
\begin{proof} Let $\varphi\rulezerosub\varphi'$, where $\varphi'=\varphi[\psi/\psi']$ and every $\psi''\in\sub(\psi)$ is in zero normal form. It suffices to show that $\psi\equiv\psi'$.  Then, from Lemma~\ref{lem:substitution_bs}, $\varphi\equiv\varphi'$. We prove by mutual induction that (i) $\psi\equiv\psi'$ and (ii) for every $\phi\in\varphi$ such that $\zero\models\phi$, either $\phi\rulezerostar\zero$ or $\phi\rulezerostar\zero\vee\phi'$, where $\zero\not\models\phi'$. For part (i), the only interesting cases are the following two.
\begin{description}
    \item[$\psi=\zero\wedge\psi''$.] In this case $\psi\rulezero \zero$. Note that since $\zero\wedge\psi''$ is satisfiable, there is $p$ such that $p\models \zero\wedge\psi''$, which is equivalent to $p\models\zero$ and $p\models\psi''$. Since $\mathtt{0}$ is the only process satisfying $\zero$, $\mathtt{0}$ is also the only process satisfying both $\zero$ and $\psi''$, which means that  $\psi\equiv\zero$ and (i) holds. 
    \item[$\psi=(\zero\vee\psi_1)\wedge\psi_2$,] where $\psi_2\neq \zero$ and $\psi_2\neq\zero\vee\psi_2'$. In this case, $\psi\rulezero\psi_1\wedge\psi_2$. Since $\psi_2$ is in zero normal form, from inductive hypothesis of (ii), $\zero\not\models\psi_2$.  
    Let $p$ be a process such that $p\models\psi$. It holds that $p\models(\zero\vee\psi_1)\wedge\psi_2$ iff ($p\models\zero$ or $p\models\psi_1$) and $p\models\psi_2$ iff ($p\models \zero$ and $p\models \psi_2$) or ($p\models\psi_1$ and $p\models\psi_2$). Since $\zero\not\models\psi_2$, ($p\models \zero$ and $p\models \psi_2$) is not true. Thus, we have that ($p\models\psi_1$ and $p\models\psi_2$), and $\psi\models\psi_1\wedge\psi_2$. Since $\psi_1\wedge\psi_2\models (\zero\vee\psi_1)\wedge\psi_2$ is also true,  $\psi\equiv\psi_1\wedge\psi_2$ holds. 
    \end{description} 
To prove part (ii), let $\phi\in\sub(\varphi)$ such that $\zero\models\phi$. Note that $\phi$ cannot be $\true$. Therefore, $\phi$ can have one of the following forms.
\begin{description}
    \item[$\phi=\zero$.] Trivial.
    \item[$\phi=\phi_1\vee\phi_2$,] where $\zero\models\phi_i$ for some $i=1,2$. Assume that w.l.o.g.\ $\zero\models\phi_1$ and $\zero\not\models\phi_2$.  Let $\phi_2\rulezerostar\phi_2'$. From inductive hypothesis of (i) and Lemma~\ref{lem:substitution_bs}, $\phi_2\equiv\phi_2'$ and so $\zero\not\models\phi_2'$. From inductive hypothesis, either $\phi_1\rulezerostar\zero$  or $\phi_1\rulezerostar\zero\vee \phi_1'$, where $\zero\not\models\phi_1'$. Thus, there is either a sequence $\phi_1\vee\phi_2\rulezerosub\dots\rulezerosub \zero\vee\phi_2'$, where $\zero\not\models\phi_2'$, or $\phi_1\vee\phi_2\rulezerosub\dots\rulezerosub (\zero\vee\phi_1')\vee\phi_2'\rulezero \zero\vee (\phi_1'\vee\phi_2')$, where $\zero\not\models\phi_1'\vee\phi_2'$, respectively. If $\zero\models\phi_1$ and $\zero\models\phi_2$, the proof is analogous.
    \item[$\phi=\phi_1\wedge\phi_2$,] where $\zero\models\phi_i$ for both $i=1,2$. From inductive hypothesis, for both $i=1,2$, either $\phi_i\rulezerostar\zero$ or $\phi_i\rulezerostar\zero\vee \phi_i'$, where $\zero\not\models\phi_i'$. Assume that for both $i=1,2$, $\phi_i\rulezerostar\zero\vee \phi_i'$, where $\zero\not\models\phi_i'$. Then, there is a sequence $\phi_1\wedge\phi_2\rulezerosub \dots\rulezerosub (\zero\vee \phi_1')\wedge(\zero\vee \phi_2')\rulezero \zero\vee(\phi_1'\wedge\phi_2')$, where $\zero\not\models\phi_1'\wedge\phi_2'$. The other cases can be similarly proven.
\end{description}
 Let $\varphi=\varphi_1\rulezerosub\dots\rulezerosub\varphi_n=\varphi^0$. From (i), for every $1\leq i\leq n-1$, $\varphi_i\equiv\varphi_{i+1}$, and so it holds that $\varphi\equiv\varphi^0$. Part (ii) proven above implies that if $\zero\models\varphi$, then either $\varphi^0=\zero$ or $\varphi^0=\zero\vee\varphi'$, where $\zero\not\models\varphi'$. Conversely, if $\varphi^0=\zero$ or $\varphi^0=\zero\vee\varphi'$, then $\zero\models\varphi^0$ is immediate. From (a), $\zero\models\varphi$ holds as well. Consequently, $\zero\models\varphi$ iff $\varphi^0=\zero$ or $\varphi^0=\zero\vee\varphi'$, where $\zero\not\models\varphi'$.
\end{proof}

\begin{remark}\label{rem:cs-rules}
    Note that since we follow an innermost reduction strategy, when we apply a rule on $(\zero\vee\varphi_1)\wedge \varphi_2$, $\varphi_2$ is already in zero normal form. If $\zero\models\varphi_2$, Lemma~\ref{lem:guarded-zeros-property} guarantees that either $\varphi_2=\zero$ or $\varphi_2=\zero\vee\varphi_2'$, where $\zero\not\models \varphi_2'$, and so either rule 2 or rule 4 is applied, respectively. Otherwise, if $\zero\not\models\varphi_2$, rule 3 is applied. An alternative way to check whether rule 3 is to be applied on $(\zero\vee\varphi_1)\wedge \varphi_2$ is to compute $J(\varphi_2)$ and check that $\emptyset\not\in J(\varphi_2)$, which from Claims 1 and 4 in the proof of Lemma~\ref{lem:J(phi)-property},  is equivalent to $\zero\not\models\varphi_2$.
\end{remark}

Lemma~\ref{lem:conjunction_lemma_simulation} takes the following form in the case of complete simulation.

\begin{lemma}\label{lem:conjunction_lemma_cs}
Let $\varphi_1\wedge\varphi_2\in\mathL_{CS}$ such that for every $\psi\in\sub(\varphi_1\wedge\varphi_2)$, $\psi$ is satisfiable, if $\psi=\true$, then $\true$ occurs in the scope of some $\langle a\rangle$, and $\psi$ is in zero normal form. Then, 
$\varphi_1\wedge\varphi_2\models\langle a\rangle\psi \text{ iff } \varphi_1\models\langle a\rangle\psi \text{ or } \varphi_2\models\langle a\rangle\psi.$
\end{lemma}
\begin{proof}
    $(\Leftarrow)$ If $\varphi_1\models\langle a\rangle\psi$ or $\varphi_2\models\langle a\rangle\psi$, then $\varphi_1\wedge\varphi_2\models\langle a\rangle\psi$ immediately holds.\\
    $(\Rightarrow)$ Let $\varphi_1\wedge\varphi_2\models\langle a\rangle\psi$. We distinguish between the following cases.
    \begin{description}
        \item[$\zero\not\models\varphi_1$ and $\zero\not\models\varphi_2$.] Assume that $\varphi_1\not\models\langle a\rangle\psi$. Then, there is $p_1\neq\mathtt{0}$ such that $p_1\models\varphi_1$ and $p_1\not\models\langle a\rangle \psi$. Let $p_2\models\varphi_2$. Since $\varphi_2$ is satisfiable there is such a process $p_2$ and $p_2\neq\mathtt{0}$ because of $\zero\not\models\varphi_2$. The fact that $p_1,p_2\neq\mathtt{0}$ implies that $p_1\curle_{CS} p_1+p_2$ and $p_2\curle_{CS} p_1+p_2$. From Proposition~\ref{logical_characterizations}, $p_1+p_2\models\varphi_1\wedge\varphi_2$, and so $p_1+p_2\myarrowa p'$ such that $p'\models\psi$. Thus, either $p_1\myarrowa p'$ or $p_2\myarrowa p'$. Since $p_1\not\models\langle a\rangle \psi$, it holds that $p_2\myarrowa p'$. As a result, $\varphi_2\models\langle a\rangle\psi$.
         \item[$\zero\models\varphi_1$ and $\zero\models\varphi_2$.] Then, $\zero\models\varphi_1\wedge\varphi_2$ and $\zero\not\models\langle a\rangle\psi$, which  contradicts our assumption that $\varphi_1\wedge\varphi_2\models\langle a\rangle\psi$. This case is therefore not possible.
        \item[W.l.o.g.\ $\zero\models\varphi_1$ and $\zero\not\models\varphi_2$.] Since $\varphi_1$ is in zero normal form, from Lemma~\ref{lem:guarded-zeros-property}, $\varphi_1=\zero$ or $\varphi_1=\zero\vee\varphi_1'$, where $\zero\not\models\varphi_1'$. Then, either $\varphi_1\wedge\varphi_2=\zero\wedge\varphi_2$ or $\varphi_1\wedge\varphi_2=(\zero\vee\varphi_1')\wedge\varphi_2$, respectively, which implies that either $\varphi_1\wedge\varphi_2$ is unsatisfiable, since $\zero\not\models\varphi_2$, or $\varphi_1\wedge\varphi_2\rulezero \varphi_1'\wedge\varphi_2$, since $\varphi_2\neq\zero$ and $\varphi_2\neq\zero\vee\varphi_2'$. The former case contradicts the satisfiability of $\varphi_1\wedge\varphi_2$ and the latter case contradicts the fact that  $\varphi_1\wedge\varphi_2$ is in zero normal form. So this case is not possible either.\qedhere
    \end{description}
\end{proof}

\begin{lemma}\label{lem:cs-satisfiable-dnf}
    Let $\varphi\in \mathL_{CS}$ be in zero normal form and for every $\psi\in\sub(\varphi)$, $\psi$ is satisfiable, and if $\psi=\true$, then $\true$ occurs in the scope of some $\langle a\rangle$; let also $\bigvee_{i=1}^k\varphi_i$ be $\varphi$ in DNF. Then, $\varphi_i$ is satisfiable for every $1\leq i\leq k$.
\end{lemma}
\begin{proof}
    Consider Algorithm~\ref{algo:dnf} that takes $\varphi$ and returns $\varphi$ in DNF. We prove the lemma by induction on the structure of $\varphi$. 
    \begin{description}
        \item[$\varphi$ does not contain disjunctions.] Trivial.
        \item[$\varphi=\varphi_1\vee\varphi_2$.] The DNF of $\varphi$ is $\varphi_1'\vee\varphi_2'$, where $\varphi_i'$ is the DNF of $\varphi_i$ for both $i=1,2$. By the inductive hyppothesis, the claim is true for $\varphi_i$, $i=1,2$, and so the lemma immediately holds for $\varphi$.
        \item[$\varphi=\langle a\rangle\varphi'$.] The DNF of $\varphi$ is $\bigvee_{i=1}^k\langle a\rangle\varphi'_i$, where $\bigvee_{i=1}^k\varphi'_i$ is the DNF of $\varphi'$. Formula $\varphi'$ satisfies the hypothesis of the lemma, and so from inductive hypothesis, every $\varphi_i'$ is satisfiable, which implies that every $\langle a\rangle\varphi'_i$ is also satisfiable.
        \item[$\varphi=\varphi_1\wedge(\varphi_2\vee\varphi_3)$.] The DNF of $\varphi$ is $\bigvee_{i=1}^{k_{1}}\varphi_{12}^i\vee\bigvee_{i=1}^{k_{2}}\varphi_{13}^i$, where $\bigvee_{i=1}^{k_{1}}\varphi_{12}^i,\bigvee_{i=1}^{k_{2}}\varphi_{13}^i$ are the DNFs of $\varphi_1\wedge\varphi_2$ and $\varphi_1\wedge\varphi_3$, respectively. We show that, for both $j=2,3$, the formula $\varphi_1\wedge\varphi_j$   satisfies the hypothesis of the lemma. Suppose w.l.o.g.\ that $\varphi_1\wedge\varphi_2$ is not satisfiable. Since $\varphi_1,\varphi_2$ are satisfiable, we have that w.l.o.g.\ $\mathtt{0}\models\varphi_1$ and $\mathtt{0}\not\models\varphi_2$. Then, $\mathtt{0}\models\varphi_3$ holds, since otherwise, $\varphi$ is unsatisfiable. Then, either $\varphi_1=\zero$ or $\varphi_1=\zero\vee\varphi_1'$, where $\zero\not\models\varphi_1'$, $\varphi_2\neq\zero$ and $\varphi_2\neq\zero\vee\varphi_j'$, and either $\varphi_3=\zero$ or $\varphi_3=\zero\vee\varphi_3'$,  where $\zero\not\models\varphi_3'$, because of Lemma~\ref{lem:guarded-zeros-property} and the fact that $\varphi_1,\varphi_2$, and $\varphi_3$ are in zero normal form. Any combination of these forms leads to contradiction. For example, assume that $\varphi_1=\zero\vee\varphi_1'$ and $\varphi_3=\zero\vee\varphi_3'$. Then, $\varphi=(\zero\vee\varphi_1')\wedge(\varphi_2\vee(\zero\vee\varphi_3'))\rulezerostar \zero \vee (\varphi'\wedge\varphi_2\wedge\varphi_3')$, which contradicts the fact that $\varphi$ is in zero normal form. Every other case can be addressed in a similar way and proven to lead to contradiction. Consequently, every $\psi\in\sub(\varphi_1\wedge\varphi_j)$ is either a subformula of some $\varphi_i$, $i\in\{1,2,3\}$, or $\varphi_1\wedge\varphi_j$, and so $\psi$ is satisfiable. The other parts of the hypothesis of the lemma immediately hold for both $\varphi_1\wedge\varphi_j$, where $j=2,3$. From inductive hypothesis, for every $1\leq n\leq k_{1}$ and $1\leq m\leq k_{2}$, $\varphi_{12}^m$ and $\varphi_{13}^n$ are satisfiable. This implies that every $\varphi_i$, $1\leq i\leq k$, is satisfiable.\qedhere
    \end{description}
\end{proof}

Finally, we consider the rule $\langle a\rangle \true\rulediam\true$ and rules $\true\vee\psi\rulett\true$, $\true\wedge\psi\rulett\psi$ modulo commutativity. As before, we write  $\varphi\rulediamsub\varphi'$ if $\varphi'=\varphi[\psi/\psi']$, where $\psi\rulediam\psi'$ or $\psi\rulett\psi'$ for some $\psi\in\sub(\varphi)$, and $\varphi\rulediamstar\varphi'$ to denote that there is a sequence $\varphi\rulediamsub\varphi_1\cdots\rulediamsub\varphi'$ and there is no $\varphi''$ such that $\varphi'\rulediamsub\varphi''$. 

\begin{lemma}\label{lem:cs_phi_diamond_poly_time}
    Let $\varphi\in\mathL_{CS}$ be satisfiable and $\varphi\rulediamstar\varphi^{\diamond}$. Then, $\varphi^{\diamond}$ is unique and can be computed in polynomial time.
\end{lemma}

\begin{lemma}\label{lem:cs_true_substitutions}
    Let $\varphi\in\mathL_{CS}$ be satisfiable and $\varphi\rulediamstar\varphi^{\diamond}$. Then, either $\true\not\in\sub(\varphi^{\diamond})$ or $\varphi^{\diamond}=\true$.
\end{lemma}
\begin{proof}
    Immediate from the definition of $\varphi\rulediamstar\varphi^{\diamond}$.
\end{proof}

We prove Lemma~\ref{lem:phitt-vs-phi}, which is one of the main results of this subsection. We first provide some definitions and statements needed in its proof.

\begin{definition}\label{def:cs-associated-process-extended}
    Let $\varphi\in\mathL_{CS}$ be a  
    formula given by the grammar $\varphi::=\zero ~|~ \true ~|~ \varphi\wedge\varphi~\mid ~ \langle a \rangle \varphi $. We define process $p_\varphi$ inductively as follows.
\begin{itemize}
    \item If either $\varphi=\zero$ or $\varphi=\true$ , then $p_\varphi=\mathtt{0}$.
     \item If $\varphi=\langle a\rangle \varphi'$, then $p_\varphi=a. p_{\varphi'}$.
     \item If $\varphi=\varphi_1\wedge \varphi_2$, then $p_\varphi=p_{\varphi_1}+p_{\varphi_2}$.
\end{itemize}
\end{definition}

\begin{lemma}\label{lem:cs-associated-process-extended}
     Let $\varphi\in\mathL_{CS}$ be a  satisfiable
    formula given by the grammar $\varphi::=\zero ~|~ \true ~\mid ~ \varphi\wedge\varphi ~|~\langle a \rangle \varphi  $. Then, $p_\varphi\models\varphi$.
\end{lemma}
\begin{proof}
    We prove the lemma by structural induction on $\varphi$ and limit ourselves to presenting the case when $\varphi = \varphi_1\wedge\varphi_2$.   In the remainder of our argument, we will use the following claim, which can be easily shown by induction on the structure of formulae: 
    \begin{quote}
       Let $\varphi\in\mathL_{CS}$ be a  satisfiable formula given by the grammar $\varphi::=\zero ~|~ \true ~\mid ~ \varphi\wedge\varphi ~|~\langle a \rangle \varphi  $. Then,  $\mathtt{0}\models\varphi$ iff $\varphi\equiv\zero$ or $\varphi \equiv\true$.
    \end{quote}
    Our goal is to show that 
    $p_\varphi=p_{\varphi_1}+p_{\varphi_2} \models \varphi_1\wedge\varphi_2 = \varphi$. 
    
    By the inductive hypothesis, we have that $p_{\varphi_1}\models\varphi_1$ and $p_{\varphi_2}\models\varphi_2$.
    We now proceed by considering the following cases: 
    \begin{enumerate}
        \item Neither $p_{\varphi_1}$ nor $p_{\varphi_2}$ is equivalent to $\mathtt{0}$, 
        \item Both $p_{\varphi_1}$ and $p_{\varphi_2}$ are equivalent to $\mathtt{0}$, and 
        \item W.l.o.g.  $p_{\varphi_1}$ is equivalent to $\mathtt{0}$ and $p_{\varphi_2}$ is not. 
    \end{enumerate}
    In the first case, $p_{\varphi_i} \curle_{CS} p_{\varphi_1}+p_{\varphi_2}$, for $i=1,2$. Therefore,  by Proposition~\ref{logical_characterizations}, $p_{\varphi_1}+p_{\varphi_2}\models \varphi_1\wedge\varphi_2$ and we are done. 

    In the second case, observe, first of all, that $p_{\varphi_1}+p_{\varphi_2}$ is equivalent to $\mathtt{0}$. By the aforementioned claim, we infer that $\varphi_i\equiv \zero$ or $\varphi_i \equiv\true$, for $i=1,2$. Thus, $\varphi_1\wedge\varphi_2 \equiv \zero$ or $\varphi_1\wedge\varphi_2 \equiv \true$.  In both cases, 
    $p_{\varphi_1}+p_{\varphi_2}\models \varphi_1\wedge\varphi_2$ and we are done. 

    In the third case, we use the aforementioned claim to infer that $\varphi_1\equiv \zero$ or $\varphi_1 \equiv\true$. Moreover, $p_{\varphi_1}+p_{\varphi_2}$ is equivalent to $p_{\varphi_2}$. Since $p_{\varphi_2}$  is not equivalent to $\mathtt{0}$, the formula $\varphi_1\wedge\varphi_2$ is satisfiable and $p_{\varphi_2}\models\varphi_2$, it follows that $\varphi_1 \equiv\true$. Thus, $\varphi_1\wedge\varphi_2 \equiv \varphi_2$. Since $p_{\varphi_2} \curle_{CS} p_{\varphi_1}+p_{\varphi_2}$, by Proposition~\ref{logical_characterizations}, $p_{\varphi_1}+p_{\varphi_2}\models \varphi_2 \equiv \varphi_1\wedge\varphi_2$ and we are done. 
\qedhere
\end{proof}

\begin{lemma}\label{lem:phitt-vs-phi}
    Let $\varphi\in \mathL_{CS}$ be in zero normal form and for every $\psi\in\sub(\varphi)$, $\psi$ is satisfiable, and if $\psi=\true$, then $\true$ occurs in the scope of some $\langle a\rangle$; let also $\varphi\rulediamstar\varphi^{\diamond}$. Then, 
     $\varphi$ is prime iff $\varphi^{\diamond}\models \varphi$ and $\varphi^{\diamond}$ is prime.
\end{lemma}
\begin{proof}
   ($\Leftarrow$)   Let $\varphi^{\diamond}\models \varphi$ and $\varphi^{\diamond}$ be prime. It holds that  $\langle a\rangle \true \models\true$, $\true\wedge\psi\models\psi$, and $\true\vee\psi \models\true$, for every $\psi\in\mathL_{CS}$. Thus, from Lemma~\ref{lem:substitution_bs} and the definition of $\varphi^{\diamond}$, $\varphi\models\varphi^{\diamond}$. As a result $\varphi\equiv\varphi^{\diamond}$ and so $\varphi$ is prime.\\
   ($\Rightarrow$) Assume that $\varphi$ is prime. As we just showed,  $\varphi\models \varphi^{\diamond}$. Thus, it suffices to show that $\varphi^{\diamond}\models \varphi$. Then, we have that $\varphi\equiv\varphi^{\diamond}$ and $\varphi^{\diamond}$ is prime.  The proof of $\varphi^{\diamond}\models \varphi$ is by induction on the type of the rules $\psi\rulediamsub\psi'$.
        \begin{itemize}
            \item Let $\varphi^{\diamond}$ be the result of substituting only one occurrence of $\langle a\rangle\true$  with $\true$ in $\varphi$, and $p\models\varphi^{\diamond}$; let also $\bigvee_{i=1}^{k}\varphi^{\diamond}_i$ and $\bigvee_{i=1}^{k'}\varphi_i$ be the DNFs  of $\varphi^{\diamond}$ and $\varphi$, respectively.
            It holds that $p\models \varphi^{\diamond}_i$ for some $1\leq i\leq k$. Formula $\varphi$ is satisfiable and prime, so from Proposition~\ref{prop:charact-via-primality} there is a process $p_{min}$ for which $\varphi$ is characteristic.
            We prove that $p_{min}\curle_{CS} p$, which combined with  Corollary~\ref{cor:characteristic} implies that $p\models\varphi$. If $\varphi^{\diamond}_i=\varphi_j$ for some $1\leq j\leq k'$, then $p\models \varphi$. Otherwise, $\varphi^{\diamond}_i$ coincides with $\varphi_j$ for some $1\leq j\leq k'$, where an occurrence of $\langle a\rangle \true$ has been substituted with $\true$. 
            Consider the process $p_{\varphi^{\diamond}_i}$ constructed from $\varphi^{\diamond}_i$ according to Definition~\ref{def:cs-associated-process-extended}, so that $p_{\varphi^{\diamond}_i}\models\varphi^{\diamond}_i$ as stated in Lemma~\ref{lem:cs-associated-process-extended}. The construction of $p_{\varphi^{\diamond}_i}$ implies that there is some $p_{tt}$ such that $p_{\varphi^{\diamond}_i}\myarrowtau p_{tt}$, $p_{tt}=\mathtt{0}$ and $t\in A^*$, and $p_{tt}$ corresponds to subformula $\true$ that substituted $\langle a\rangle\true$ in $\varphi$. Define process $p_{\varphi^{\diamond}_i}^1$ to be a copy of $p_{\varphi^{\diamond}_i}$ extended with $p_{tt}^1\myarrowa p_1=\mathtt{0}$, and $p_{\varphi^{\diamond}_i}^2$  to be a copy of $p_{\varphi^{\diamond}_i}$ extended with $p_{tt}^2\myarrowa p_2\myarrowa p_3=\mathtt{0}$. From Lemma~\ref{lem:cs-satisfiable-dnf}, $\varphi_j$ is satisfiable and note that $p_{\varphi^{\diamond}_i}^1$ is $p_{\varphi_j}$, which implies that $p_{\varphi^{\diamond}_i}^1\models\varphi_j$ because of Lemma~\ref{lem:cs-associated-process-extended}. Moreover, it immediately holds that $p_{\varphi^{\diamond}_i}^2\models\varphi_j$ as well. Therefore, both $p_{\varphi^{\diamond}_i}^1$ and $p_{\varphi^{\diamond}_i}^2$ satisfy $\varphi$. This means that $p_{min}\curle_{CS} p_{\varphi^{\diamond}_i}^j$ for both $j=1,2$. Let  $p_{min}\myarrowtau q$ for some $t\in A^*$. Then, there are some $q_j$, $j=1,2$, such that $p_{\varphi^{\diamond}_i}^j\myarrowtau q_j$ and $q\curle_{CS} q_j$.  
            \begin{itemize}
                \item Assume that there is some $q$ such that $p_{min}\myarrowtau q$ and $q\curle_{CS} p_1$. Thus, $q=p_1=\mathtt{0}$. On the other hand, $q\curle_{CS} p_2$ does not hold, since $p_2\neq \mathtt{0}$. So there is $p'$ such that $p_{\varphi^{\diamond}_i}^2\myarrowtau p'$, $p'\neq p_2$ and $q\curle_{CS} p'$. Moreover, $p'\neq p_3$, since $p_{\varphi^{\diamond}_i}^2\notmyarrowtau p_3$. But then, $p'$ is a copy of a process $p''$ such that $p_{\varphi^{\diamond}_i}\myarrowtau p''$ and $q\curle_{CS} p''$.
                \item Assume that there is some $q$ such that $p_{min}\myarrowtau q$ and $q\curle_{CS} p'$, where $p' \neq p_1$. Similar arguments can show that there is some $p''$ such that  $p_{\varphi^{\diamond}_i}\myarrowtau p''$ and $q\curle_{CS} p''$.
            \end{itemize} 
            As a result, $p_{min}\curle_{CS} p_{\varphi^{\diamond}_i}$. In a similar way, we can prove that $p_{min}$ is complete-simulated by any process that satisfies $\varphi^{\diamond}_i$, and so $p_{min}\curle_{CS} p$.
            \item Let $\varphi^{\diamond}=\varphi[\true\wedge\psi/\psi]$. From Lemma~\ref{lem:substitution_bs} and the fact that $\varphi^{\diamond}\models\varphi$, as $\true\wedge\psi\equiv\psi$.
            \item Let $\varphi^{\diamond}=\varphi[\true\vee\psi/\true]$. Similarly to the previous case, from Lemma~\ref{lem:substitution_bs} and the fact that  $\varphi^{\diamond}\models\varphi$, since $\true\vee\psi\equiv\true$.\qedhere
        \end{itemize}
\end{proof}

\begin{corollary}\label{cor:phitt-vs-phi}
    Let $\varphi\in\mathL_{CS}$ be a formula such that  every $\psi\in\sub(\varphi)$ is satisfiable; let also $\varphi\rulettstar\varphi^{tt}\rulezerostar\varphi^0\rulediamstar\varphi^{\diamond}$. Then, every $\psi\in\sub(\varphi^\diamond)$ is satisfiable and $\varphi$ is prime iff $\varphi^\diamond\models\varphi$ and $\varphi^\diamond$ is prime.
\end{corollary}
\begin{proof}
    Let $\psi\in \sub(\varphi^{tt})$. Then, there is some $\psi'\in\sub(\varphi)$, such that $\psi'\rulettstar\psi$, and so $\psi'\models\psi$. This implies that $\psi$ is satisfiable. Analogously, we can show that every $\psi\in\sub(\varphi^\diamond)$ is satisfiable.  It holds that $\varphi\equiv\varphi^{tt}\equiv\varphi^0$ from Lemmas~\ref{lem:phitt-poly-time-complexity} and~\ref{lem:guarded-zeros-property}(a). Formula $\varphi^0$ satisfies the hypothesis of Lemma~\ref{lem:phitt-vs-phi} and so $\varphi^0$ is prime iff $\varphi^\diamond\models\varphi^0$ and $\varphi^\diamond$ is prime. Combining the aforementioned facts, we have that $\varphi$ is prime iff $\varphi^\diamond\models\varphi$ and $\varphi^\diamond$ is prime.
\end{proof}

\begin{corollary}\label{cor:primality-phitt}
     Let $\varphi\in\mathL_{CS}$ be a formula such that  every $\psi\in\sub(\varphi)$ is satisfiable; let also $\varphi\rulettstar\varphi^{tt}\rulezerostar\varphi^0\rulediamstar\varphi^{\diamond}$ and $\bigvee_{i=1}^k \varphi^{\diamond}_i$ be $\varphi^{\diamond}$ in DNF. Then, 
     $\varphi^{\diamond}$ is prime iff $\varphi^{\diamond}\models \varphi^{\diamond}_j$ for some $1\leq j \leq k$, such that $\varphi^{\diamond}_j\neq \true$.
\end{corollary}
\begin{proof} Let $\varphi^{\diamond}$ be prime. By the definition of primality and the fact that $\varphi^{\diamond}\models \bigvee_{i=1}^k \varphi^{\diamond}_i$, we have that $\varphi^{\diamond}\models \varphi^{\diamond}_j$ for some $1\leq j \leq k$. Suppose that $\varphi^{\diamond}_j=\true$. Since $\true \vee \psi$,  $\psi\in\mathL_{CS}$,  is not prime, $\bigvee_{i=1}^k \varphi^{\diamond}_i$ is also not prime. From Lemma~\ref{lem:DNF-equiv}, $\varphi^{\diamond}$ is not prime, which contradicts our assumption. So $\varphi^{\diamond}_j\neq\true$. Conversely, let $\varphi^{\diamond} \models \varphi^{\diamond}_j$ for some $1\leq j \leq k$, such that $\varphi^{\diamond}_j\neq\true$. From Lemma~\ref{lem:cs_true_substitutions}, $\varphi^{\diamond}$ and $\varphi_j^{\diamond}$ do not contain $\true$. To prove that $\varphi^{\diamond}$ is prime, let $\varphi^{\diamond}\models \bigvee_{l=1}^m \phi_l$. From Lemmas~\ref{lem:DNF-equiv} and~\ref{lem:disjunction_lemma}, $\varphi^{\diamond}_i\models \bigvee_{l=1}^m \phi_l$, for every $1\leq i\leq k$. In particular, $\varphi^{\diamond}_j\models \bigvee_{l=1}^m \phi_l$. Since $\varphi^{\diamond}_j$ does not contain disjunctions and $\true$, from Corollary~\ref{cor:grammar_prime_cs}, $\varphi^{\diamond}_j$ is prime. Consequently, $\varphi^{\diamond}_j\models \phi_s$ for some $1\leq s \leq m$. Finally, since $\varphi^{\diamond}\models\varphi^{\diamond}_j$, it holds that $\varphi^{\diamond}\models\phi_s$. 
\end{proof}

\begin{example}  
Formula $\varphi=\langle a\rangle\langle a\rangle\true\wedge\langle a\rangle \zero$ is not prime and $\varphi^\diamond=\langle a\rangle \zero \not\models\varphi$, whereas the prime formula $\psi=(\langle a\rangle\true\wedge\langle a\rangle\zero)\vee\langle a\rangle\true$ has $\psi^\diamond=\langle a\rangle\zero$, which logically implies $\psi$.
\end{example}

We can prove now the following main proposition.

\begin{proposition}\label{prop:phitt-primality}
  Let $\varphi\in\mathL_{CS}$ be a formula such that every $\psi\in\sub(\varphi)$ is satisfiable; let also  $\varphi\rulettstar\varphi^{tt}\rulezerostar\varphi^0\rulediamstar\varphi^{\diamond}$. There is a polynomial-time algorithm that decides whether $\varphi^{\diamond}$ is prime.
\end{proposition}
\begin{proof}
    We describe algorithm \algott which takes $\varphi^\diamond$ and decides whether $\varphi^\diamond$ is prime. If $\varphi^\diamond=\true$, \algott rejects. Otherwise, from Lemma~\ref{lem:cs_true_substitutions}, $\true\not\in\varphi^\diamond$. Then, \algott constructs the alternating graph $G_{\varphi^{\diamond}}=(V,E,A,s,t)$ by starting with vertex $(\varphi^\diamond,\varphi^\diamond\Rightarrow\varphi^\diamond)$ and repeatedly applying the rules for complete simulation, i.e.\ the rules from Table~\ref{tab:S-rules}, where rule (tt) is replaced by the following one:
    \begin{prooftree}
    \AxiomC{$\zero,\zero\Rightarrow\zero$ }
    \RightLabel{\scriptsize(0)}
    \UnaryInfC{\textsc{True}}
\end{prooftree}
Then, the algorithm solves \reacha on $G_{\varphi^\diamond}$, where $s$ is $(\varphi^\diamond,\varphi^\diamond\Rightarrow\varphi^\diamond)$ and $t=\textsc{True}$, and accepts $\varphi^\diamond$ iff there is an alternating path from $s$ to $t$.  From Corollary~\ref{cor:phitt-vs-phi}, every $\psi\in\sub(\varphi^\diamond)$ is satisfiable. Correctness of \algott  is  immediate  from the following claim.\\
\noindent\textbf{\textcolor{darkgray}{Claim A.}} $\varphi^\diamond$ is prime iff there is an alternating path in $G_{\varphi^\diamond}$ from $s$ to $t$.\\
\noindent\textbf{\textcolor{darkgray}{Proof of Claim A.}} ($\Leftarrow$) The proof of this implication is similar to the proof of Lemma~\ref{lem:sim-algo-correct-1}. If $(\varphi_1,\varphi_2\Rightarrow\psi)$ is a vertex in the alternating path from $s$ to $t$, then property $P_1$ is true for $\varphi_1,\varphi_2,\psi$ and this can be proven by induction on the type of the rules. We only include two cases that are different here. 
\begin{description}
    \item[Case (0).] $P_1$ trivially holds for $\zero,\zero,\zero$.
    \item[Case (L$\wedge_1$).] Let $\bigvee_{i=1}^{k_{12}}\varphi_{12}^i$ be $\varphi_1\wedge\varphi_2$ in DNF. The argument is the same as in the proof of Lemma~\ref{lem:sim-algo-correct-1}. In particular, if $P_1$ is true either for $\varphi_1,\varphi,\langle a\rangle\psi$ or for $\varphi_2,\varphi,\langle a\rangle\psi$, then either $\varphi_1^{i_1},\varphi^{j}\models\langle a\rangle \psi^{k}$ for some $i_1,j,k$, or $\varphi_2^{i_2},\varphi^j\models\langle a\rangle \psi^k$ for some $i_2,j,k$. From the easy direction of Lemma~\ref{lem:conjunction_lemma_cs}, $\varphi_1^{i_1}\wedge\varphi_2^{i_2},\varphi^j\models\langle a\rangle \psi^k$, where $\varphi_1^{i_1}\wedge\varphi_2^{i_2}=\varphi_{12}^i$ for some $1\leq i\leq k_{12}$.
\end{description}
As a result, $P_1$ is true for $\varphi^\diamond,\varphi^\diamond,\varphi^\diamond$, and from Corollary~\ref{cor:primality-LCS-3}, $\varphi^\diamond$ is prime.\\
($\Rightarrow$) This implication can be proven similarly to Lemma~\ref{lem:sim-algo-correct-2}. We prove the parts that exhibit some differences from those in Lemma~\ref{lem:sim-algo-correct-2}.\\
\noindent\textbf{\textcolor{darkgray}{Claim 1(a).}} For every vertex $x=(\varphi_1,\varphi_2\Rightarrow \psi)$  in $G_{\varphi^{\diamond}}$ such that $\varphi_1,\varphi_2,\psi\neq\zero$ and $\varphi_1,\varphi_2,\psi$ satisfy $P_2$, one of the rules for complete simulation can be applied on $x$.\\
\noindent\textbf{\textcolor{darkgray}{Proof of Claim 1(a).}} If a rule cannot be applied on $x$, then it must be the case that:
\begin{itemize}
    \item either $\varphi_1=\langle a\rangle \varphi_1'$, $\varphi_2=\langle b\rangle \varphi_2'$, and $\psi=\langle c\rangle \psi'$, where $a=b=c$ is not true, which leads to contradiction as we have already proven in the proof of Lemma~\ref{lem:sim-algo-correct-2},
    \item or all of $\varphi_1,\varphi_2,\psi$ are $\langle a\rangle \varphi$ or $\zero$, and there is at least one of each kind. For example, if $\varphi_1=\langle a\rangle \varphi_1'$, $\varphi_2=\langle a\rangle \varphi_2'$, and $\psi=\zero$, then $P_2$ does not hold for $\varphi_1,\varphi_2,\psi$, contradiction. All other cases can be proven similarly.
\end{itemize}
\noindent\textbf{\textcolor{darkgray}{Claim 1(b).}} If an existential rule is applied on $x=(\varphi_1,\varphi_2\Rightarrow \psi)\in V$, where $\varphi_1,\varphi_2,\psi\neq\zero$ and $\varphi_1,\varphi_2,\psi$ satisfy $P_2$, then there is some $z=(\varphi_1',\varphi_2'\Rightarrow \psi')\in V$ such that $(x,z)\in E$ and $\varphi_1',\varphi_2',\psi'$ satisfy $P_2$. \\
\noindent\textbf{\textcolor{darkgray}{Proof of Claim 1(b).}} All cases of the induction proof can be proven in the same manner as in Lemma~\ref{lem:sim-algo-correct-2}. In particular, consider (L$\wedge_i$). The hypothesis of Lemma~\ref{lem:conjunction_lemma_cs} holds for $\varphi_1\wedge\varphi_2$. So, Lemma~\ref{lem:conjunction_lemma_cs} can be used in place of Lemma~\ref{lem:conjunction_lemma_simulation}.\\
\noindent\textbf{\textcolor{darkgray}{Claim 2.}} If $x$ is a vertex $(\varphi_1,\varphi_2\Rightarrow\psi)$ in $G_{\varphi^{\diamond}}$ such that $\varphi_1,\varphi_2,\psi$ satisfy $P_2$, then there is an alternating path from $x$ to \textsc{True}.\\
\noindent\textbf{\textcolor{darkgray}{Proof of Claim 2.}} All cases of the induction proof are the same except for the case $x=(\zero,\zero\Rightarrow\zero)$. Then, $P^G(x,\textsc{True})$ trivially holds.\\
Claim 1(c) needs no adjustment. As $\phi^{\diamond}$ is prime, from Corollary~\ref{cor:primality-phitt}, $\phi^{\diamond},\phi^{\diamond},\phi^{\diamond}$ satisfy $P_2$ and there is an alternating path from $s$ to $t$. This completes the proof of Claim A. 

The polynomial-time complexity of \algott relies on the polynomial size of $G_{\varphi^{\diamond}}$ and linear-time solvability of \reacha.
\end{proof}

\begin{remark}\label{rem:type-ordering-rules}
  At this point, we comment on the type of the rules and the ordering in which they are applied on a given formula $\varphi$ in this subsection. Note that formulae which are satisfied by $\zero$ have a simple zero normal form, i.e.\ their zero normal form is either $\zero$ or $\zero\vee\varphi'$, where $\zero\not\models\varphi'$. This is possible since we initially applied rules $\true\vee\psi\rulett \true$ and $\true\wedge\psi\rulett \psi$ and we obtained $\varphi^{tt}$, such that the zero normal form of every tautology in $\varphi^{tt}$ is also $\zero\vee\varphi'$, where $\zero\not\models\varphi'$. Next, we apply rules that result in the equivalent formula $\varphi^0$, which is in zero normal form. Formula $\varphi^0$ has a DNF $\bigvee_{i=1}^{k}\varphi_i^0$, where every $\varphi_i^0$ is satisfiable as shown in Lemma~\ref{lem:cs-satisfiable-dnf}, which is a crucial property in the proof of Lemma~\ref{lem:phitt-vs-phi}. A formula that is not in zero normal form can have a DNF where some disjuncts are unsatisfiable. For example, the DNF of $\psi=(\langle a\rangle\true\vee\langle b\rangle \true)\wedge(\zero\vee\langle a\rangle \zero)$ is $(\langle a\rangle\true\wedge\zero)\vee(\langle a\rangle\true\wedge\langle a\rangle\zero)\vee(\langle b\rangle \true\wedge\zero)\vee(\langle b\rangle \true\wedge\langle a\rangle\zero)$, where $\langle a\rangle\true\wedge\zero$ and $\langle b\rangle\true\wedge\zero$ are unsatisfiable. However, the zero normal form of $\psi$ is $\psi^0=\zero\vee(\langle a\rangle \zero\wedge\langle a\rangle\true\wedge\langle b\rangle\true)$ which is in DNF and every disjunct is satisfiable. Moreover, Lemma~\ref{lem:conjunction_lemma_cs}, which is necessary for proving our main result, does not hold for formulae that are not in zero normal form. For instance, $(\zero\vee\langle a\rangle\zero)\wedge(\langle a\rangle\true\vee\langle b\rangle\true)\models\langle a\rangle\zero$, but $\zero\vee\langle a\rangle\zero\not\models\langle a\rangle\zero$ and $\langle a\rangle\true\vee\langle b\rangle\true\not \models\langle a\rangle\zero$. Finally, we apply rules to obtain $\varphi^\diamond$, which, in the case that is not $\true$, does not contain $\true$, so it has a DNF the disjuncts of which are satisfiable and prime. As a result, $\varphi^\diamond$ satisfies various desired properties that allow us to use a variant of \algos that checks primality of $\varphi^\diamond$.
\end{remark}

\begin{proposition}\label{prop:find-p-cs-algo-phitt}
    Let $\varphi\in\mathL_{CS}$ be a formula such that every $\psi\in\sub(\varphi)$ is satisfiable; let also  $\varphi\rulettstar\varphi^{tt}\rulezerostar\varphi^0\rulediamstar\varphi^\diamond$. If $\varphi^{\diamond}$ is prime, there is a polynomial-time algorithm that constructs a process for which  $\varphi^{\diamond}$ is characteristic within $\mathL_{CS}$.
\end{proposition}
\begin{proof}
    As in the case of simulation and the proof of Corollary~\ref{cor:find-p-algo}, there is an algorithm that finds an alternating path $\mathP_a$ in $G_{\varphi^\diamond}$ from $s$ to $t$ and associates a process to every vertex of $\mathP_a$ so that the process associated to $s$ is a process for which $\varphi$ is characteristic within $\mathL_{CS}$. 
\end{proof}

\csprime*
\begin{proof}
    We describe algorithm \algocs that decides whether $\varphi$ is prime in polynomial time. On input $\varphi$, \algocs computes $\varphi^{tt},\varphi^0,$ and $\varphi^\diamond$ such that $\varphi\rulettstar\varphi^{tt}\rulettstar\varphi^{0}\rulediamstar\varphi^\diamond$. Then, it checks whether $\varphi^{\diamond}$ is prime by calling \algott($\varphi^{\diamond}$). If $\varphi^{\diamond}$ is not prime, \algocs rejects. Otherwise, \algocs computes $p$ for which $\varphi^{\diamond}$ is characteristic within $\mathL_{CS}$. Finally, it checks whether $p\models\varphi$. \algocs accepts iff $p\models\varphi$.

     If $p\models\varphi$, from Lemma~\ref{def:characteristic}, for every $q$ such that $q\models\varphi^{\diamond}$, it holds that $q\models\varphi$. Thus, $\varphi^{\diamond}\models\varphi$. Correctness of \algocs now follows immediately from Corollary~\ref{cor:phitt-vs-phi}. The polynomial-time complexity of the algorithm is a corollary of Lemmas~\ref{lem:phitt-poly-time-complexity}, \ref{lem:phizero-poly-time-complexity}, and~\ref{lem:cs_phi_diamond_poly_time}, which state that $\varphi^{tt}$, $\varphi^0$, and $\varphi^\diamond$, respectively, can be computed in polynomial time, Proposition~\ref{prop:phitt-primality} that shows polynomial-time complexity of \algott, and Propositions~\ref{prop:find-p-cs-algo-phitt} and~\ref{model-checking-complexity}, which demonstrate that $p$ can be efficiently computed and $p\models\varphi$ can be also solved in polynomial time, respectively.
\end{proof}

\begin{corollary}\label{cor:cs-decide-characteristic}
   Deciding characteristic formulae within $\mathL_{CS}$ is polynomial-time solvable.
\end{corollary}
\begin{proof}
Let $\varphi\in\mathL_{CS}$. Consider the algorithm $\mathrm{Char}_{CS}$ that on input $\varphi$ proceeds as follows: it checks whether $\varphi$ is satisfiable by calling $\conscs(\varphi)$. If $\varphi$ is unsatisfiable, $\mathrm{Char}_{CS}$  rejects $\varphi$. Otherwise, it calls $\subcs(\varphi)$ to compute $\varphi'$ which is logically equivalent to $\varphi$ and contains no unsatisfiable subformulae. Then, it calls \algocs on input $\varphi'$ to decide whether $\varphi'$ is prime. It accepts iff $\algocs(\varphi')$ accepts.
The correctness and the polynomial-time complexity of $\mathrm{Char}_{CS}$ is immediate from Propositions~\ref{prop:charact-via-primality}, \ref{cor:cs-satisfiability}, and~\ref{prop:cs-satisfiability-satisfiable}, and~\ref{prop:cs-primality}.
\end{proof}

\begin{corollary}\label{cor:find-p-cs-algo}
     Let $\varphi\in\mathL_{CS}$. If $\varphi$ is satisfiable and prime, then there is a polynomial-time algorithm that constructs a process for which  $\varphi$ is characteristic within $\mathL_{CS}$.
\end{corollary}

\subsection{Primality in $\mathL_{RS}$}\label{subsection:rs-primality-appendix}

To examine the complexity of deciding prime formulae in $\mathL_{RS}$, we distinguish between  bounded  and unbounded action sets as in the case of satisfiability. We present a polynomial-time algorithm that decides primality when $|A|$ is bounded, and show that the problem is \conp-complete when $|A|$ is unbounded.

\subsubsection{The case of $|A|=k$, $k\geq 1$}\label{subsection:rs-primality-bounded-appendix}

In the case that $|A|=1$, ready simulation coincides with complete simulation. So, we assume that $|A|= k$, $k\geq 2$. We first introduce the notion of \emph{saturated formulae}, which intuitively captures the following  property: if a saturated formula $\varphi$ is satisfied in $p$, then $\varphi$ describes exactly which actions label the outgoing edges of $p$.

\begin{definition}\label{def:saturated-fromula}
    Let $\varphi\in\mathL_{RS}$ such that if $\psi\in\sub(\varphi)$ is unsatisfiable, then $\psi=\ff$ and occurs in the scope of some $[a]$.
    We say that $\varphi$ is saturated if $I(\varphi)$ is a singleton.
\end{definition}

\begin{remark}\label{rem:saturated}
    From now on, when we say that $\varphi$ is saturated, we imply that if $\psi\in\sub(\varphi)$ is unsatisfiable, then $\psi=\ff$ and occurs in the scope of some $[a]$.
\end{remark}

\begin{lemma}\label{lem:rs-not-saturated-property}
    Let $\varphi\in\mathL_{RS}$ such that if $\psi\in\sub(\varphi)$ is unsatisfiable, then $\psi=\ff$ and occurs in the scope of some $[a]$. If $\varphi$ is not saturated, there are two processes $p_1$ and $p_2$, such that $p_i\models\varphi$ for both $i=1,2$ and $I(p_1)\neq I(p_2)$.
\end{lemma}
\begin{proof}
    From the assumptions of the lemma, Definition~\ref{def:saturated-fromula}, and Lemma~\ref{lem:I(phi)-property}, $|I(\varphi)|>1$. Let $S_1,S_2\in I(\varphi)$, such that $S_1\neq S_2$. From Lemma~\ref{lem:I(phi)-property}, there are $p_1$, $p_2$, such that $I(p_i)=S_i$ and $p_i\models\varphi$, where $i\in\{1,2\}$.
\end{proof}

\begin{corollary}\label{cor:rs-not-saturated-property}
    Let $\varphi\in\mathL_{RS}$ such that if $\psi\in\sub(\varphi)$ is unsatisfiable, then $\psi=\ff$ and occurs in the scope of some $[a]$. If $\varphi$ is prime, then $\varphi$ is saturated.
\end{corollary}
\begin{proof}
    Suppose that $\varphi$ is satisfiable, prime, and not saturated. Let $p$ be a process for which $\varphi$ is characteristic within $\mathL_{RS}$. Consider two processes $p_1$, $p_2$ that satisfy $\varphi$ and $I(p_1)\neq I(p_2)$, the existence of which is guaranteed by Lemma~\ref{lem:rs-not-saturated-property}. Then, $p\curle_{RS} p_1$ and $p\curle_{RS} p_2$, contradiction.
\end{proof}

Note that a saturated formula $\varphi$ might not describe exactly the labels of the outgoing edges of processes reachable from $p$, where $p\models\varphi$. To focus on this first level of edges that start from $p$, we construct a propositional formula corresponding to $\varphi$, where any formula of the form $\langle a\rangle \varphi'$ that requires an edge labelled with $a$ leaving from $p$, corresponds to a propositional variable $x_a$.

\begin{definition}\label{def:rs-propositional-formula}
    Let $\varphi\in\mathL_{RS}$ such that if $\psi\in\sub(\varphi)$ is unsatisfiable, then $\psi=\ff$ and occurs in the scope of some $[a]$. The mapping $\sat(\varphi):\mathL_{RS}\rightarrow\mathL_{prop}$, where $\mathL_{prop}$ is the set of propositional formulae, is inductively defined as follows:
    \begin{itemize}
        \item $\sat(\true)=\mathtt{TRUE}$,
        \item $\sat([a]\ff)=\neg x_a$, 
        \item $\sat(\langle a\rangle\varphi')= x_a$,
        \item $\sat(\varphi_1\wedge\varphi_2)= \sat(\varphi_1)\wedge\sat(\varphi_2)$,
        \item $\sat(\varphi_1\vee\varphi_2)= \sat(\varphi_1)\vee\sat(\varphi_2)$,
    \end{itemize}
    where $\mathtt{TRUE}$ denotes a propositional tautology.
\end{definition}

\begin{remark}\label{rem:rs-propositional-formula}
When we construct $\sat(\varphi)$, if $\varphi=\langle a\rangle \varphi'$, we can attach  $\varphi'$ as a label to $x_a$ by setting $\sat(\langle a\rangle\varphi')=x_a^{\varphi'}$, where $\varphi'$ acts as a label for this occerrence of $x_a$. Then, given a propositional formula $\psi$ over the set of variables $\mathrm{VAR}_k=\{x_{a_1},\dots,x_{a_k}\}$, together with labels for each positive occurrence of the variables, we can construct the formula in $\mathL_{RS}$ that corresponds to $\psi$. 
\end{remark}

We prove below that in the case of a saturated formula $\varphi$, $\sat(\varphi)$ has a unique satisfying assignment $s$ and $I(\varphi)$ determines $s$.

\begin{lemma}\label{lem:rs-propositional-formula-sat-assignment}
   Let $\varphi\in\mathL_{RS}$ such that if $\psi\in\sub(\varphi)$ is unsatisfiable, then $\psi=\ff$ and occurs in the scope of some $[a]$. If  $p$ is a process such that $p\models\varphi$, the truth assignment $s:\mathrm{VAR}_k\rightarrow\{\mathrm{true,false}\}$ such that $s(x_a)=\mathrm{true}$ iff $p\myarrowa p'$ for some $p'$ is satisfying for $\sat(\varphi)$. Inversely, if $t$ is a satisfying truth assignment for $\sat(\varphi)$, there is a process $p$ such that $p\models\varphi$ and $p\myarrowa p'$ iff $t(x_a)=\mathrm{true}$.
\end{lemma}
\begin{proof}
    The proof is by induction on the structure of $\varphi$.
 \begin{itemize}
     \item If either $\varphi=\true$  or $\varphi=[a]\ff$, then the lemma can be easily proven.
     \item If $\varphi=\langle a\rangle\varphi'$, then let $p\models\varphi$. It holds that $p\myarrowa p'$, where $p'\models\varphi'$, and perhaps there is  $b\neq a$ such that $p\myarrowb p''$. We have that $\sat(\varphi)=x_a$ and assignment $s$ defined in the lemma is satisfiable for $\sat(\varphi)$. Inversely, if there is a satisfying truth assignment $t$ for $\sat(\varphi)$, then $\sat(\varphi)=x_a$, and so $t(x_a)=\mathrm{true}$. Since $\varphi'$  is satisfiable, there is a process $p'$ such that $p'\models\varphi'$. Consider process and  $p$ such that $p\myarrowa p'$ and $p\myarrowb \mathtt{0}$ for every $x_b\neq x_a$ such that $t(x_b)=\mathrm{true}$. Then, $p\models\varphi$.
     \item Assume that  $\varphi=\varphi_1\wedge\varphi_2$ and let $p\models\varphi$. Since $p\models\varphi_1$ and $p\models\varphi_2$, assignment $s$ defined in the lemma is satisfying for both $\sat(\varphi_1)$ and $\sat(\varphi_2)$ from inductive hypothesis. So, $s$ is also satisfying for $\sat(\varphi_1)\wedge\sat(\varphi_2)=\sat(\varphi_1\wedge\varphi_2)$. Inversely, assume that we have a satisfying assignment $t$ for $\sat(\varphi_1\wedge\varphi_2)=\sat(\varphi_1)\wedge\sat(\varphi_2)$. So, $t$ is satisfying for both $\sat(\varphi_1)$ and $\sat(\varphi_2)$. By inductive hypothesis, there are $p_1$ and $p_2$ such that $p_1\models\varphi_1$ and $p_2\models \varphi_2$, respectively, and $p_1\myarrowa p_1'$ iff $p_2\myarrowa p_2'$ iff $t(x_a)=\mathrm{true}$, which implies that $I(p_1)=I(p_2)$. Consider process $p_1+p_2$. It holds that $p_i\curle_{RS}p_1+p_2$ for both $i=1,2$, and so we have that $p_1+p_2\models \varphi_1\wedge\varphi_2$.
     \item Assume that  $\varphi=\varphi_1\vee\varphi_2$ and let $p\models\varphi$. Since $p\models\varphi_1$ or $p\models\varphi_2$, assignment $s$ defined in the lemma is satisfying for  $\sat(\varphi_1)$ or $\sat(\varphi_2)$ from inductive hypothesis. So, $s$ is also satisfying for $\sat(\varphi_1)\vee\sat(\varphi_2)=\sat(\varphi_1\vee\varphi_2)$. Inversely, let $t$ be a satisfying assignment for $\sat(\varphi_1\vee\varphi_2)=\sat(\varphi_1)\vee\sat(\varphi_2)$. Assume w.l.o.g.\ that $t$ is satisfying for $\sat(\varphi_1)$. By inductive hypothesis, there is $p_1\models \varphi_1$ and $p_1\myarrowa p'$ iff $t(x_a)=\mathrm{true}$. It also holds that $p_1\models\varphi$.\qedhere
 \end{itemize}
\end{proof}

\begin{corollary}\label{cor:rs-propositional-formula-saturated-sat-assignment}
   Let $\varphi\in\mathL_{RS}$ be saturated and $I(\varphi)=\{S\}$. Then, $s:\mathrm{VAR}_k\rightarrow\{\mathrm{true,false}\}$ is a satisfying truth assignment for $\sat(\varphi)$ iff $s(x_a)=\mathrm{true}\Longleftrightarrow a\in S$.
\end{corollary}
\begin{proof}
$(\Rightarrow)$ Let $s$ be a satisfying assignment for $\sat(\varphi)$. Then, from Lemma~\ref{lem:rs-propositional-formula-sat-assignment}, there is $p$ such that $p\models\varphi$ and $a\in I(p)\Longleftrightarrow s(x_a)=\mathrm{true}$. From Lemma~\ref{lem:I(phi)-property}, $I(p)=S$ and so $a\in S\Longleftrightarrow s(x_a)=\mathrm{true}$.\\
$(\Leftarrow)$ Let $s$ be a truth assignment such that $s(x_a)=\mathrm{true}\Longleftrightarrow a\in S$. Then, there is  $p$ such that $p\models\varphi$, which means that $I(p)=S$ from Lemma~\ref{lem:I(phi)-property}.  Thus, $s(x_a)=\mathrm{true}\Longleftrightarrow a\in I(p)$. From Lemma~\ref{lem:rs-propositional-formula-sat-assignment}, $s$ is satisfying for $\sat(\varphi)$.
\end{proof}

Next, if $\varphi$ is saturated, we can simplify $\sat(\varphi)$---and consequently, $\varphi$ as well---so that the resulting propositional formula has a DNF with only satisfiable disjuncts. 

\begin{definition}\label{def:rs-simplification}
    Let $\varphi\in\mathL_{RS}$ be saturated, $I(\varphi)=\{S\}$, and $\psi=\sat(\varphi)$. We denote by $\simpl(\psi)$ the formula we obtain by making the following substitutions in $\psi$:
    \begin{itemize}
        \item for every $a\in S$, substitute $\neg x_a$ with $\mathtt{FALSE}$ in $\psi$,
        \item for every $a\not\in S$, substitute $x_a$ with $\mathtt{FALSE}$ in $\psi$, and
        \item apply rules $\mathtt{FALSE}\vee\psi\rightarrow \psi$ and $\mathtt{FALSE}\wedge\psi\rightarrow\mathtt{FALSE}$ modulo commutativity,
    \end{itemize}
    where $\mathtt{FALSE}$ denotes a propositional contradiction. 
\end{definition}

\begin{lemma}\label{lem:rs-simplification}
    Let $\varphi\in\mathL_{RS}$ be saturated, $I(\varphi)=\{S\}$, and $\psi=\sat(\varphi)$. Then, $\simpl(\psi)$ is logically equivalent to $\psi$. Moreover, $\mathtt{FASLE}$ does not occur in $\simpl(\psi)$, if $a\in S$, then $\neg x_a$ does not occur in $\simpl(\psi)$, and if $a\not\in S$, then $x_a$ does not occur in $\simpl(\psi)$.
\end{lemma}
\begin{proof}
    Let $I(\varphi)=\{S\}$. From Corollary~\ref{cor:rs-propositional-formula-saturated-sat-assignment}, $\psi$ has a unique assignment $s$ and it holds that $s(x_a)=\mathrm{true}\Longleftrightarrow a\in S$. So, if any of the substitutions presented in Definition~\ref{def:rs-simplification}, is made on $\psi$, we obtain a formula that is only satisfied by  truth assignment $s$. It is easy to see that the second part of the lemma is also true.
\end{proof}

\begin{lemma}\label{lem:rs-simplification-dnf}
    Let $\varphi\in\mathL_{RS}$ be saturated, $I(\varphi)=\{S\}$, and $\psi=\sat(\varphi)$. If $\bigvee_{i=1}^k\psi_i$ denotes the DNF of $\simpl(\psi)$, then $\psi_i$ is satisfiable for every $1\leq i\leq k$.
\end{lemma}
\begin{proof}
Let $s:\mathrm{VAR}_k\rightarrow\{\mathrm{true,false}\}$ be such that $s(x_a)=\mathrm{true}$ iff $a\in S$. From Corollary~\ref{cor:rs-propositional-formula-saturated-sat-assignment} and Lemma~\ref{lem:rs-simplification}, $s$ is the only satisfying assignment for $\simpl(\psi)$. The lemma is immediate from the following two claims, which we prove below.
\begin{description}
    \item[Claim 1.] Let $\psi'\in\sub(\psi)$. Then, $s$ is satisfying for $\psi'$.
    \item[Claim 2.] Let $\phi$ be a propositional formula and $\bigvee_{i=1}^n\phi_i$ denote the DNF of $\phi$. If there is a truth assignment $t$ such that is satisfying for every subformula of $\phi$, then $t$ is satisfying for $\phi_i$, for every $1\leq i\leq n$.
\end{description}
\noindent\textbf{\textcolor{darkgray}{Proof of Claim 1.}}  If $s(x_a)=\mathrm{false}$, then $a\not\in S$, and from Lemma~\ref{lem:rs-simplification}, $x_a$ does not appear in $\psi'$. Analogously, if $s(x_a)=\mathrm{true}$, $\neg x_a$ does not appear in $\psi'$. Hence, $s$ assigns the false value only to literals that do not appear in $\psi'$, and so $s$ is satisfying for $\psi'$.\\
\noindent\textbf{\textcolor{darkgray}{Proof of Claim 2.}} We prove the claim by structural induction on $\phi$.
\begin{itemize}
    \item If $\phi$ does not contain disjunctions, then  $\phi$ is already in DNF, and $t$ is satisfying for $\phi$.
    \item If $\phi=\phi_1\vee\phi_2$, then  $t$ is satisfying for  every subformula of $\phi_i$, where $i=1,2$. Let $\bigvee_{i=1}^{k_1}\phi_1^i$ and $\bigvee_{i=1}^{k_2}\phi_2^i$ denote the DNFs of $\phi_1$ and $\phi_2$, respectively. From inductive hypothesis, $t$ is satisfying for $\phi_j^i$, for every $j=1,2$ and $1\leq i\leq k_j$. This is sufficient since $\bigvee_{i=1}^n\phi_i=\bigvee_{i=1}^{k_1}\phi_1^i\vee\bigvee_{i=1}^{k_2}\phi_2^i$.
    \item If $\phi=\phi_1\wedge\phi_2$, then  $t$ is satisfying for  every subformula of $\phi_i$, where $i=1,2$. Let $\bigvee_{i=1}^{k_1}\phi_1^i$ and $\bigvee_{i=1}^{k_2}\phi_2^i$ denote the DNFs of $\phi_1$ and $\phi_2$, respectively.  It holds that for every $1\leq i\leq k$, $\phi_i=\phi_1^{i_1}\wedge \phi_2^{i_2}$ for some $1\leq i_1\leq k_1$ and $1\leq i_2\leq k_2$. Since from inductive hypothesis, $t$ is satisfying for $\phi_j^i$, for every $j=1,2$ and $1\leq i\leq k_j$, $t$ is satisfying for $\phi_i$, for every $1\leq i\leq k$.\qedhere
\end{itemize}
\end{proof}

Properties of the simplified version of $\sat(\varphi)$ can be transferred to $\varphi$.

\begin{definition}\label{def:rs-saturated-simplified}
    Let $\varphi\in\mathL_{RS}$ be saturated and $\psi=\sat(\varphi)$. We denote by $\simpl(\varphi)$ the formula in $\mathL_{RS}$ that corresponds to $\simpl(\psi)$ as described in Remark~\ref{rem:rs-propositional-formula}. We say that $\varphi$ is simplified if $\varphi=\simpl(\varphi)$.
\end{definition}

\begin{corollary}\label{cor:rs-simplification}
    Let $\varphi\in\mathL_{RS}$ be saturated and $I(\varphi)=\{S\}$. Then, $\simpl(\varphi)\equiv\varphi$; if $a\in S$, then any $[a]\ff$ occurs in $\simpl(\varphi)$ only in the scope of some $\langle b\rangle$, $b\in A$; and if $a\not\in S$, then any $\langle a\rangle\varphi'$, where $\varphi'\in\mathL_{RS}$, occurs in $\simpl(\varphi)$ only in the scope of some $\langle b\rangle$, $b\in A$.
\end{corollary}

\begin{corollary}\label{cor:rs-simplification-dnf}
    Let $\varphi\in\mathL_{RS}$ be saturated, $I(\varphi)=\{S\}$, and $\psi=\sat(\varphi)$; let also  $\bigvee_{i=1}^k\psi_i$ denote the DNF of $\simpl(\psi)$ and $\bigvee_{i=1}^k\varphi_i$ denote the formula in $\mathL_{RS}$ that corresponds to $\bigvee_{i=1}^k\psi_i$  as described in Remark~\ref{rem:rs-propositional-formula}. Then, every $\varphi_i$ is satisfiable and disjunctions occur in $\varphi_i$ only in the scope of some $\langle a\rangle$, where $a\in A$.
\end{corollary}



Given a formula $\varphi\in\mathL_{RS}$ such that if $\psi\in\sub(\varphi)$ is unsatisfiable, then $\psi=\ff$ and occurs in the scope of some $[a]$, we process the formula by running Algorithm~\ref{algo:satur} on input $\varphi$. We show that the resulting formula, denoted by $\varphi^s$, has properties that allow us to use a variant of \algos to check its primality. In the case that $\varphi^s$ is prime and logically implies $\varphi$, then $\varphi$ is also prime. As the reader may have already notice, the strategy is similar to the case of complete simulation.

\begin{algorithm}
\caption{Saturation of a formula in $\mathL_{RS}$}
\begin{algorithmic}[1]
\DontPrintSemicolon
\Procedure{Satur}{$\varphi$}
     \State \Repeat{$\varphi'=\varphi$}{
     {$\varphi'\gets\varphi$}\;
    {compute $\varphi^{tt}$ such that $\varphi\rulettstar\varphi^{tt}$\;
      $\varphi\gets\varphi^{tt}$}\;
       \lIf{$|I(\varphi)|\neq 1$}{$\varphi\gets\true$} 
     \lElse{$\varphi\gets\simpl(\varphi)$}
      \ForAll{occurrences of $\langle a\rangle\psi$ in $\varphi$ not in the scope of some $\langle b\rangle$, $b\in A$}
    {\lIf{$|I(\psi)|\neq 1$}{substitute $\langle a\rangle\psi$ with $\true$ in $\varphi$}
     \Else{$\psi\gets\simpl(\psi)$\;
           substitute $\langle a\rangle\psi$ with $\langle a \rangle\Call{Satur}{\psi}$ in $\varphi$}}
    }
     \State {$\varphi\gets\simpl(\varphi)$}
     \State {return $\varphi$\;}
\EndProcedure
\end{algorithmic}
\label{algo:satur}
\end{algorithm}

\begin{lemma}\label{lem:satur-poly-time}
    Let $\varphi\in\mathL_{RS}$ be such that if $\psi\in\sub(\varphi)$ is unsatisfiable, then $\psi=\ff$ and occurs in the scope of some $[a]$. Then, $\Call{Satur}{\varphi}$ runs in polynomial time.
\end{lemma}
\begin{proof}
  It is not hard to see that all steps of Algorithm~\ref{algo:satur} run in polynomial time. Specifically, $I(\varphi)$ can be computed in polynomial time since $|A|=k$, where $k$ is a constant. 
\end{proof}

\begin{lemma}\label{lem:saturated-algo}
    Let $\varphi\in\mathL_{RS}$ be such that if $\psi\in\sub(\varphi)$ is unsatisfiable, then $\psi=\ff$ and occurs in the scope of some $[a]$. Then, $\Call{Satur}{\varphi}$ returns either $\true$ or a saturated and simplified formula $\varphi^s$ such that $\true\not\in\sub(\varphi^s)$ and for every $\langle a\rangle \psi\in\sub(\varphi^s)$, $\psi$ is saturated and simplified. 
\end{lemma}
\begin{proof}
    Immediate from the substitutions made by Algorithm~\ref{algo:satur}.
\end{proof}

\begin{lemma}\label{lem:rs-non-saturated-substitution}
Let $\varphi\in\mathL_{RS}$ be prime, saturated, and simplified; let also $\langle a\rangle \varphi'\in\sub(\varphi)$ such that there is an occurrence of $\langle a\rangle \varphi'$ in $\varphi$, which is not in the scope of any $\langle b\rangle$, $b\in A$, and $\varphi'$ is not saturated. If $\varphi^s$ denotes $\varphi$ where this occurrence of $\langle a\rangle \varphi'$ has been substituted with $\true$, then  $\varphi^s\models\varphi$.
\end{lemma}
\begin{proof}
Consider the propositional formulae $\psi=\sat(\varphi)$ and $\psi^s=\sat(\varphi^s)$. Let $\bigvee_{i=1}^k\psi_i$ and $\bigvee_{i=1}^{k'} \psi^s_i$ denote the DNFs of $\psi$ and $\psi^s$, respectively. Formula $\psi$ satisfies the assumptions of Lemma~\ref{lem:rs-simplification-dnf} and so every $\psi_i$ is satisfiable. 
 Let $\bigvee_{i=1}^k\varphi_i$ and $\bigvee_{i=1}^{k'} \varphi^s_i$ denote the formulae in $\mathL_{RS}$ that correspond to $\bigvee_{i=1}^k\psi_i$ and $\bigvee_{i=1}^{k'} \psi^s_i$,  respectively, as described in Remark~\ref{rem:rs-propositional-formula}. From Corollary~\ref{cor:rs-simplification-dnf}, every $\varphi_i$, $1\leq i\leq k$, is satisfiable and disjunctions occur in $\varphi_i$ and $\varphi^s_j$, $1\leq j\leq k'$, only in the scope of some $\langle b\rangle$, $b\in A$. Moreover, $\bigvee_{i=1}^k\varphi_i\equiv\varphi$ and $\bigvee_{i=1}^{k'} \varphi^s_i\equiv\varphi^s$. 

 Let $p\models \varphi^s$ and $p_{min}$ be a process for which $\varphi$ is characteristic within $\mathL_{RS}$. Such a $p_{min}$ exists since $\varphi$ is satisfiable and prime. We show that $p_{min}\curle_{RS} p$, which combined with  Corollary~\ref{cor:characteristic} implies that $p\models\varphi$. Note that $p\models \varphi^s_i$ for some $1\leq i\leq k'$. If $\varphi^s_i=\varphi_j$ for some $1\leq j\leq k$, then $p\models\varphi$. Otherwise, $\varphi^s_i$ coincides with some $\varphi_j$, where an occurrence of $\langle a\rangle \varphi'$ has been substituted with $\true$. We describe now how to construct a process $p_{\varphi^s_i}$ that satisfies $\varphi^s_i$. 
 We define $p_{\varphi^s_i}$ to be such that for every $\langle a_i\rangle\varphi_{a_i}\in\sub(\varphi^s_i)$, such that $\langle a_i\rangle\varphi_{a_i}$ is not in the scope of some $\langle b\rangle$, $b\in A$, it holds that $p\myarrowa p_{\varphi_{a_i}}$, where $p_{\varphi_{a_i}}$ is some process that satisfies $\varphi_{a_i}$, and $p_{\varphi^s_i}$ has no other outgoing edges. We also consider two copies of $p_{\varphi^s_i}$, namely $p_{\varphi^s_i}^1$ and $p_{\varphi^s_i}^2$, that are as follows:  $p_{\varphi^s_i}^1=p_{\varphi^s_i}+a.p_1$ is  and $p_{\varphi^s_i}^2=p_{\varphi^s_i}+a.p_2$, where $p_1\models\varphi'$ and $p_2\models\varphi'$ and $I(p_1)\neq I(p_2)$. Such processes exist because $\varphi'$ is not saturated and Lemma~\ref{lem:rs-not-saturated-property} holds. Note that $p_{\varphi^s_i}^j\models\varphi_j$ for both $j=1,2$, which means that $p_{\varphi^s_i}^j\models\varphi$ for both $j=1,2$, and $p_{min}\curle_{RS} p_{\varphi^s_i}^j$ for both $j=1,2$.
 Let  $p_{min}\myarrowa q$. Then, there are some $q_j$, $j=1,2$, such that $p_{\varphi^s_i}^j\myarrowa q_j$ and $q\curle_{RS} q_j$.  
            \begin{itemize}
                \item Assume that there is some $q$ such that $p_{min}\myarrowa q$ and $q\curle_{RS} p_1$. Thus, $I(q)=I(p_1)$. On the other hand, $q\curle_{RS} p_2$ does not hold, since $I(p_2)\neq I(p_1)=I(q)$. So there is $p'$ such that $p_{\varphi^{\diamond}_i}^2\myarrowa p'$, $p'\neq p_2$, and $q\curle_{RS} p'$. This means that $p'$ is a copy of a process $p''$ such that $p_{\varphi^s_i}\myarrowa p''$ and $q\curle_{RS} p''$.
                \item Assume that there is some $q$ such that $p_{min}\myarrowtau q$ and $q\curle_{RS} p'$, where $p' \neq p_1$. Simpler arguments can show that there is some $p''$ such that  $p_{\varphi^s_i}\myarrowtau p''$ and $q\curle_{RS} p''$.
            \end{itemize} 
            As a result, $p_{min}\curle_{RS} p_{\varphi^s_i}$. Any process that satisfies $\varphi^s_i$ has the form of $p_{\varphi^s_i}$, so this completes the proof of the lemma.
\end{proof}

\begin{lemma}\label{lem:rs-non-saturated-substitution-2}
Let $\varphi\in\mathL_{RS}$ be prime, saturated, and simplified; let also $\langle a\rangle \varphi'\in\sub(\varphi)$ such that there is an occurrence of $\langle a\rangle \varphi'$ in $\varphi$ in the scope of some $\langle b\rangle$, $b\in A$, and $\varphi'$ is not saturated. If $\varphi^s$ denotes $\varphi$ where this occurrence of $\langle a\rangle \varphi'$ has been substituted with $\true$, then  $\varphi^s\models\varphi$.
\end{lemma}
\begin{proof}
 We provide a  proof sketch. Formally, the proof is by induction on the number of $\langle b\rangle$ in the scope of which $\langle a\rangle\varphi'$ occurs in $\varphi$. Let $\langle b\rangle \psi\in\sub(\varphi)$ be such that $\langle a\rangle\varphi'$ occurs in $\psi$ not in the scope of any $\langle c\rangle$, $c\in A$.   From Corollary~\ref{cor:rs-simplification-dnf}, every saturated formula $\varphi$ corresponds to an equivalent $\bigvee_{i=1}^k\varphi_i$ with the properties outlined in the corollary. By combining all these formulae corresponding to  the saturated formulae examined before this occurrence of $\langle a\rangle\varphi'$ by procedure \textsc{Satur}, we can prove that there is $\bigvee_{i=1}^m\varphi_i\equiv\varphi$, every $\varphi_i$ is satisfiable, $\langle a\rangle \varphi'$ occurs in some $\varphi_i$'s and the structure of $\varphi_i$'s is such that a similar argument to the one in the proof of Lemma~\ref{lem:rs-non-saturated-substitution} works. 
\end{proof}

Next, we show one of the main results of this subsection. 

\begin{proposition}\label{prop:primality-LRS-algo}
    Let $\varphi\in\mathL_{RS}$ be a formula such that  if $\psi\in\sub(\varphi)$ is unsatisfiable, then $\psi=\ff$ and occurs in the scope of some $[a]$; let also $\varphi^s$ denote the output of $\Call{Satur}{\varphi}$. Then, 
    $\varphi$ is prime iff $\varphi^s\models \varphi$ and $\varphi^s$ is prime.
\end{proposition}
\begin{proof}
($\Leftarrow$)   Let $\varphi^s\models \varphi$ and $\varphi^s$ be prime. It holds that  $\langle a\rangle \psi \models\true$, $\true\wedge\psi\models\psi$, and $\true\vee\psi \models\true$, for every $\psi\in\mathL_{RS}$. Thus, from Lemma~\ref{lem:substitution_bs} and the type of substitutions made to compute $\varphi^s$, $\varphi\models\varphi^s$. As a result $\varphi\equiv\varphi^s$ and so $\varphi$ is prime.\\
($\Rightarrow$) Assume that $\varphi$ is prime. As we just showed,  $\varphi\models \varphi^s$. Thus, it suffices to show that $\varphi^s\models \varphi$. Then, we have that $\varphi\equiv\varphi^s$ and $\varphi^s$ is prime.  The proof of $\varphi^s\models \varphi$ is by induction on the type of substitutions made to derive $\varphi^s$ from $\varphi$. 
If either $\varphi^s=\varphi[\true\vee\psi/\true]$ or $\varphi^s=\varphi[\true\wedge\psi/\psi]$, then $\varphi^s\equiv\varphi$ as already shown in the proof of Lemma~\ref{lem:phitt-vs-phi}. If \textsc{Satur} is called on input a prime formula $\varphi$, then $\varphi$ is saturated from Corollary~\ref{cor:rs-not-saturated-property} and substitution in line 4 is not made. Moreover, every recursive call is on saturated formulae and again this type of substitution is not made.
The only interesting case is when $\langle a\rangle \psi$ occurs in a saturated subformula of $\varphi$ not in the scope of some $\langle b\rangle$ and $\psi$ is not saturated. Then, this occurrence of $\langle a\rangle \psi$ is substituted with $\true$ in $\varphi$. Then, Lemmas~\ref{lem:rs-non-saturated-substitution} and~\ref{lem:rs-non-saturated-substitution-2} show that $\varphi^s\models\varphi$.
\end{proof}

 If a formula $\varphi^s\neq\true$ is the output of $\Call{satur}{\varphi}$, where the only unsatisfiable subformulae that $\varphi\in\mathL_{RS}$ contains are occurrences of $\ff$ in the scope of some $[a]$, then $\varphi^s$ has properties shown in the following statements. To start with, a variant of Lemma~\ref{lem:conjunction_lemma_cs} holds for saturated formulae.

\begin{lemma}\label{lem:conjunction_lemma_rs}
Let $\varphi\in\mathL_{RS}$ such that if $\psi\in\sub(\varphi)$ is unsatisfiable, then $\psi=\ff$ and occurs in the scope of some $[a]$; let also $\varphi^s$ be the output of $\Call{satur}{\varphi}$. For every $\varphi_1\wedge\varphi_2,\langle a\rangle\psi\in\sub(\varphi^s)$, the following are true: 
\begin{enumerate}[(a)]
    \item $\varphi_1\wedge\varphi_2\models\langle a\rangle\psi$ iff $\varphi_1\models\langle a\rangle\psi$ or $\varphi_2\models\langle a\rangle\psi$, and
    \item $\varphi_1\wedge\varphi_2\models[a]\ff$ iff $\varphi_1\models[a]\ff$ or $\varphi_2\models[a]\ff$.
\end{enumerate}
\end{lemma}
\begin{proof} Note that in the case of $\varphi^s=\true$, the lemma is trivial. Assume that $\true\not\in\sub(\varphi^s)$. The direction from left to right is easy for both cases. We prove the inverse direction for (a) and (b). Let $\langle a\rangle \varphi'\in\sub(\varphi)$ such that $\varphi_1\wedge\varphi_2$ occurs in $\varphi'$ not in the scope of some $\langle b\rangle$, where $b\in A$. Since $\varphi'$ is saturated, it holds that $I(\varphi')=\{S\}$ for some $S\subseteq A$. \\
(a) ($\Rightarrow$) Since $\true\not\in\sub(\varphi^s)$, then $\psi$ is not a tautology. Let $\varphi_1\wedge\varphi_2\models\langle a\rangle\psi$. Assume that $\varphi_1\not\models\langle a\rangle\psi$, and let $p_1$ such that $p_1\models\varphi_1$ and $p_1\not\models\langle a\rangle \psi$ and $p_2$  be such that $p_2\models\varphi_2$. Then, we construct process $p_i'$ by modifying $p_i$,  so that $I(p_i')=S$ and $p_i'\models\varphi_i$, for $i=1,2$. First, we set $p_i'=p_i+\sum_{a\in S\setminus I(p_i)} a.q$, where $q$ is any process that does not satisfy $\psi$. Second, for every $a\in I(p_i)\setminus S$, we remove every transition $p_i'\myarrowa p'$ from $p_i'$. We argue that $p_i'\models\varphi_i$. If  $a\in S\setminus I(p_i)$, from Corollary~\ref{cor:rs-simplification} and the fact that $\varphi_1\wedge\varphi_2$ occurs in $\varphi'$ not in the scope of some $\langle b\rangle$, if $[a]\ff$  occurs in $\varphi_i$ then it occurs in the scope of some $\langle b\rangle$, $b\in A$. So, $p_i+a.q\models\varphi_i$. Similarly, if $a\in I(p_i)\setminus S$, then any $\langle a\rangle\psi$ can  occur in $\varphi_i$ only in the scope of some $\langle b\rangle$, $b\in A$. Therefore, if $p_i'\models \varphi_i$ and we remove all transitions $p_i'\myarrowa p'$, the resulting process still satisfies $\varphi_i$. We now consider process $p_1'+p_2'$. As $I(p_1'+p_2')=I(p_1')=I(p_2')$, we have that $p_i'\curle_{RS} p_1'+p_2'$, and so $p_1'+p_2'\models\varphi_1\wedge\varphi_2$, which in turn implies that $p_1'+p_2'\models\langle a\rangle\psi$. This means that for some $i=1,2$, $p_i'\myarrowa p'$ such that $p'\models\psi$. Since $p_1\not\models\langle a\rangle \psi$, if $p_1'\myarrowa p''$, then $p_1\myarrowa p''$ and $p''\not\models \psi$, or $p''=q$ and $q\not\models\psi$. So, $p_2\models\langle a\rangle\psi$.\\
(b) ($\Rightarrow$) The proof is along the lines of the previous proof. Assume that $\varphi_1\not\models[a]\psi$, and let $p_1$ such that $p_1\models\varphi_1$ and $p_1\not\models[a] \psi$. Suppose that the same holds for $p_2$ and let $p_2$  such that $p_2\models\varphi_2$ and $p_2\not\models[a] \psi$. So, both $p_1$ and $p_2$ have an $a$-transition. As in the case of (a), we can construct $p_i'$ by removing and adding transitions. In this case,  for every $b\in S\setminus I(p_i)$, we add transitions $b.q$ as above, where $q$ can be any process, and we remove all transitions $p_i\myarrowb p'$ for every $b\in I(p_i)\setminus S$ except for the $a$-transitions. At the end, $p_i'\models\varphi_i$ and $I(p_i)=S\cup\{a\}$. As in the proof of (a), $p_1'\wedge p_2'\models\varphi_1\wedge\varphi_2$ and so $p_1'\wedge p_2'\models [a]\ff$. But  $I(p_1)=I(p_2)=S\cup\{a\}$, contradiction. As a result, $\varphi_2\models [a]\ff$.
\end{proof}

If  $\varphi^s\neq\true$ and it does not contain disjunctions, then $\varphi^s$ is characteristic within $\mathL_{RS}$.

\begin{lemma}\label{lem:rs-associated-process}
    Let $\varphi\in\mathL_{RS}$ be a formula given by the grammar $\varphi::=\varphi\wedge\varphi~\mid ~\langle a\rangle\varphi~\mid ~[a]\ff$. If $\varphi$ is saturated and for every $\langle a\rangle \psi\in\sub(\varphi)$, $\psi$ is saturated, then $\varphi$ is prime and a process $p_\varphi$ for which $\varphi$ is characteristic within $\mathL_{RS}$ can be constructed in polynomial time.
\end{lemma}
\begin{proof}
   Since a saturated formula is satisfiable, $\varphi$ is prime iff $\varphi$ is characteristic within $\mathL_{RS}$. We describe a polynomial-time recursive algorithm that constructs $p_\varphi$ and we prove that $\varphi$ is characteristic for $p_\varphi$. Since $\varphi$ is saturated, $I(\varphi)=\{S\}$, for some $S\subseteq A$. Moreover, w.l.o.g.\ $\varphi=\bigwedge_{i=1}^k\varphi_i$, where $k\geq 1$, every $\varphi_i$ is of the form $\langle a\rangle\varphi'$ or $[a]\ff$, and $\varphi'$ is given by the same grammar as $\varphi$. We construct $p_\varphi$ such that for every $\varphi_i=\langle a\rangle\varphi'$, $p_\varphi\myarrowa p_{\varphi'}$, where  $p_{\varphi'}$ is constructed recursively, and $p_{\varphi}$ has no other outgoing edge. First, we show that $p_\varphi\models\varphi$.  Let $\varphi_i=\langle a\rangle\varphi'$. Then, $p_\varphi\myarrowa p_{\varphi'}$ and from inductive hypothesis, $p_{\varphi'}\models\varphi'$. So, $p_\varphi\models\varphi_i$. If $\varphi_i=[a]\ff$, then there is no $\varphi_i$ such that $\varphi_i=\langle a\rangle\varphi'$, since $\varphi$ is satisfiable. So $p_\varphi\notmyarrowa$ and $p_\varphi\models \varphi_i$. As a result, $p_\varphi\models \varphi_i$ for every $1\leq i\leq k$. To prove that $\varphi$ is characteristic for $p_\varphi$, we show that for every $p$, $p\models\varphi$ iff $p_\varphi\curle_{RS} p$. Let $p\models\varphi$. Since $p_\varphi\models\varphi$ is also true, from Lemma~\ref{lem:I(phi)-property}, $I(p)=I(p_\varphi)=S$. If $p_\varphi\myarrowa p'$, then from construction, $p'=p_{\varphi'}$, where $\varphi_i=\langle a\rangle\varphi'$ for some $1\leq i\leq k$, and as we showed above, $p_{\varphi'}\models\varphi'$. So,  $p\models\varphi_i$ and $p\myarrowa p''$ such that $p''\models\varphi'$. From inductive hypothesis, $p_{\varphi'}\curle_{RS} p''$. So $p_\varphi\curle_{RS} p$. Inversely, assume that $p_\varphi\curle_{RS} p$. Let $\varphi_i=\langle a\rangle \varphi'$. We have that $p_\varphi\myarrowa p_{\varphi'}$ and $p_{\varphi'}\models\varphi'$. Hence, $p\myarrowa p''$ such that $p_{\varphi'}\curle_{RS} p''$, and so $p''\models\varphi'$. So, $p\models\langle a\rangle \varphi'$. Let $\varphi_i=[a]\ff$. Then, $p_\varphi\notmyarrowa$, and so $p\notmyarrowa$, which implies that $p\models [a]\ff$. So, $p\models\varphi_i$, for every $1\leq i\leq k$.
\end{proof}

\begin{lemma}\label{lem:rs-dnf-disjuncts}
    Let $\varphi\in\mathL_{RS}$ be a saturated formula given by the grammar $\varphi::=\varphi\wedge\varphi~\mid~\varphi\vee\varphi~\mid~\langle a\rangle\varphi~\mid~[a]\ff$ such that for every $\langle a\rangle\varphi'\in\sub(\varphi)$, $\varphi'$ is saturated; let also $\bigvee_{i=1}^k\varphi_i$ be the DNF of $\varphi$. Then, $\varphi_i$ is saturated and for every $\langle a\rangle\varphi'\in\sub(\varphi_i)$, $\varphi'$ is saturated.
\end{lemma}
\begin{proof} 
From Lemma~\ref{lem:DNF-equiv}, $\varphi\equiv\bigvee_{i=1}^k\varphi_i$, and so $I(\varphi)=\bigcup_{i=1}^k I(\varphi_i)$. Consequently, $I(\varphi_i)=I(\varphi)$ for every $1\leq i\leq k$, which implies that $I(\varphi_i)$ is a singeton and $\varphi_i$ is saturated. If $\langle a\rangle \varphi'\in\sub(\varphi_i)$ for some $1\leq i\leq k$, then $\langle a\rangle \varphi'\in\sub(\varphi)$ and so $\varphi'$ is saturated.
\end{proof}

 Let $\varphi\in\mathL_{RS}$ be a saturated formula given by the grammar $\varphi::=\varphi\wedge\varphi~\mid~\varphi\vee\varphi~\mid~\langle a\rangle\varphi~\mid~[a]\ff$ such that for every $\langle a\rangle\varphi'\in\sub(\varphi)$, $\varphi'$ is saturated; let also $\bigvee_{i=1}^k\varphi_i$ be the DNF of $\varphi$. The proofs of Proposition~\ref{prop:primality-LRS}--Corollary~\ref{cor:primality-LRS-3} are analogous to the proofs of Proposition~\ref{prop:primality-LS}--Corollary~\ref{cor:primality-LS-3}, where here we also make use of the fact that every $\varphi_i$ is characteristic for $p_{\varphi_i}$ from Lemmas~\ref{lem:rs-associated-process} and~\ref{lem:rs-dnf-disjuncts}.

\begin{proposition}\label{prop:primality-LRS}
    Let $\varphi\in\mathL_{RS}$ be a saturated formula given by the grammar $\varphi::=\varphi\wedge\varphi~\mid~\varphi\vee\varphi~\mid~\langle a\rangle\varphi~\mid~[a]\ff$ such that for every $\langle a\rangle\varphi'\in\sub(\varphi)$, $\varphi'$ is saturated; let also $\bigvee_{i=1}^k\varphi_i$ be the DNF of $\varphi$. Then, 
     $\varphi$ is prime iff $\varphi\models \varphi_j$ for some $1\leq j \leq k$.
\end{proposition}

\begin{lemma}\label{lem:rs-common-divisor-pairs}
    Let $\varphi\in\mathL_{RS}$ be a saturated formula given by the grammar $\varphi::=\varphi\wedge\varphi~\mid~\varphi\vee\varphi~\mid~\langle a\rangle\varphi~\mid~[a]\ff$ such that for every $\langle a\rangle\varphi'\in\sub(\varphi)$, $\varphi'$ is saturated; let also $\bigvee_{i=1}^k\varphi_i$ be the DNF of $\varphi$. If for every pair $p_{\varphi_i},p_{\varphi_j}$, $1\leq i,j\leq k$, there is some process $q$ such that $q\curle_{RS} p_{\varphi_i}$, $q\curle_{RS} p_{\varphi_j}$, and $q\models\varphi$, then there is some process $q$ such that $q\curle_{RS} p_{\varphi_i}$ for every $1\leq i\leq k$, and $q\models\varphi$. 
\end{lemma}

\begin{corollary}\label{cor:rs-common-divisor-pairs}
   Let $\varphi\in\mathL_{RS}$ be a saturated formula given by the grammar $\varphi::=\varphi\wedge\varphi~\mid~\varphi\vee\varphi~\mid~\langle a\rangle\varphi~\mid~[a]\ff$ such that for every $\langle a\rangle\varphi'\in\sub(\varphi)$, $\varphi'$ is saturated; let also $\bigvee_{i=1}^k\varphi_i$ be the DNF of $\varphi$. If for every pair $p_{\varphi_i},p_{\varphi_j}$, $1\leq i,j\leq k$, there is some process $q$ such that $q\curle_{RS} p_{\varphi_i}$, $q\curle_{RS} p_{\varphi_j}$, and $q\models\varphi$, then there is some $1\leq m\leq k$, such that $p_{\varphi_m}\curle_{RS} p_{\varphi_i}$ for every $1\leq i\leq k$. 
\end{corollary}

\begin{proposition}\label{prop:primality-LRS-2}
    Let $\varphi\in\mathL_{RS}$ be a saturated formula given by the grammar $\varphi::=\varphi\wedge\varphi~\mid~\varphi\vee\varphi~\mid~\langle a\rangle\varphi~\mid~[a]\ff$ such that for every $\langle a\rangle\varphi'\in\sub(\varphi)$, $\varphi'$ is saturated; let also $\bigvee_{i=1}^k\varphi_i$ be the DNF of $\varphi$. Then, $\varphi$ is prime iff for every pair $p_{\varphi_i},p_{\varphi_j}$, $1\leq i,j\leq k$, there is some process $q$ such that $q\curle_{RS} p_{\varphi_i}$, $q\curle_{RS} p_{\varphi_j}$, and $q\models \varphi$.
\end{proposition} 

\begin{corollary}\label{cor:primality-LRS-3}
    Let $\varphi\in\mathL_{RS}$ be a saturated formula given by the grammar $\varphi::=\varphi\wedge\varphi~\mid~\varphi\vee\varphi~\mid~\langle a\rangle\varphi~\mid~[a]\ff$ such that for every $\langle a\rangle\varphi'\in\sub(\varphi)$, $\varphi'$ is saturated; let also $\bigvee_{i=1}^k\varphi_i$ be the DNF of $\varphi$. Then, $\varphi$ is prime iff for every pair $\varphi_i$, $\varphi_j$ there is some $1\leq m \leq k$ such that $\varphi_i\models \varphi_m$ and $\varphi_j\models\varphi_m$.
\end{corollary}

\begin{corollary}\label{cor:primality-phis}
     Let $\varphi\in\mathL_{RS}$ be a formula such that if $\psi\in\sub(\varphi)$ is unsatisfiable, then $\psi=\ff$ and occurs in the scope of some $[a]$; let also $\varphi^s$ be the output of $\Call{satur}{\varphi}$  and $\bigvee_{i=1}^k \varphi^s_i$ be $\varphi^s$ in DNF. Then, 
     $\varphi^s$ is prime iff $\varphi^s\models \varphi^s_j$ for some $1\leq j \leq k$, such that $\varphi^s_j\neq \true$.
\end{corollary}
\begin{proof} The proof is analogous to the proof of Corollary~\ref{cor:primality-phitt}, where here we need that from Lemmas~\ref{lem:saturated-algo}, \ref{lem:rs-associated-process}, and~\ref{lem:rs-dnf-disjuncts}, $\varphi^s_j$ is prime, for every $1\leq i\leq k$.
\end{proof}

\begin{proposition}\label{prop:phis-primality}
  Let $\varphi\in\mathL_{RS}$ be a formula such that if $\psi\in\sub(\varphi)$ is unsatisfiable, then $\psi=\ff$ and occurs in the scope of some $[a]$; let also  $\varphi^s$ be the output of $\Call{satur}{\varphi}$. There is a polynomial-time algorithm that decides whether $\varphi^s$ is prime.
\end{proposition}
\begin{proof}
    We describe algorithm \algosat, which takes $\varphi^s$ and decides whether $\varphi^s$ is prime. If $\varphi^s=\true$, \algosat rejects. Otherwise, from Lemma~\ref{lem:saturated-algo}, $\true\not\in\varphi^s$. Then, \algosat constructs the alternating graph $G_{\varphi^s}=(V,E,A,s,t)$ by starting with vertex $(\varphi^s,\varphi^s\Rightarrow\varphi^s)$ and repeatedly applying the rules for ready simulation, i.e.\ the rules from Table~\ref{tab:S-rules}, where rule (L$\wedge_i$), where $i=1,2$, is replaced by the following two rules:\\
    \begin{minipage}{0.45 \textwidth}
        \begin{prooftree}
       \AxiomC{$\varphi_1\wedge\varphi_2,\varphi\Rightarrow \langle a\rangle\psi$}
       \RightLabel{\scriptsize(L$\wedge_i^\diamond$)}
        \UnaryInfC{$\varphi_1,\varphi\Rightarrow \langle a\rangle \psi~|_{\exists}~ \varphi_2,\varphi\Rightarrow \langle a\rangle \psi$}
        \end{prooftree}
    \end{minipage}
    \begin{minipage}{0.45 \textwidth}
        \begin{prooftree}
       \AxiomC{$\varphi_1\wedge\varphi_2,\varphi\Rightarrow [a]\ff$}
       \RightLabel{\scriptsize(L$\wedge_i^\Box$)}
        \UnaryInfC{$\varphi_1,\varphi\Rightarrow [a]\ff~|_{\exists}~ \varphi_2,\varphi\Rightarrow [a]\ff$}
        \end{prooftree}
    \end{minipage}\\
    
    \noindent and (tt) is replaced by the rule ($\Box$) as given below:
     \begin{prooftree}
        \AxiomC{$[a]\ff,[a]\ff\Rightarrow[a]\ff$ }
        \RightLabel{\scriptsize($\Box$)}
         \UnaryInfC{\textsc{True}}
        \end{prooftree}
Then, \algosat calls $\reacha(G_{\varphi^s})$, where $s$ is $(\varphi^s,\varphi^s\Rightarrow\varphi^s)$ and $t=\textsc{True}$, and accepts $\varphi^s$ iff there is an alternating path from $s$ to $t$.  The correctness of \algosat can be proven similarly to the correctness of \algott, by using analogous results proven in this subsection for ready simulation, i.e.\ Lemma~\ref{lem:conjunction_lemma_rs} and Corollaries~\ref{prop:primality-LRS-2} and~\ref{cor:primality-phis}.
\end{proof}

Similarly to the cases of the simulation and complete simulation preorders, \algosat can be modified to return a process for which the input formula is characteristic.

\begin{proposition}\label{prop:find-p-rs-algo-phitt}
   Let $\varphi\in\mathL_{RS}$ be a formula such that if $\psi\in\sub(\varphi)$ is unsatisfiable, then $\psi=\ff$ and occurs in the scope of some $[a]$; let also  $\varphi^s$ be the output of $\Call{satur}{\varphi}$. If $\varphi^s$ is prime, there is a polynomial-time algorithm that constructs a process for which  $\varphi^s$ is characteristic within $\mathL_{RS}$.
\end{proposition}
\begin{proof}
   A process for which $\varphi^s$ is characteristic within $\mathL_{RS}$ can be constructed by calling algorithm \algosat on $\varphi^s$ and following steps analogous to the ones described for the simulation preorder in Corollary~\ref{cor:find-p-algo}.
\end{proof}

We can now prove that prime formulae can be decided in polynomial time.

\rsprime*
\begin{proof}
  It suffices to compute $\varphi^s:=\Call{satur}{\varphi}$ and check whether $\varphi^s$ is prime and $\varphi^s\models\varphi$. Checking whether $\varphi^s\models\varphi$ holds is equivalent to constructing a process $p$ for which $\varphi^s$ is characteristic within $\mathL_{RS}$ and checking whether $p\models\varphi$. The correctness and polynomial-time complexity of this procedure is an immediate corollary of Lemma~\ref{lem:satur-poly-time} and Propositions~\ref{prop:primality-LRS-algo}, \ref{prop:phis-primality}, and~\ref{prop:find-p-rs-algo-phitt}.
\end{proof}

Finally, deciding characteristic formulae in $\mathL_{RS}$ can be done in polynomial time.

\begin{corollary}\label{cor:rs-decide-characteristic}
Let $|A|=k$, $k\geq 1$.  Deciding characteristic formulae within $\mathL_{RS}$ is polynomial-time solvable.
\end{corollary}
\begin{proof}
    The corollary follows from Propositions~\ref{prop:rs-satisfiability}, \ref{prop:rs-satisfiability-satisfiable}, and~\ref{prop:rs-primality}.
\end{proof}

\begin{corollary}\label{cor:find-p-rs-algo}
Let $|A|=k$, $k\geq 1$ and $\varphi\in\mathL_{RS}$. If $\varphi$ is satisfiable and prime, then there is a polynomial-time algorithm that constructs a process for which  $\varphi$ is characteristic within $\mathL_{RS}$.
\end{corollary}

\subsubsection{The case of $|A|$ being unbounded}\label{subsection:rs-primality-unbounded-appendix}

The \conp-hardness of deciding pimality in $\mathL_{RS}$ was demonstrated in the proof of Proposition~\ref{prop:decide-prime-rs-infinite-actions-hard} in the main body of the paper. To prove that the problem belongs to \conp, we first provide some properties of prime formulae. 

\begin{definition}\label{def:rs-associated-process-extended}
    Let $\varphi\in\mathL_{RS}$ be a  
    formula given by the grammar $\varphi::=\true ~\mid~ \varphi\wedge\varphi~\mid ~ \langle a \rangle \varphi ~\mid~ [a]\ff$. We define process $p_\varphi$ inductively as follows.
\begin{itemize}
    \item If either $\varphi=[a]\ff$ or $\varphi=\true$ , then $p_\varphi=\mathtt{0}$.
     \item If $\varphi=\langle a\rangle \varphi'$, then $p_\varphi=a. p_{\varphi'}$.
     \item If $\varphi=\varphi_1\wedge \varphi_2$, then $p_\varphi=p_{\varphi_1}+p_{\varphi_2}$.
\end{itemize}
\end{definition}

\begin{lemma}\label{lem:rs-associated-process-extended}
    Let $\varphi\in\mathL_{RS}$ be a satisfiable 
    formula given by the grammar $\varphi::=\true ~\mid~ \varphi\wedge\varphi~\mid ~ \langle a \rangle \varphi ~\mid~ [a]\ff$. Then,
    \begin{enumerate}[(a)]
        \item $p_\varphi\models\varphi$, and
        \item if $\true\not\in\sub(\varphi)$, $\varphi$ is saturated and for every $\langle a\rangle\varphi'\in\sub(\varphi)$, $\varphi'$ is saturated, then $\varphi$ is characteristic within $\mathL_{RS}$ for $p_\varphi$.
    \end{enumerate} 
\end{lemma}
\begin{proof}
    The proof of (a) is similar to the proof of Lemma~\ref{lem:cs-associated-process-extended}, whereas the proof of (b) is analogous to the proof of Lemma~\ref{lem:rs-associated-process}.
\end{proof}


\begin{lemma}\label{lem:rs-prime-property-prelemma-1}
    Let $\varphi$ be a satisfiable and saturated formula given by the grammar $\varphi::=\true ~\mid~ \varphi\wedge\varphi~\mid ~ \langle a \rangle \varphi ~\mid~ [a]\ff$. If $\varphi\models\psi$ for some prime $\psi$ and $\Call{Satur}{\varphi}\neq\true$, then $\Call{Satur}{\varphi}\models\psi$.
\end{lemma}
\begin{proof}
    We denote $\Call{Satur}{\varphi}$ by $\varphi^s$. Note that $\psi$ is characteristic within $\mathL_{RS}$ for some process, which we denote by $p_\psi$. For every process $p$, $p\models\varphi\implies p\models\psi$ by assumption, and from Corollary~\ref{cor:logical-char-implication}, $p_\psi\curle_{RS} p$. Since $\varphi^s\neq \true$ and $\varphi^s$ does not contain disjunctions, $\varphi^s$ is characteristic within $\mathL_{RS}$ from Lemmas~\ref{lem:saturated-algo} and~\ref{lem:rs-associated-process}. Thus, $\varphi^s$ is characteristic within $\mathL_{RS}$ for $p_{\varphi^s}$, where $p_{\varphi^s}$ is the process that corresponds to $\varphi^s$ and is constructed as described in Definition~\ref{def:rs-associated-process-extended}. We prove that $p_\psi\curle_{RS} p_{\varphi^s}$ by induction on the type of substitutions made by procedure \textsc{Satur} in $\varphi$.
    \begin{itemize}
        \item Let $\varphi^s=\varphi[\true\wedge\varphi'/\varphi']$. Then $\varphi^s\equiv\varphi$ and as was shown above $p_\psi\curle_{RS} p$ for every $p$ that satisfies $\varphi$, or equivalently, for every $p$ that satisfies $\varphi^s$. In particular, $p_{\varphi^s}$ described in Definition~\ref{def:rs-associated-process-extended}, satisfies $\varphi^s$ from Lemma~\ref{lem:rs-associated-process-extended}.
        \item  Let $\varphi^s$ be derived from $\varphi$ by substituting an occurrence of $\langle a\rangle\varphi'$ with $\true$, where $\varphi'$ is a non-saturated formula. Then, there is a process $p_{tt}=\mathtt{0}$ such that $p_{\varphi^s}\myarrowtau p_{tt}$, $t\in A^*$, and $p_{tt}$ corresponds to this occurrence of $\true$ that substituted $\langle a\rangle\varphi'$ in $\varphi$. We consider two copies of $p_{\varphi^s}$, namely $p_{\varphi^s}^1$ and $p_{\varphi^s}^2$, that are as follows: for $i=1,2$, $p_{\varphi^s}^i$ is $p_{\varphi^s}$ where $p_{tt}$ is substituted with $p_{tt}^i=a.p_i$, where $p_1,p_2$ are two processes that satisfy $\varphi'$ and $I(p_1)\neq I(p_2)$. The existence of $p_1,p_2$ is guaranteed by Lemma~\ref{lem:rs-not-saturated-property}. Then, $p_{\varphi^s}^i\models\varphi$, and by assumption, $p_{\varphi^s}^i\models\psi$ and $p_\psi\curle_{RS}p_{\varphi^s}^i$ from Corollary~\ref{cor:logical-char-implication}, for both $i=1,2$. If there is $p_\psi\myarrowtau p'$, such that $p_{\varphi^s}^1\myarrowtau p_1$ and $p'\curle_{RS} p_1$, then $p'\not\curle_{RS} p_2$, since $I(p_1)\neq I(p_2)$, and so there is $p_{\varphi^s}^2\myarrowtau q_2$ such that $p'\curle_{RS} q_2$ and $q_2\neq p_2$. Note that $q_2$ is the copy of some $q$ such that $p_{\varphi^s}\myarrowtau q$ and $p'\curle_{RS} q$ by the definition of $p_{\varphi^s}^2$. As a result, for every $p_\psi\myarrowa p'$, there is $p_{\varphi^s}\myarrowa p''$ such that $p'\curle_{RS} p''$. So, $p_\psi\curle_{RS}p_{\varphi^s}$. 
    \end{itemize}
    Since $\psi,\varphi^s$ are characteristic within $\mathL_{RS}$ for $p_\psi$ and $p_{\varphi^s}$, respectively, we have that $\varphi^s\models\psi$ from Corollary~\ref{cor:logical-char-implication}.
\end{proof}

\begin{lemma}\label{lem:rs-prime-property-prelemma-2}
    Let $\varphi$ be a satisfiable and saturated formula given by the grammar $\varphi::=\true ~\mid~ \varphi\wedge\varphi~\mid ~ \langle a \rangle \varphi ~\mid~ [a]\ff$. If $\varphi\models\psi$ for some prime $\psi$, then $\Call{Satur}{\varphi}\neq\true$.
\end{lemma}
\begin{proof}
   Let $\varphi^s$ denote $\Call{Satur}{\varphi}$ and suppose that  $\varphi^s=\true$. Let also $p_\psi$ and $p_{\varphi^s}$ be as in the proof of Lemma~\ref{lem:rs-prime-property-prelemma-1}. In the proof of Lemma~\ref{lem:rs-prime-property-prelemma-1}, we showed that $p_\psi\curle_{RS} p_{\varphi^s}$. In this case, from Definition~\ref{def:rs-associated-process-extended}, $p_{\varphi^s}=\mathtt{0}$, which means that $p_\psi=\mathtt{0}$ and $\psi\equiv\zero$. Since $\varphi\models\psi$, $\varphi\equiv\zero$ as well. Then, it is not possible that $\varphi^s=\true$ because of the type of the substitutions made by \textsc{Satur}, contradiction.
\end{proof}

\begin{lemma}\label{lem:rs-prime-property}
    Let $\varphi\in\mathL_{RS}$ and $\bigvee_{i=1}^k\varphi_i$ be the DNF of $\varphi$. Then, $\varphi$ is prime iff there is some prime $\varphi_i$, where $1\leq i\leq k$, such that for every satisfiable $\varphi_j$, where $1\leq j\leq k$, 
    $\Call{satur}{\varphi_j}\models\varphi_i$.
\end{lemma}
\begin{proof} For every $\varphi\in\mathL_{RS}$, let $\varphi^s$ denote $\Call{satur}{\varphi}$. \\
    ($\Leftarrow$) Assume that for every satisfiable $\varphi_j$, $\varphi^s_j\models\varphi_i$, for some prime $\varphi_i$. For every $\varphi\in\mathL_{RS}$, it holds that $\varphi\models\varphi^s$ as was shown in the proof of Proposition~\ref{prop:primality-LRS-algo}. Hence, $\varphi_j\models\varphi^s_j$ and so $\varphi_j\models\varphi_i$. Consequently, for every $\varphi_j$, $\varphi_j\models\varphi_i$. From Lemmas~\ref{lem:DNF-equiv} and~\ref{lem:disjunction_lemma} and the fact that $\varphi_i\models\varphi$, $\varphi\equiv\varphi_i$, which implies that $\varphi$ is prime.\\
    ($\Rightarrow$) Let $\varphi$ be prime. Then, from Lemma~\ref{lem:DNF-equiv} and Definition~\ref{def:prime-formula}, there is some $\varphi_i$ such that $\varphi\models\varphi_i$. This means that $\varphi\equiv\varphi_i$ and so $\varphi_i$ is prime. From Lemma~\ref{lem:disjunction_lemma}, $\varphi_j\models\varphi_i$ for every $1\leq j\leq k$. Assume that there is some $1\leq j\leq k$ such that $\varphi_j$ is satisfiable. Then, from Lemmas~\ref{lem:rs-prime-property-prelemma-1} and~\ref{lem:rs-prime-property-prelemma-2}, $\varphi_j^s\neq\true$ and $\varphi^s_j\models\varphi_i$.
\end{proof}

\begin{lemma}\label{lem:rs-prime-property-2}
    Let $\varphi\in\mathL_{RS}$ and $\bigvee_{i=1}^k\varphi_i$ be the DNF of $\varphi$.  Then, $\varphi$ is prime iff (a) for every satisfiable  $\varphi_i$, $1\leq i\leq k$, $\Call{Satur}{\varphi_i}\neq\true$, and (b) for every satisfiable $\varphi_i$, $\varphi_j$, $1\leq i,j\leq k$, there is some $\varphi_m$, $1\leq m\leq k$, such that  $\Call{Satur}{\varphi_i}\models\varphi_m$ and $\Call{Satur}{\varphi_j}\models\varphi_m$. 
\end{lemma}
\begin{proof}  For every $\varphi\in\mathL_{RS}$, let $\varphi^s$ denote $\Call{satur}{\varphi}$. \\
    ($\Rightarrow$)  This direction is immediate from Lemmas~\ref{lem:rs-prime-property-prelemma-2} and~\ref{lem:rs-prime-property}.\\
    ($\Leftarrow$) If $\varphi$ is unsatisfiable, then $\varphi$ is prime and we are done. Assume that $\varphi$ is satisfiable and (a) and (b) are true. It suffices to show that there is some $\varphi_m$, $1\leq m\leq k$, such that for every $1\leq i\leq k$, $\varphi_i^s\models\varphi_m$. Then, $\varphi_m^s\models\varphi_m$ and consequently, $\varphi_m\equiv\varphi_m^s\neq\true$, which implies that $\varphi_m$ is prime. From Lemma~\ref{lem:rs-prime-property}, $\varphi$ is prime. Let $\varphi_1^s,\dots,\varphi_k^s$ be $k$ satisfiable formulae such that for every pair $\varphi_i^s$, $\varphi_j^s$ there is some $\varphi_m$ such that  $\varphi_i^s\models\varphi_m$ and $\varphi_j^s\models\varphi_m$. We prove by strong induction on $k$ that there is $\varphi_m$ such that for every $\varphi_i^s$, $\varphi_i^s\models\varphi_m$. For $k=2$, the argument is trivial. Let the argument hold for $k\leq n-1$ and assume we have $n$ satisfiable formulae $\varphi_1^s,\dots,\varphi_n^s$. From assumption, we have  that for every pair $\varphi_i^s,\varphi_n^s$, $1\leq i\leq n-1$, there is $\varphi_{in}$ such that $\varphi_i^s\models\varphi_{in}$ and $\varphi_n^s\models\varphi_{in}$.  Then, since $\varphi_{1n},\dots,\varphi_{n-1,n}$ are at most $n-1$ formulae, from inductive hypothesis there is some $\varphi_m$, $1\leq m\leq n$, such that $\varphi_{in}^s\models\varphi_m$, for every $1\leq i\leq n-1$. As a result,  $\varphi_i^s,\varphi_n^s\models\varphi_{in}\models\varphi_{in}^s\models\varphi_m$, which means that for every $1\leq i\leq n$, $\varphi_i^s\models\varphi_m$, which was to be shown.
\end{proof}

\begin{lemma}\label{lem:check-saturated-without-disjunctions}
    Let $\varphi\in\mathL_{RS}$ be a satisfiable formula given by the grammar $\varphi::=\varphi\wedge\varphi~\mid~\langle a\rangle\varphi~\mid~ [a]\ff$. Then, deciding whether $\varphi$ is saturated can be done in polynomial time.  
\end{lemma}
\begin{proof}
     W.l.o.g.\ $\varphi=\bigwedge_{i=1}^m\varphi_i$, where $\varphi_i$ is either $[a]\ff$ or $\langle a\rangle\varphi'$, where $\varphi'$ is given by the same grammar as $\varphi$. Let $\psi=\sat(\varphi)$. Then, $\psi=\bigwedge_{i=1}^m\psi_i$, where $\psi_i$ is either $x_a$ or $\neg x_a$.
      It is not hard to see that $\varphi$ is saturated iff for every $x_{a}$, exactly one of $x_{a}$, $\neg x_{a}$ occurs in $\psi$.
\end{proof}


\begin{proposition}\label{prop:conp-algo-prime-rs-unbounded}
Let $|A|$ be unbounded. Deciding prime formulae in $\mathL_{RS}$ is in \conp.
\end{proposition}
\begin{proof}
 We now describe algorithm \algorsu that decides primality of $\varphi\in \mathL_{RS}$. Let $\bigvee_{i=1}^k\varphi_i$ be the DNF of $\varphi$. Given $\varphi\in\mathL_{RS}$, \algorsu calls the \conp algorithm for unsatisfiability of $\varphi$. If a universal guess of that algorithm accepts, \algorsu accepts. Otherwise,  \algorsu universally guesses a pair $\varphi_i$, $\varphi_j$ as described in Lemma~\ref{lem:DNF-poly-guess}. Since these formulae do not contain disjunctions, \algorsu decides in polynomial time whether $\varphi_i$, $\varphi_j$ are satisfiable as explained in the proof of Lemma~\ref{lem:RS-poly-sat-no-disj}. If at least one of them is unsatisfiable, \algorsu accepts. Otherwise, it computes $\varphi_i^s=\Call{Satur}{\varphi_i}$ and $\varphi_j^s=\Call{Satur}{\varphi_j}$, which can be done in polynomial time from Lemma~\ref{lem:check-saturated-without-disjunctions}. If at least one of $\varphi_i^s,\varphi_j^s$ is $\true$, then it rejects. Otherwise, \algorsu needs to check whether there is some $\varphi_m$ such that  $\varphi_i^s\models\varphi_m$ and $\varphi_j^s\models\varphi_m$. It does that by constructing a DAG $G_\varphi^{ij}$ similarly to the proof of  Proposition~\ref{prop:phis-primality}. It starts with vertex $s=(\varphi_i,\varphi_j\Rightarrow\varphi)$ and applies the rules that were introduced in the proof of  Proposition~\ref{prop:phis-primality} for ready simulation. Next, it solves \reacha on $G_\varphi^{ij}$, where $t=\textsc{True}$. \algorsu accepts iff there is an alternating path from $s$ to $t$. The correctness of these last steps can be proven similarly to the case of $\mathL_{RS}$ with a bounded action set. In particular, a variant of Lemma~\ref{lem:conjunction_lemma_rs} holds here since $\varphi_i$, $\varphi_j$ are satisfiable and prime.  \algorsu  is correct due to Lemma~\ref{lem:rs-prime-property-2}.
 \end{proof}

\rsprimehard*
\begin{proof}
    The \conp-hardness of the problem was shown in the proof of the current proposition in the main body of the paper. Membership in \conp follows from Proposition~\ref{prop:conp-algo-prime-rs-unbounded}.
\end{proof}

\begin{corollary}\label{cor:char-rs-unbounded-complexity}
    Let $|A|$ be unbounded. Deciding characteristic formulae within $\mathL_{RS}$ (a) is \us-hard, and (b) belongs to \dpc.   
\end{corollary}
\begin{proof}
(a)  Let $\varphi$ be an instance of $\textsc{UniqueSat}$. We construct $\varphi'\in\mathL_{RS}$, which is $\varphi$, where $x_i$ is substituted with $\langle a_i\rangle \zero$ and $\neg x_i$ is substituted with $[a_i]\ff$. Note that $\varphi$ has exactly one satisfying assignment iff $\varphi'$ is characteristic within $\mathL_{RS}$. Let $s$ denote the unique satisfying assignment of $\varphi$. A process for which $\varphi'$ is characteristic within $\mathL_{RS}$ is the process $\displaystyle p=\sum_{i:s(x_i)=\mathrm{true}}\langle a_i\rangle \mathtt{0}$. Inversely, let $\varphi'$ be characteristic for a process $p$ within $\mathL_{RS}$. The truth assignment that maps to true exactly the $x_i$'s for which $a_i\in I(p)$ is a unique satisfying assignment for $\varphi$.\\
(b) This is immediate from Proposition~\ref{prop:charact-via-primality} and membership of $\mathL_{RS}$-satisfiability and $\mathL_{RS}$-primality to \NP and \conp, respectively.
\end{proof}

 \subsection{Primality in $\mathL_{TS}$}\label{subsection:ts-primality}

\subsubsection{The case of $|A|>1$}\label{subsection:ts-primality-unbounded}

\tsprimehard*
\begin{proof}
    We show that the complement of the problem is \NP-hard. We describe a polynomial-time reduction from \SAT to deciding non-prime formulae within $\mathL_{TS}$, which is based on the proofs of Propositions~\ref{prop:decide-prime-rs-infinite-actions-hard} and~\ref{prop:sat-TS}. However, we have to encode the literals more carefully here.
    
    Let $A=\{0,1\}$ and $\varphi$ be an instance of \SAT over the variables $x_0,\dots, x_{n-1}$. Similarly to the proof of Proposition~\ref{prop:sat-TS}, the initial idea is to associate every variable $x_i$ with the binary representation of $i$. In more detail, for every $0\leq i\leq n-1$, we associate  $x_i$ with $0\,{b_i}_1\dots{b_i}_k$, where every ${b_i}_j\in\{0,1\}$ and $k=\lceil\log n\rceil$.
    This means that $x_i$ is associated with the binary string ${b_i}_0{b_i}_1\dots{b_i}_k$, where ${b_i}_0=0$ and ${b_i}_1\dots{b_i}_k$ is the binary representation of $i$. The binary string ${b_i}_0{b_i}_1\dots {b_i}_k$ can now be mapped to formula $\enc(x_i)= [\overline{{b_i}_0}] \wedge \langle {b_i}_0\rangle( [\overline{{b_i}_1}]\ff\wedge \langle {b_i}_1\rangle (\dots ([\overline{{b_i}_k}]\ff\wedge \langle {b_i}_k\rangle\zero)\dots))$, where $\overline{b}=\begin{cases}
        1, &\text{if } b=0,\\
        0, &\text{if } b=1
    \end{cases}$. We map a negative literal $\neg x_i$ to $\enc(\neg x_i)=[{b_i}_1] [{b_i}_2]\dots [{b_i}_k]\ff$. 
    
    Define formula $\varphi'$ to be $\varphi$ where every literal $l$ has been substituted with $\enc(l)$ and formula $\varphi'':=(\varphi'\wedge [1]\ff)\vee (\langle 1\rangle ([0]\ff\wedge [1]\ff)\wedge [1][0]\ff\wedge[1][1]\ff\wedge[0]\ff)$. As in the proof of Proposition~\ref{prop:decide-prime-rs-infinite-actions-hard}, we can prove that if $\varphi$ is satisfiable, then $\varphi''$ is not prime and not characteristic within $\mathL_{TS}$. If $\varphi$ is not satisfiable, then $\varphi''$ is prime and characteristic for $p=a.\mathtt{0}$ within $\mathL_{TS}$, where $a=1$.
\end{proof}

\subsubsection{The case of bounded $|A|$ and bounded modal depth}


\tsboundeddepth*
\begin{proof}
    We describe algorithm $\mathrm{Char}_{TS}^{k,d}$ that decides characteristic formulae within $\mathL_{TS}$ and analyze its complexity. $\mathrm{Char}_{TS}^{k,d}$ computes all processes over $A$ that have depth less than or equal to $d+1$. We denote the set of processes over $A$ with modal depth less than or equal to $i$  by $P^i$. 
    $\mathrm{Char}_{TS}^{k,d}$ also computes the set of processes in $P^{d+1}$ that satisfy $\varphi$, which we denote by $P^{d+1}_{\mathrm{sat}}$. It does so by solving model checking for every process in $P^{d+1}$. Then, $\mathrm{Char}_{TS}^{k,d}$ checks whether there is a process $p\in P^d$ such that $p\models\varphi$ and for every $q\in P^{d+1}$, $q\models\varphi\implies p\curle_{TS} q$. The algorithm accepts iff such a $p\in P^d$ exists. To show the correctness of  $\mathrm{Char}_{TS}^{k,d}$, we prove the following claim.\\
    \textbf{\textcolor{darkgray}{Claim.}} $\varphi$ is satisfiable and prime iff there is $p\in P^d$ such that $p\models\varphi$ and for every $q\in P^{d+1}$, $q\models\varphi\implies p\curle_{TS} q$.\\
    \textbf{\textcolor{darkgray}{Proof of Claim.}} Assume that $\varphi$ is satisfiable and prime. Since $\varphi$ is satisfiable, then it must be satisfied in a process with depth less than or equal to $d$. So, there is $q$ that satisfies $\varphi$ and $\depth(q)=m\leq d$. Since $\varphi$ is prime, and so characteristic within $\mathL_{TS}$, there is $p$ such that for every $p'$, $p'\models\varphi\Leftrightarrow p\curle_{TS} p'$. In particular,  $p'\curle_{TS} q$, and since $\mathrm{traces}(p')=\mathrm{traces}(q)$, it holds that $\depth(p')=\depth(q)=m$. So, $\depth(p')\in P^d$. Inversely, assume that there is $p\in P^d$ that satisfies $\varphi$ and for every $q\in P^{d+1}$, $q\models\varphi\implies p\curle_{TS} q$. Then, $\depth(p)=m$, for some $m\leq d$. This implies that if $q\in P^{d+1}$ with $\depth(q)=d+1$, then $q\not \models \varphi$, because if such a $q$ satisfies $\varphi$, $p\curle_{TS} q$ and $\mathrm{traces}(p)\neq\mathrm{traces}(q)$, which is impossible. Consequently, for every $q\in P$, if $\depth(q)\geq d+1$, then $q\not\models\varphi$, since $\md(\varphi)=d$. This, in turn, implies that for every $q\in P$, $q\models \varphi\implies p\curle_{TS} q$. For the other direction, if $p\curle_{TS} q$, then $q\models \varphi$ from Proposition~\ref{logical_characterizations}. As a result $\varphi$ is characteristic within $\mathL_{TS}$ for $p$.

   We analyze its complexity. By an easy induction on the depth of processes, we can show that the cardinality of $P^{d+1}$ is $\mathcal{O}(2^{k^{d+1}})$ and the size of a process in $P^{d+1}$ is $\mathcal{O}(k^{d+1})$. Trace simulation needs to be checked on processes of constant size, and this can be done in constant time from Corollary~\ref{cor:trace-simulation-constant-input-size}. Model checking can be done in linear time in $|p|$ and $|\varphi|$ by Proposition~\ref{model-checking-complexity}. Overall the algorithm requires linear time in $|\varphi|$ if $k$ and $d$ are considered constants.
\end{proof}

\begin{corollary}
 Let $|A|= k$, where $k\geq 1$ is a constant and $\varphi\in\mathL_{TS}$ with $\md(\varphi)=d$. Deciding whether $\varphi$ is characteristic within $\mathL_{TS}$ can be done in $\mathcal{O}(2^{2k^{d+1}}\cdot k^{d+1}\cdot |\varphi|)$.
\end{corollary}

\begin{corollary}\label{cor:char-ts-complexity}
     Let $|A|= k$, where $k\geq 1$ is a constant. Deciding characteristic formulae within $\mathL_{TS}$ or $\mathL_{2S}$ is \us-hard.
\end{corollary}
\begin{proof}
    There is a polynomial-time reduction from \textsc{UniqueSat} to deciding characteristic formulae within $\mathL_{TS}$. Given an instance $\varphi$ of $\textsc{UniqueSat}$, we construct $\varphi''\in\mathL_{TS}$ such that $\varphi$ has a unique satisfying assignment iff $\varphi''$ is characteristic within $\mathL_{TS}$. We first define $\varphi'\in\mathL_{TS}$ to be $\varphi$ where every literal $l$ is substituted with $\enc(l)$ as described in the proof of Proposition~\ref{prop:decide-prime-ts-infinite-actions-hard} in Appendix~\ref{subsection:ts-primality-unbounded}. We denote $\lceil \log n \rceil$ by $k$. Formula $\varphi'$ is one conjunct of $\varphi''$. To complete $\varphi''$, the idea is that in the case that $\varphi$ has a unique satisfying assignment, any process satisfying $\varphi''$ must have only traces which correspond to variables that are set to true in the satisfying assignment (and perhaps prefixes of such traces). So, we have to forbid all other traces. For every sequence of actions $a_1\dots a_l$, of length $2\leq l\leq k$, we define $\displaystyle\mathrm{fb}^{a_1\dots a_l}:=\bigvee_{t\in A^{k+1-l}} \langle a_1\rangle \dots \langle a_l\rangle t\vee[a_1]\dots [a_l]\ff$. Intuitively, $\mathrm{fb}^{a_1\dots a_l}$ says that if $a_1\dots a_l$ has not been imposed by a variable that was set to true, then it is forbidden. For a specific sequence $a_1\dots a_l$, formula $\mathrm{fb}^{a_1\dots a_l}$ contains $2^{k+1-l}+1$ disjuncts and there are $2^{l-1}$ different sequences of length $l$ that we have to take care of since the first symbol is always $0$ (see the definition of $\enc(l)$ for a literal $l$). Thus, we need to include $n+2^{l-1}$ disjuncts for every $1\leq l\leq k$. If we sum up over all $l$ we get $\mathcal{O}(n \log n)$ disjuncts of length at most $\log n$ each. Finally, we define $\displaystyle\varphi'':=\varphi'\wedge \bigwedge_{2\leq l\leq k}\bigwedge_{a_1\dots a_l}\mathrm{fb}^{a_1\dots a_l} \wedge [1]\ff$. From the analysis above, $\varphi''$  is of polynomial size. Moreover, $\varphi$ has a unique satisfying assignment iff $\varphi''$ is characteristic for $p$, where $p$ is the process whose traces are the ones corresponding to the variables set to true in the unique satisfying assignment.

    The reduction can be modified to work for deciding characteristic formulae within $\mathL_{2S}$ as well. In this case, formulae of the form $[a_1]\dots[a_l] ([0]\ff\vee [1]\ff)$ must be included as conjuncts in $\varphi''$, for every sequence of actions $a_1\dots a_l$ and  every $2\leq l\leq k$.
\end{proof}



\subsection{Primality in $\mathL_{3S}$}\label{subsection:threes-primality}

\subsubsection{The case of $|A|>1$}

We prove here \pspace-hardness of deciding primality in $\mathL_{3S}$ when $|A|>1$, by reducing validity of $\compL_{3S}$ to it.

\decidethreesprime*
\begin{proof}
    $\compL_{3S}$-validity is \pspace-hard from Theorem~\ref{thm:2s-validity} and the fact that $\mathL_{2S}\subseteq\compL_{3S}$. The proof of Theorem~\ref{thm:2scompl-prime} can be slightly adjusted to show that $\compL_{3S}$-validity efficiently reduces to deciding whether a formula $\varphi\in\mathL_{3S}$ is prime in $\mathL_{3S}$.
\end{proof}

Below we prove the main theorem of this paper.

\maintheorem*
\begin{proof}
    (a) This follows from Corollaries~\ref{cor:simulation-decide-characteristic}, \ref{cor:cs-decide-characteristic}, and~\ref{cor:rs-decide-characteristic}.

    (b) This is Corollary~\ref{cor:char-rs-unbounded-complexity}.

    (c) This is Corollary~\ref{cor:char-ts-complexity}.

    (d) The proof of Theorem~\ref{thm:2scompl-prime} implies that deciding whether $\varphi\in\mathL_{3S}$ is characteristic within $\mathL_{3S}$ is \pspace-hard.
\end{proof}

\section{The complexity of deciding preorders in van Glabbeek's branching time spectrum}\label{section:decide-preorders-appendix}

In part~\ref{section:decide-preorders-appendix} of the appendix, we demonstrate the gap between deciding $p\curle_X q$, for any $X\in\{S,CS,RS,2S,BS\}$, and deciding $p\curle_{TS} q$, where $p,q$ are finite processes. This is closely related to the gap shown in Section~\ref{section:find}.

\begin{proposition} \label{equivalence-relations-easy}
The following equivalence relations are decidable in polynomial time over finite processes:
\begin{itemize}
    \item simulation equivalence~\cite{KS90, HT94},
    \item complete simulation equivalence~\cite{HT94},
    \item ready simulation equivalence~\cite{KS90, HT94}, 
    \item 2-nested simulation equivalence~\cite{GV92, ShuklaRHS96}, and
    \item bisimilarity~\cite{PT87,KS90}.
\end{itemize}
The same result holds true for the preorders underlying those equivalence relations. 
\end{proposition}

On the contrary, we prove here that deciding the $TS$ preorder on finite processes is \pspace-complete under polytime Turing reductions. The problem becomes  \conp-complete under polytime Turing reductions on finite loop-free processes. The hardness results are by reduction from trace equivalence of two processes. Therefore, we first define trace equivalence (\treq), show the relationship between the $TS$ preorder and \treq, and then establish the complexity of deciding $\curle_{TS}$.

\begin{definition}[\cite{Glabbeek01}]\label{def:trace-equivalence}
We say that $p,q\in P$ are trace equivalent, denoted by $p \treq q$, if  $\mathrm{traces}(p)=\mathrm{traces}(q)$.
\end{definition}

The logic that characterizes $\treq$ contains formulae that are formed using only $\true$ and diamond operators.

\begin{proposition}[\cite{Glabbeek01}]\label{prop:trace-equivalence-logic}
    $p\treq q$ iff $\mathL_{trace}(p)=\mathL_{trace}(q)$, where $\mathL_{trace}$ is defined by the grammar $\varphi_{trace}::= \true~\mid~\langle a\rangle \varphi_{trace}$.
\end{proposition}

Unlike the equivalences of Proposition~\ref{equivalence-relations-easy}, checking whether two finite processes are trace equivalent is \pspace-complete.

\begin{proposition}\emph{(\cite[Lemma 4.2]{KS90}).}\label{prop:trace-complexity-a}
Given two finite processes $p$ and $q$, deciding whether $p \treq q$ is \pspace-complete.
\end{proposition}

First, we prove that deciding $\treq$ reduces to deciding $\curle_{TS}$ on two different instances, hence we obtain a hardness result for the latter.


\begin{proposition}\label{prop:trace-sim-pspace-hard}
Given two finite processes $p$ and $q$, deciding $p\curle_{TS} q$ is \pspace-hard under polytime Turing reductions.
\end{proposition}
\begin{proof}
From Lemma~\ref{lem:trace-equiv-ts-preorder}, there is a polynomial-time algorithm that decides $\treq$ using two oracle calls to $\curle_{TS}$. Since $\treq$ is \pspace-hard by Proposition~\ref{prop:trace-complexity-a}, $\curle_{TS}$ is also \pspace-hard under polytime Turing reductions.
\end{proof}

To prove that deciding $\curle_{TS}$ on finite processes can be done in polynomial space, we first show in Lemma~\ref{recursion_trace_simulation} how to define it recursively using $\treq$.

\begin{definition}
Let $T=\{(p,q)\mid p\treq q\}$. We define $\leq_k^T$ recursively as follows:
    \begin{itemize}
        \item $p\leq_0^T q$ for all $(p,q)\in T$.
        \item $p\leq_{i+1}^T q$ iff $(p,q)\in T$ and $p\myarrowa p'$ implies that there is $q\myarrowa q'$ and $p'\leq_i^T q'$.
    \end{itemize}
\end{definition}

\begin{lemma}\label{recursion_trace_simulation}
    For finite processes, $\displaystyle\curle_{TS}=\bigcap_{i=0}^{+\infty} \leq_i^T$.
\end{lemma}
\begin{proof}
The proof is analogous to the proof of Proposition 3.3 in \cite{HT94}.
 
$\curle_{TS}\subseteq\bigcap_{i=0}^{+\infty} \leq_i^T$: The proof is by induction on $i$. By definition of $\curle_{TS}$, $p\curle_{TS} q$ implies that $p\treq q$, and so $p\leq_0^T q$. For the inductive step,  assume that $\curle_{TS}\subseteq \leq_k^T$. Let $p\curle_{TS} q$ and $p\myarrowa p'$. Then, there is $q\myarrowa q'$ with $p'\curle_{TS} q'$. By inductive hypothesis, $p'\leq_k^T q'$. Thus, $p\leq_{k+1}^T q$.

$\bigcap_{i=0}^{+\infty} \leq_i^T\subseteq\curle_{TS}$: Let $R=\bigcap_{i=0}^{+\infty} \leq_i^T$ and $(p,q)\in R$. The following claim holds.
\begin{quote}
    From the fact that the processes are finite, there is  $m\geq 0$, such that for all $(s,t)$,
$(s,t)\in R$ iff $s\leq_m^T t$.
\end{quote}
Let $m$ be as stated in the claim. Since $(p,q)\in R$, we have that $p\leq_{m+1}^T q$. By definition of $\leq_{m+1}^T$, for every $p\myarrowa p'$, there is $q\myarrowa q'$, such that $p'\leq_m^T q'$. Then $(p',q')\in R$ from the claim.
Therefore, $R$ is a $T$-simulation, i.e.\ a simulation that is a subset of $T=\{(p,q)\mid p\treq q\}$. The latter holds because $(p,q)\in R$ implies $(p,q)\in \leq_0^T$. By definition, $\curle_{TS}$ is the largest $T$-simulation, so $R\subseteq \curle_{TS}$.
\end{proof}

The following corollary comes also from \cite{HT94}.

\begin{corollary}[\cite{HT94}]\label{cor:trace-simulation-finite-proc}
    For finite processes, let $n=|\{(p,q)\mid p\treq q\}|$. It holds that $\displaystyle\curle_{TS}=\bigcap_{i=0}^{n^2} \leq_i^T$.
\end{corollary}

Corollary~\ref{cor:trace-simulation-finite-proc} means that to check whether $p\curle_{TS} q$, with $|p|+|q|= m$, it suffices to check whether $p\leq_i^T q$, for every $0\leq i\leq m^4$, since $|\{(p,q)\mid p\treq q\}|\leq m^2$. The following lemma implies that it actually suffices to check only $p\leq_{m^4}^T q$. In Proposition~\ref{prop:trace-sim-pspace-mem}, we show that this can be done in polynomial space.

\begin{lemma}\label{lem:inclusions}
    For every $i\in\mathbb{N}$, it holds that $\leq_{i+1}^T\subseteq \leq_i^T$.
\end{lemma}
\begin{proof}
    We prove the lemma by induction. If $i=0$, then $p\leq_1^T q$ implies that $(p,q)\in T$, which means that $p\leq_0^T q$. Suppose that $\leq_{k}^T \subseteq \leq_{k-1}^T$ and let $p\leq_{k+1}^T q$. Then, $(p,q)\in T$ and for every $p\myarrowa p'$ there is $q\myarrowa q'$ and $p' \leq_k q'$. By inductive hypothesis, $p'\leq_{k-1}^T q'$, which implies that $p\leq_{k}^T q$, and so $\leq_{k+1}^T \subseteq \leq_{k}^T$.
\end{proof}

\begin{proposition}\label{prop:trace-sim-pspace-mem}
Given two finite processes $p$ and $q$, deciding $p\curle_{TS} q$ is in \pspace.
\end{proposition}
\begin{proof}
Let two processes $p,q$ with $|p|+|q|=m$. Then, $|\{(p,q)\mid p\treq q\}|\leq m^2$. Given $p$ and $q$, and $k\in\mathbb{N}$, Algorithm~\ref{alg:trace-simulation} decides whether $p\leq_k^T q$. By Corollary~\ref{cor:trace-simulation-finite-proc} and Lemma~\ref{lem:inclusions}, it holds that $p\curle_{TS} q$ iff $p\leq_{m^4}^T q$. Thus, Algorithm~\ref{alg:trace-simulation} on input $(p,q,m^4)$ solves the problem. The space complexity of Algorithm~\ref{alg:trace-simulation} is determined by the polynomial space needed to check whether two processes have the same traces. 
\end{proof}

\begin{algorithm}
\caption{Procedure $\mathrm{Check}_{\leq_i^T}$ on input $(p,q,k)$ checks whether $p\leq_k^T q$ holds}
\begin{algorithmic}[1]
\DontPrintSemicolon
\Procedure{$\mathrm{Check}_{\leq_i^T}$}{$p,q,k$}
\State \lIf{$\mathrm{traces}(p)=\mathrm{traces}(q)$ and $k=0$}{accept}
\State \If{$\mathrm{traces}(p)=\mathrm{traces}(q)$ and $k>0$}{
$nextp \gets 0$\;
  \For{$a\in Act$}{
 \For{state $p'$}{
 \lIf{$(p,a,p')\in\longrightarrow$}{$nextp\gets nextp + 1$}
}}
 \lFor{$j \gets 1$ to $nextp$}{$tr[j]\gets\mathrm{false}$}{
{$j \gets 1$}\;
 \For{$a\in A$}{
 \For{$p\myarrowa p'$}{
 \For{$q\myarrowa q'$}{
 \lIf{\Call{$\mathrm{Check}_{\leq_i^T}$}{$p',q',k-1$} accepts}
{$tr[j]\gets\mathrm{true}$}
}
$j\gets j+1$
}}}
\For{$j \gets 1$ to $nextp$}{
\lIf{$tr[j]=\mathrm{false}$}{reject}
}
}
\State {accept}
\EndProcedure
\end{algorithmic}
\label{alg:trace-simulation}
\end{algorithm}

In the sequel, we prove that deciding $\curle_{TS}$ on finite loop-free processes is \conp-complete under polytime Turing reductions. To this end, we first show that deciding trace equivalence of two finite loop-free processes is \conp-complete. The proof follows the proof of \conp-completeness of deciding equivalence of two star-free regular expressions~\cite[Theorem 2.7(1)]{HRS76}. Then, the hardness result of our problem is immediate from the reduction given in Proposition~\ref{prop:trace-sim-pspace-hard}. Finally, we describe an \NP algorithm that decides $\not\curle_{TS}$ on finite loop-free processes, hence its complement $\curle_{TS}$ lies in \conp.

\begin{proposition}\label{prop:trace-complexity-b}
Given two finite loop-free processes $p$ and $q$, deciding $p\treq q$ is \conp-complete.
\end{proposition}
\begin{proof}
The \conp-complete problem \dnftaut reduces to deciding $\treq$ on finite loop-free processes. Let $\varphi=c_1\vee\dots\vee c_m$ be a DNF formula with variables $x_1,\dots, x_n$.
We construct a process $p_0$ with $A=\{0,1\}$ as follows: for every clause $c_i$,
    we add a path $p_{i0},p_{i1},\dots, p_{in}$, where $p_{i0}=p_0$, for every $1\leq i\leq m$, such that: 
    \begin{itemize}
        \item $(p_{ij-1},p_{ij})\in\myarrowo$ if $x_j$ is a literal in $c_i$,
        \item $(p_{ij-1},p_{ij})\in\myarrowz$ if $\neg x_j$ is a literal in $c_i$,
        \item $(p_{ij-1},p_{ij})\in\myarrowo$ and $(p_{ij-1},p_{ij})\in\myarrowz$ if none of $x_j, \neg x_j$ is a literal in $c_i$.
    \end{itemize}
We can also easily construct process $q$ with $\mathrm{traces}(q)=\{0,1\}^n$ in polynomial time. Formula $\varphi$ is a tautology iff $\mathrm{traces}(p_0)=\mathrm{traces}(q)$. Thus, deciding $\treq$ is \conp-hard. 

To check whether $p\treq q$ holds, for finite loop-free processes $p$ and $q$, it suffices to verify that every trace of $p$ is also a trace of $q$ and vice versa. Each such verification fails iff one can guess a trace $t$ of one of the two processes that is not a trace of the other. Moreover, since $p$ and $q$ are finite and loop-free, the length of $t$ is at most the maximum depth of those two processes. This implies that checking whether $p\not\treq q$ is in {\NP} and therefore deciding whether $p\treq q$ holds is in \conp.
\end{proof}

\begin{proposition}\label{prop:conp-hardness-trace-sim}
 Given two finite loop-free processes $p$ and $q$, deciding $p\curle_{TS} q$ is \conp-hard under polytime Turing reductions.
\end{proposition}
\begin{proof}
The proposition follows from the reduction described in the proof of Proposition~\ref{prop:trace-sim-pspace-hard} and the \conp-hardness of deciding $\treq$ on  finite loop-free processes that was proven in Proposition~\ref{prop:trace-complexity-b}.
\end{proof}

The algorithm for deciding $\not\curle_{TS}$ on finite loop-free processes described in Proposition~\ref{prop:trace-simulation-complexity-a} uses the algorithm given in the following lemma as a subroutine.

\begin{lemma}\label{lem:many-pairs-non-trace-equiv}
    Given $n$ pairs of finite loop-free processes $(p_1,q_1),\dots, (p_n,q_n)$, deciding whether $p_i\not\treq q_i$, for every $i\in\{1,\dots,n\}$, is \NP-complete.
\end{lemma}
\begin{proof}
Membership in \NP: Consider the non-deterministic polynomial-time Turing machine that, for every $i\in\{1,\dots,n\}$, guesses a trace $t_i$ and verifies that $t_i$ belongs to exactly one of $\mathrm{traces}(p_i)$ and $\mathrm{traces}(q_i)$. The problem is \NP-hard even for $i=1$ from Proposition~\ref{prop:trace-complexity-b}.
\end{proof}

\begin{proposition}\label{prop:trace-simulation-complexity-a}
Given finite loop-free processes $p$ and $q$, deciding $p\not\curle_{TS} q$ is in \NP.
\end{proposition}
\begin{proof}
Let $p$, $q$ be two  finite loop-free processes with $|p|+|q| = m$. By Corollary~\ref{cor:trace-simulation-finite-proc} and Lemma~\ref{lem:inclusions}, it holds that $p\curle_{TS} q$ iff $p\leq_{m^4}^T q$. Therefore, $p\not\curle_{TS} q$ iff Algorithm~\ref{alg:trace-simulation} rejects on input $(p,q,m^4)$.\\ 
\noindent\textbf{\textcolor{darkgray}{Claim 1.}} Algorithm~\ref{alg:trace-simulation} rejects on input $(p,q,m^4)$ iff either $\mathrm{traces}(p)\neq \mathrm{traces}(q)$ or at some point it examines a  pair $(p^*,q^*)\in S_p\times S_q$, such that $\mathrm{traces}(p^*)= \mathrm{traces}(q^*)$ and there is $p^*\myarrowa p^{**}$ such that for every $q^*\myarrowa q^{**}$, $\mathrm{traces}(p^{**})\neq \mathrm{traces}(q^{**})$.\\
Claim 1 states that Algorithm~\ref{alg:trace-simulation} rejects on input $(p,q,m^4)$ iff it finds two states in $S_p$ and $S_q$, respectively, that are not trace equivalent. We call these states \emph{non-equivalent witnesses}.  

Let $M$ denote the non-deterministic polynomial-time Turing machine that does the following: Guess a set of pairs $Bad\subseteq S_p\times S_q$ and for all $(p',q')\in Bad$, verify that $\mathrm{traces}(p')\neq \mathrm{traces}(q')$. Call Algorithm~\ref{alg:trace-simulation} on input $(p,q,m^4)$, where for every $(p',q')\in S_p\times S_q$ consider that $\mathrm{traces}(p')\neq \mathrm{traces}(q')$ iff $(p',q')\in Bad$. Accept iff Algorithm~\ref{alg:trace-simulation} rejects.\\
\noindent\textbf{\textcolor{darkgray}{Claim 2.}} $M$ has an accepting path iff $p\not\curle_{TS} q$.\\
\noindent\textbf{\textcolor{darkgray}{Proof of Claim 2.}} If $M$ has an accepting path, then $M$ guesses a set $Bad$ on this path that makes Algorithm~\ref{alg:trace-simulation} reject on $(p,q,m^4)$. By Claim 1 and the fact that Algorithm~\ref{alg:trace-simulation} considers that two processes are not trace equivalent iff their pair belongs to $Bad$, this means that Algorithm~\ref{alg:trace-simulation} finds two non-equivalent witnesses in $Bad$. Since $M$ verifies that for all $(p',q')\in Bad$, $\mathrm{traces}(p')\neq\mathrm{traces}(q')$, Algorithm~\ref{alg:trace-simulation} correctly rejects. Conversely, if $p\not\curle_{TS} q$, then Algorithm~\ref{alg:trace-simulation} rejects on input $(p,q,m^4)$, which in turn implies that Algorithm~\ref{alg:trace-simulation} on $(p,q,m^4)$ finds two non-equivalent witnesses $p^{**}$ and $q^{**}$ by Claim 1. Hence, there is a path of $M$ on which $M$ guesses $Bad$ such that $(p^{**},q^{**})\in Bad$, and so the path is accepting.
\end{proof}

The following two corollaries are  immediately derived from Propositions~\ref{prop:trace-complexity-b}, \ref{prop:conp-hardness-trace-sim}, and~\ref{prop:trace-simulation-complexity-a}.

\begin{corollary}
Given two finite loop-free processes $p$ and $q$, deciding $p\curle_{TS} q$  is \conp-complete under polytime Turing reductions.
\end{corollary}

\begin{corollary}
If deciding $\treq$ on  finite loop-free processes is in \cP, then $\cP=\NP$. The same holds for deciding $\curle_{TS}$.
\end{corollary}

\begin{corollary}\label{cor:trace-simulation-constant-input-size}
    Let $p,q$ be finite loop-free processes. If the size of $p$ and $q$ is bounded by a constant, then $p\curle_{TS} q$ can be solved in constant time.
\end{corollary}
\begin{proof}
    The \NP algorithm for deciding $p\not\curle_{TS} q$ that was provided in the proof of Proposition~\ref{prop:trace-simulation-complexity-a} can give a deterministic algorithm for the same problem running in exponential time w.r.t.\ $|p|$ and $|q|$. If $|p|$ and $|q|$ are bounded by a constant, then the running time of the algorithm is also bounded by a constant.
\end{proof}

\section{Finding characteristic formulae: the gap between trace simulation preorder and the rest of the preorders}\label{section:find-appendix}

Full proofs of results of Section~\ref{section:find} can be found in this part of the appendix.

\subsection{The simulation, complete simulation, ready simulation, n-nested simulation, and bisimilarity}\label{subsection:find-easy-appendix}

Let $X\in\{S,CS\}$ or $X=RS$ and $|A|$ is bounded by a constant. If formulae are given in explicit form, then finding the characteristic formula for a finite loop-free process is not in $\cP$, unless $\cP=\NP$.

\begin{lemma}\label{lem:generate-all-traces}
Let $X\in\{S,CS\}$ or $X=RS$ and $|A|=k$, where $k\geq 1$ is a constant. Assume that $\varphi$ is characteristic for a finite loop-free process $p$ within $\mathL_X$ and is given in explicit form. Then there is an algorithm that generates all traces of $p$ in polynomial time with respect to $|\varphi|$.
\end{lemma}
\begin{proof}
Let $X=S$ and let $\varphi$ be a formula in explicit form that is characteristic for some finite, loop-free process $p$ within $\mathL_S$. Note that if $\varphi$ is also characteristic for some process $q$, then by Corollary~\ref{cor:unique-process}, $p\equiv_S q$. Then, by Definition~\ref{def:characteristic},  for every $\psi\in\mathL_S$, $p\models \psi$ iff $q\models \psi$, which  implies that for every $\psi\in\mathL_T$, $p\models \psi$ iff $q\models \psi$, since $\mathL_T\subseteq \mathL_S$. Then, $\mathrm{traces}(p)=\mathrm{traces}(q)$ by Definition~\ref{def:trace-equivalence} and Proposition~\ref{prop:trace-equivalence-logic}. To generate all traces of $p$, we find process $q$ for which $\varphi$ is characteristic within $\mathL_S$ as described in the proof of Corollary~\ref{cor:find-p-algo}. Process $q$ is of polynomial size w.r.t.\ $|\varphi|$. By traversing all paths of $q$, which can be done using the Depth-First Search algorithm, we can generate all traces of $q$ in time polynomial in $|q|$, and so polynomial in $|\varphi|$. Using Corollaries~\ref{cor:find-p-cs-algo} and~\ref{cor:find-p-rs-algo} and the fact that $\mathL_T\subseteq \mathL_X$, $X\in\{CS,RS\}$, we can prove the lemma for $X=CS$ and $X=RS$ with a bounded action set. 
\end{proof}

\begin{remark}\label{trace-example}
Lemma~\ref{lem:generate-all-traces} does not imply that given a finite loop-free process $p$, there is an algorithm that generates all traces of $p$ in polynomial time. For example, let $p_{2}$ be the process shown in  Figure~\ref{fig:exp_chip}. The characteristic formula for $p_{2}$ within $\mathL_S$ is $\chi(p_{2})=\langle a \rangle(\langle a \rangle \true \wedge \langle b \rangle \true)\wedge\langle b \rangle(\langle a \rangle \true \wedge \langle b \rangle \true)$. 
In general, consider processes $p_{n}$, $n\in\mathbb{N}$, defined inductively as follows: $p_0=\mathtt{0}$ and for every $n>0$, $p_n=a.p_{n-1}+b.p_{n-1}$.
We have that $|\chi(p_{n})|=2^n\cdot l_0+2^{n-1}\cdot 3+3$, where $l_0=|\chi(p_0)|=|\true|=1$. Lemma~\ref{lem:generate-all-traces} guarantees that all traces of $p_{n}$ can be generated in polynomial time with respect to $|\chi(p_{n})|$, which is exponential in $|p_{n}|$.
\begin{figure}
    \centering
    \includegraphics[scale=1]{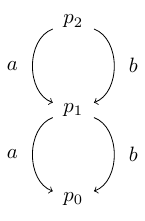}
    \caption{Process $p_2$ for which $\langle a \rangle(\langle a \rangle \true \wedge \langle b \rangle \true)\wedge\langle b \rangle(\langle a \rangle \true \wedge \langle b \rangle \true)$ is characteristic within $\mathL_S$.}
    \label{fig:exp_chip}
\end{figure}
\end{remark}


\findingcharhard*
\begin{proof}
Suppose that there is an algorithm $\mathcal{A}$ that finds the characteristic formula within $\mathL_S$ for a given finite loop-free process in polynomial time. We show that trace equivalence between finite loop-free processes can be solved in polynomial time, which implies $\cP=\NP$ by Propositions~\ref{prop:conp-complete-under-turing} and~\ref{prop:trace-complexity-b}. Given an input pair of finite loop-free processes $(p,q)$ for the problem of deciding trace equivalence, consider algorithm $\mathcal{B}$ which finds the characteristic formulae $\chi(p)$ and $\chi(q)$ for $p$ and $q$, respectively, by using $\mathcal{A}$ as a subroutine. Then, it generates the traces of $p$ and $q$ in polynomial time w.r.t.\ $|\chi(p)|$ and $|\chi(q)|$, respectively, as described in Lemma~\ref{lem:generate-all-traces}, and checks whether $\mathrm{traces}(p)=\mathrm{traces}(q)$. It is not hard to see that $\mathcal{B}$ decides trace equivalence of $p$ and $q$ in polynomial time. The proof is analogous in the case of $X=CS$ and $X=RS$ with a bounded action set.
\end{proof}

However, we prove in Proposition~\ref{prop:find-char-complexity-easy} that if the output of the problem is given in declarative form, then finding characteristic formulae within $\mathL_X$, $X\in\{CS,RS,BS,nS\}$, $n\geq 1$, can be done in polynomial time. The proof is based on  Propositions~\ref{prop:find-characteristic} and~\ref{prop:char-formula-3S} and Definition~\ref{def:char-formula-3S}.

\begin{proposition}[\cite{IngolfsdottirGZ87,AcetoILS12}]\label{prop:find-characteristic}
     Given a finite loop-free process $p$, the characteristic formula within $\mathL_X$, $X\in\{S,CS,RS,2S,BS\}$, denoted by $\chi_X(p)$, is inductively defined as follows:
     \vspace{2mm}
     \begin{enumerate}\setlength\itemsep{2mm}
         \item $\mathL_S$: $\displaystyle\chi_S(p)= \bigwedge_{a\in A}~\bigwedge_{p\myarrowa p'}\langle a\rangle \chi_S(p')$.
          \item $\mathL_{CS}$:
           $\displaystyle\chi_{CS}(p)=\begin{cases}
           \zero, &\text{if } p\not\rightarrow\\
            \displaystyle\bigwedge_{a\in A}~\bigwedge_{p\myarrowa p'}\langle a\rangle \chi_{CS}(p'), &\text{otherwise}\end{cases}$.
          \item $\mathL_{RS}$:
            $\displaystyle\chi_{RS}(p)= \bigwedge_{a\in A}~\bigwedge_{p\myarrowa p'}\langle a\rangle \chi_{RS}(p')\wedge \bigwedge_{a\in A: p\notmyarrowa} [a]\ff$.
            \item $\mathL_{2S}$:
            $\displaystyle\chi_{2S}(p)=\bar{\chi}_S(p)\wedge\bigwedge_{a\in A}~\bigwedge_{p\myarrowa p'} \langle a \rangle\chi_{2S}(p')$, where
                $\displaystyle\bar{\chi}_{S}(p)=\bigwedge_{a\in A}[a]~\bigvee_{p\myarrowa p'} \bar{\chi}_S(p')$.
             \item $\mathL_{BS}$: $\displaystyle\chi_{BS}(p)=\bigwedge_{p\myarrowa p'}\langle a\rangle\chi_{BS}(p')\wedge \bigwedge_{a\in A}[a]\bigvee_{p\myarrowa p'}\chi_{BS}(p')$.
     \end{enumerate}
\end{proposition}

Similarly to Proposition~\ref{prop:find-characteristic}(4), we define $\chi_{3S}(p)$ for a  finite loop-free process $p$.

\begin{definition}\label{def:char-formula-3S}
   Given a finite loop-free process $p$, formula $\chi_{3S}(p)$ is inductively defined as follows: $$\displaystyle\chi_{3S}(p)=\bar{\chi}_{2S}(p)\wedge \bigwedge_{a\in A}\bigwedge_{p\myarrowa p'} \langle a\rangle \chi_{3S}(p'), \text{ where }\displaystyle\bar{\chi}_{2S}(p)=\bigwedge_{a\in A} [a]\big(\bigvee_{p\myarrowa p'} \bar{\chi}_{2S}(p')\big) \wedge \chi_S(p).$$
\end{definition}

We prove that $\chi_{3S}(p)$ is characteristic for $p$ within $\mathL_{3S}$ below.

\begin{lemma}\label{lem:bar-chi-2S}
    For every  finite loop-free process $q$, $q\models \bar{\chi}_{2S}(p)$ iff $q\curle_{2S} p$.
\end{lemma}
\begin{proof} We prove both implications on $\depth(q)$.\\
    ($\Rightarrow$) Assume that $q\models\bar{\chi}_{2S}(p)$ and $q\myarrowa q'$. Then, since $q\models\bar{\chi}_{2S}(p)$, we have that $q'\models\bigvee_{p\myarrowa p'} \bar{\chi}_{2S}(p')$. Thus, there is some $p'$ such that $p\myarrowa p'$ and $q'\models \bar{\chi}_{2S}(p')$. By inductive hypothesis, $q'\curle_{2S} p'$, which also implies that $q'\curle_S p'$. This means that $q\curle_S p$. Moreover, since $q\models \bar{\chi}_{2S}(p)$, we have that $q\models\chi_S(p)$, and so $p\curle_S q$. Consequently, (a) for every $q\myarrowa q'$, there is $p\myarrowa p'$ such that $q'\curle_{2S} p'$ and (b) $p\equiv_S q$, which implies that $q\curle_{2S} p$.\\
    ($\Leftarrow$) Assume that $q\curle_{2S} p$ and $q\myarrowa q'$. Since $q\curle_{2S} p$, there is $p\myarrowa p'$ such that $q'\curle_{2S} p'$. By inductive hypothesis, $q'\models\bar{\chi}_{2S}(p')$. As a result, $q\models\bigwedge_{a\in A} [a]\bigvee_{p\myarrowa p'} \bar{\chi}_{2S}(p')$. Moreover, since $q\curle_{2S} p$, we have that $p\curle_S q$, and so $q\models\chi_S(p)$. Consequently, $q\models\bar{\chi}_{2S}(p)$.
    \end{proof}

\begin{proposition}\label{prop:char-formula-3S}
 Let $p$ be a  finite loop-free process. Then, $\chi_{3S}(p)$ is characteristic for $p$ within $\mathL_{3S}$.
\end{proposition}
\begin{proof}
    We show that for every  finite loop-free process $q$,   $p\curle_{3S} q$ iff $q\models\chi_{3S}(p)$. We prove both implications by induction on $\depth(q)$.\\
    ($\Rightarrow$) We first show that $q\models\langle a\rangle\chi_{3S}(p')$ for every $p\myarrowa p'$. Let $p\myarrowa p'$. Since $p\curle_{3S} q$, there is $q\myarrowa q'$ such that $p'\curle_{3S} q'$. By inductive hypothesis, $q'\models \chi_{3S}(p')$, so $q\models\langle a\rangle\chi_{3S}(p')$ and we are done. Second, we show that $q\models\bar{\chi}_{2S}(p)$. Since $p\curle_{3S} q$, we have that $q\curle_{2S} p$. By Lemma~\ref{lem:bar-chi-2S}, $q\models\bar{\chi}_{2S}(p)$.\\
    ($\Leftarrow$) Assume that $q\models\chi_{3S}(p)$. Reasoning as above, we see that $p\curle_{3S} q$ by examining the two types of conjuncts if $\chi_{3S}(p)$.
\end{proof}

We can extend the results for preorder $3S$ to every preorder $nS$, where $n\geq 3$, as follows.

\begin{proposition}
    Given a finite loop-free process $p$, the characteristic formula within $\mathL_{nS}$, where $n\geq 3$, denoted by $\chi_{nS}(p)$, can be inductively constructed as follows:
    $$\displaystyle\chi_{nS}(p)=\bar{\chi}_{(n-1)S}(p)\wedge \bigwedge_{a\in A}\bigwedge_{p\myarrowa p'} \langle a\rangle \chi_{nS}(p'), \text{ where }$$
    $$\displaystyle\bar{\chi}_{(n-1)S}(p)=\bigwedge_{a\in A} [a]\big(\bigvee_{p\myarrowa p'} \bar{\chi}_{(n-1)S}(p')\big) \wedge \chi_{(n-2)S}(p).$$
\end{proposition}


\begin{remark}
Proposition~\ref{prop:find-char-complexity-easy}  implies alternative polynomial-time algorithms for preorder checking for preorders $CS, RS, BS,$ and $nS$, where $n\geq 1$. From Proposition~\ref{model-checking-complexity-decl}, model checking is in $\mathcal{O}(|p|\cdot \mathrm{decl}(\varphi))$, so the reduction described in Proposition~\ref{prop:preorder-char-reduction}(a) is of polynomial time complexity. Other examples of polynomial-time algorithms for deciding preorders (or equivalence relations) by computing characteristic formulae and applying a model-checking algorithm can be found in~\cite{SteffenI94,HS96}. 
\end{remark}

\subsection{The trace simulation preorder}


We provide the proof of Proposition~\ref{prop:find-char-ts-hard}.

\tsfindcharhard*
\begin{proof}
From Proposition~\ref{prop:preorder-char-reduction}(a), given two finite loop-free processes $p$ and $q$, we can decide $p\curle_{TS} q$ by finding  $\varphi_p$ of the proposition and checking whether $q\models \varphi_p$. If the hypotheses of the proposition are true, then this reduction requires polynomial time, which combined with Propositions~\ref{prop:conp-hardness-trace-sim} and~\ref{prop:conp-complete-under-turing}, implies $\cP=\NP$. 
\end{proof}

We give an example of a family  of processes, namely $(p_i)_{i\in\mathbb{N}}$, such  that  $p_i$ has a characteristic formula $\phi_i$ within $\mathL_{TS}$ such that $\mathrm{decl}(\phi_{i})$ is linear and $\eqlen(\phi_{i})$ is exponential in $|p_i|$.

\begin{example}
    Consider process $p_2$ from Figure~\ref{fig:exp_chip} with the only difference that here $A=\{a,b,c\}$. A characteristic formula  within $\mathL_{TS}$ for $p_2$ is the one given by the equations: 
    $$\begin{aligned}
        &\phi_2 =\langle a\rangle \phi_1\wedge \langle b\rangle \phi_1\wedge [c]\ff \wedge [a][c]\ff \wedge [b][c]\ff \wedge [a][a][c]\ff \wedge [a][b][c]\ff \wedge [b][a][c]\ff \wedge [b][b][c]\ff\\
        &\phi_1 =\langle a\rangle \phi_0\wedge \langle b\rangle \phi_0\wedge [c]\ff \wedge [a][c]\ff \wedge [b][c]\ff 
        \\
        &\phi_0 = [a]\ff \wedge [b]\ff \wedge [c]\ff.
    \end{aligned}$$
    In general, for the process $p_n$ defined in Remark~\ref{trace-example} with the only difference that $A=\{a,b,c\}$, there is a characteristic formula $\phi_n$ with $\mathrm{decl}(\phi_n)=n+1$ and $\eqlen(\phi_n)=(4+2^0\cdot 2+2^1\cdot 3+\dots+2^n\cdot(n+2))+2^{n+1}$, where $4$ corresponds to the symbols  $\langle a\rangle \phi_{n-1}$ and $\langle b\rangle \phi_{n-1}$, $2^0\cdot 2+2^1\cdot 3+\dots+2^n\cdot(n+2)$ is the number of symbols of all subformulae in $\phi_n$ starting with $[act]$ for some $act\in\{a,b,c\}$, and $2^{n+1}$ is the number of $\wedge$ symbols in $\phi_n$.
\end{example}

For any process $p$, let $\mathrm{ExcTraces}(p)$ be the formula that describes which traces do not belong to $\mathrm{traces}(p)$. This formula can be written in the language given by the following grammar:
$$\phi_{ET}::= \ff ~ \mid ~ [a]\phi_{ET} ~ \mid ~ 
 \phi_{ET}\wedge\phi_{ET} .$$

For example, let $A=\{a,b\}$ and $p$ be a process with $\mathrm{traces}(p)=\{\varepsilon ,a,ab,b\}$. Then, $\mathrm{ExcTraces}(p) = [b][a]\ff\wedge [b][b]\ff \wedge [a][a]\ff\wedge [a][b][a]\ff\wedge [a][b][b]\ff$. 

\begin{proposition}\label{prop:find-characteristic-TS}
    Given a finite loop-free process $p$, the characteristic formula $\chi(p)$ within $\mathL_{TS}$ can be inductively constructed as follows:
      $\displaystyle\chi(p)= \bigwedge_{a\in A}~\bigwedge_{p\myarrowa q}\langle a\rangle \chi(q)\wedge \mathrm{ExcTraces}(p)$.
\end{proposition}

\begin{corollary}\label{cor:find-char-complexity-TS}
There is an algorithm that given a finite loop-free process $p$, outputs a formula $\varphi$ in declarative form that is characteristic within $\mathL_{TS}$ for $p$ and has polynomial declaration size and exponential equational length in $|p|$. 
\end{corollary}
\begin{proof}
    Immediate from Proposition~\ref{prop:find-characteristic-TS}.
\end{proof}

\section{A note on deciding characteristic formulae modulo equivalence relations}\label{section:decide-char-mod-equiv-relations-appendix}

The reader can find the full proof of Theorem~\ref{thm:reduction-validity-char} below.

\reduction*
\begin{proof}
Given $\varphi\in\compL_X$, where $X\in\{CS,RS,TS,2S,3S,BS\}$, we construct a formula $\varphi'\in\mathL_X$ such that $\varphi$ is valid if and only if  $\varphi'$ is characteristic modulo $\equiv_X$ for some $p$. Let $\varphi_{ch}$ be a characteristic formula modulo $\equiv_X$ for process $p_{ch}$ and $\varphi_{nch}\in\mathL_X$ be a formula that is not characteristic modulo $\equiv_X$. For each $X\in\{CS,RS,TS,2S\}$, the formulae $\varphi_{ch}$ and $\varphi_{nch}$ exist by Proposition~\ref{prop:S-CS-RS-equiv-char} and the discussion in the paragraph following its proof. In the same way, we can show that characteristic formulae modulo $\equiv_{3S}$ or $\equiv_{BS}$ exist. Given $\varphi\in\compL_X$, it can be determined in linear time whether $p_{ch}\models\varphi$. We distinguish the following two cases:
\begin{itemize}
    \item Assume that $p_{ch}\not\models\varphi$. Then $\varphi$ is not valid and we set $\varphi'=\varphi_{nch}$.
    \item Assume that $p_{ch}\models\varphi$. In this case, we set $\varphi'=\neg \varphi \vee \varphi_{ch}$. We show that $\varphi$ is valid if and only if $\varphi'$ is characteristic modulo $\equiv_X$ for some process $p$. 
    \begin{itemize}
        \item For the implication from left to right, assume that $\varphi$ is valid. Then $\neg \varphi \vee \varphi_{ch}$ is equivalent to $\varphi_{ch}$, which is characteristic for $p_{ch}$ modulo $\equiv_X$.
        \item Conversely, assume that $\neg \varphi \vee \varphi_{ch}$ is characteristic for some process $p$ modulo $\equiv_X$. Let $q$ be any process. We show that $q\models \varphi$ and therefore that $\varphi$ is valid. If $p_{ch}\equiv_X q$, then $\mathL_X(p_{ch})=\mathL_X(q)$ and, since $p_{ch}\models \varphi$ by assumption, it holds that $q\models\varphi$. Suppose now that $p_{ch}\not\equiv_X q$.  Since we have that $p_{ch}\models \neg\varphi\vee \varphi_{ch}$ and $\neg\varphi\vee \varphi_{ch}$ is characteristic for $p$ modulo $\equiv_X$, it follows that $\varphi_{ch} \equiv\neg\varphi\vee \varphi_{ch}$.  Therefore, $q\not\models\neg\varphi\vee \varphi_{ch}$, which implies that $q\models\varphi$, and we are done.\qedhere
    \end{itemize}
\end{itemize}
\end{proof}

\end{document}